\documentclass[aps,pra,onecolumn,showpacs,preprintnumbers,amsmath,amssymb,nofootinbib,notitlepage]{revtex4-1}

\usepackage{aas_macros}
\usepackage{mathrsfs}
\usepackage{bm}
\usepackage{xcolor}
\usepackage{tikz}
\usepackage{pgfplots}
\usepackage{txfonts}
\usepackage[breaklinks=true,colorlinks,citecolor=blue,linkcolor=blue,urlcolor=blue]{hyperref}

\allowdisplaybreaks[1]
\newcommand{\dd}{\mathrm{d}}

\begin{document}
    
\title{Analytical ray-tracing in planetary atmospheres}
    
\author{A.~Bourgoin, M. Zannoni, P. Tortora}
\affiliation{Dipartimento di Ingegneria Industriale, University of Bologna, via fontanelle 40, Forl\`i, Italy}
\email{adrien.bourgoin@unibo.it}
    
\date{\today}
    
\begin{abstract}
\emph{Context. }Ground-based astro-geodetic observations and atmospheric occultations, are two examples of observational techniques requiring a scrutiny analysis of atmospheric refraction. In both cases, the measured changes in observables (range, Doppler shift, or signal attenuation) are geometrically related to changes in the photon path and the light time of the received electromagnetic signal. In the context of geometrical optics, the change in the physical properties of the signal are related to the refractive profile of the crossed medium. Therefore, having a clear knowledge of how the refractivity governs the photon path and the light time evolution is of prime importance to clearly understand observational features. Analytical studies usually focused on spherically symmetric atmospheres and only few aimed at exploring the effect of the non-spherical symmetry on the observables.

\emph{Aims. }In this paper, we analytically perform the integration of the photon path and the light time of rays traveling across a planetary atmosphere. We do not restrict our attention to spherically symmetric atmospheres and introduce a comprehensive mathematical framework which allows to handle any kind of analytical studies in the context of geometrical optics.

\emph{Methods. }Assuming that the index of refraction of the medium is a linear function of the Newtonian potential, we derive an exact solution to the equations of geometrical optics. The solution's arbitrary constants of integration are parametrized by the refractive response of the medium to the gravitational potential, and by the first integrals of the problem. Varying the constants, we are able to reformulate the equation of geometrical optics into a new set of osculating equations describing the constants' evolution for any arbitrary changes in the index of refraction profile.

\emph{Results. }The osculating equations are identical to the equation of geometrical optics, however, they offer a comprehensive framework to handle analytical studies which aim at exploring non-radial dependencies in the refractive profile. To highlight the capabilities of this new formalism, we carry out five realistic applications for which we derive analytical solutions. The accuracy of the method of integration is assessed by comparing our results to a numerical integration of the equations of geometrical optics in the presence of a quadrupolar moment $J_2$. This shows that the analytical solution leads to the determination of the light time and the refractive bending with relative errors at the level of one part in $10^{8}$ and one part in $10^{5}$, for typical values of the refractivity and the $J_2$ parameter at levels of $10^{-4}$ and $10^{-2}$, respectively. 
\end{abstract}

\maketitle

\section{Introduction}
\label{sec:intro}

In the context of Maxwell's theory of electromagnetism, the refraction phenomenon is understood as the superposition of an incident electromagnetic wave with wavelets generated by some electric particles being accelerated in response to a local electromagnetic field. If the medium containing the particles is sufficiently tenuous (like for a gas), the local field is proportional to the incident field itself. The waves' superposition generates a resulting signal which exhibits a phase shift with respect to the incident one. This apparent change in the primary signal's phase is parametrized thanks to a dimensionless parameter called the index of refraction of the medium. It encloses the physical properties related to the interaction between the electromagnetic signal and the material filling the medium. Therefore, any variation of the physical properties of the medium will change the index of refraction which in turn will modify the features of the incident signal. In Astronomy, those changes in the signal's characteristics are the observables, thus, it is of prime importance to understand the relationship between refraction and light propagation.

Ground-based (GB) astro-geodetic observations and atmospheric occultations (AO) are two examples of observational techniques requiring a careful analysis of refraction. For GB observations, refraction is an accuracy-limiting factor and it has to be modeled in the analysis software; for AO, it is thoroughly modeled to infer the physical properties of the refractive medium. In both cases, the refraction usually occurs at characteristic lengths which are much larger than the wavelength of the electromagnetic signal and the problem can be studied in the approximation of geometrical optics. In that context, the different observables (range, Doppler, or attenuation) can be related to the geometry of the photon path and the light time which are both controlled by the index of refraction. Thus, a clear knowledge of how the geometry of the light propagation is related to the index of refraction is important to understand the changes in observables.

For GB observations, especially for techniques operating for the realization of the International Earth Rotation and Reference Systems Service (IERS), the accuracy of the analysis is largely affected by errors in modeling the group delay during propagation of the signal through the atmosphere \cite{2010ITN....36....1P}. Usually, the atmospheric delay is evaluated at zenith and is computed with an analytical model of \cite{1973_Mar_Mur} or with the more recent model (also analytic) of \cite{2004GeoRL..3114602M}. Then, the projection to a given elevation angle is operated with the help of the mapping function developed by \cite{2002GeoRL..29.1414M}. Those models are formulated under the assumption that the atmosphere is spherically symmetric and assume a radial dependency for the index of refraction. Consequently, they leave aside some possible effects due to the presence of horizontal gradients in the atmosphere\footnote{Those horizontal gradients can arise at two different levels. Firstly, they can be generated by an atmosphere with a non-spherical shape, and secondly, they can be due to a local horizontal variation of the physical quantities entering in the computation of the index of refraction e.g. the temperature.}, which might be somehow inaccurate. This issue has only been partially overcome by \cite{1997JGR...10220489C} for Very Long Baseline Interferometry (VLBI), and by \cite{doi:10.1029/2006JB004834} for Satellite Laser Ranging (SLR), by performing numerical integration (numerical ray-tracing) across multi-layered spherical atmosphere, assuming constant refractivity inside each layer. Even if in those cases the index of refraction may possess non-radial dependencies, the shape of the atmosphere is still considered for being spheric. So, it is worth mentioning that beyond the radial dependency of the refractive index, only numerical integrations are usually used for such GB observations.

In AO experiments, the basic idea 	is to establish radio links between a transmitter and a receiver when the latter is being occulted by a planetary atmosphere as seen from the transmitter. Conversely to GB observations, AO experiments use the effect of the refraction by the atmosphere to determine the physical properties of the medium. This is often referred to as the inverse problem. In the literature, there exists different approaches devoted to the resolution of the inverse problem. For instance, the initial data processing of the Mariner~IV occultations was based primarily on model fitting \cite{1965Sci...149.1243K}. The model used for determining the index of refraction profile was approximate and only valid for a planet with a thin spherically symmetric atmosphere \cite{1965JGR....70.3217F,1968P&SS...16.1035F}. Later, it as been shown \cite{1968JGR....73.1819P} that determining directly the refractivity profile from the Doppler frequencies (or from the bending angle) is a special case of Abelian integral inversion when the index of refraction is assumed to be driven by spherical symmetry. In the past, this method has been successfully applied to the dense atmosphere of Venus \cite{1971AJ.....76..123F} using the data of the occultation of Mariner~V. More recently, Abelian integral inversion was used to process radio data of Cassini occultation by Titan \cite{2011Icar..215..460S,2012Icar..221.1020S} or GPS/MET (Global Positioning System Meteorology Experiment) occultation by the Earth atmosphere \cite{1997JGR...10223429K,1999AnGeo..17..122S}. Later, with the Voyager 2 occultation by Neptune in 1989 and the more recent Cassini mission orbiting Saturn, the problem of studying non-spherically symmetric atmosphere was raised. A first analytical model was proposed by \cite{1992AJ....103..967L}. This solution based on a Taylor series expansion is suited for a numerical evaluation, but does not provide a clear understanding of the effect of non-spherical symmetry on the photon path and the time-delay. More recently, in order to account for additional effects such the light dragging effect by the winds, \cite{2015RaSc...50..712S} proposed a purely numerical ray-tracing technique to solve the inverse problem. This method is of course more general than analytical models but necessitates a much higher computational cost.% Also, only numerical ray-tracing techniques are currently employed beyond the spherical symmetry. 

In this context, we propose a fully analytical and comprehensive description of the light path inside planetary atmospheres. The solution is expressed in terms of a free parameter $\alpha$, which characterizes the refractive response of the medium to the gravitational pull of the planet. This parameter is either an input for GB observations or an output for AO experiments. The geometric properties of the ray (as the photon path and the time-delay) are expressed in term of $\alpha$ and in return the observables, which are related to the light geometry, can also be expressed in terms of this parameter. In this paper, we present a useful mathematical framework which allows to perform analytical studies beyond the usual spherically symmetric case. The validity of the derived solution is assessed by comparing it with numerical ray-tracing performed on a simulated atmosphere. We try to keep the discussion as general as possible in order to let the possibility of applying the incoming description to any situation involving light propagation in planetary atmospheres e.g. GB observations or AO experiments.

\section{Definitions and hypothesis}

In this paper, we use the geometrical optics approximation which is characterized by neglecting the wavelength of electromagnetic waves \cite{1975ctf..book.....L}. In geometrical optics, the concept of rays is introduced as curves whose tangents at each point coincide with the direction of propagation of the electromagnetic waves. In the following, we will mainly focus on the evolution of the separation vector, locating a point along the light ray, and the unit vector tangent to the ray at that point. In this chapter, we introduce the basic principles for determining the evolution of these quantities.

\subsection{Equations of geometrical optics}
\label{sec:eq_optics}

Let us introduce \textbf{h}, the separation vector between the center of a reference frame and a point along the light ray trajectory. Also, we define \textbf{s}, the unit vector which is tangent to the light ray at \textbf{h}'s location
\begin{equation}
\label{eq:s}
\mathbf{s}\equiv \frac{\dd\mathbf{h}}{\dd s}\text{.}
\end{equation}
$\dd\mathbf{h}$ is an elemental displacement vector along the ray and $\dd s$ represents its length, which is defined as $\dd s^2=\dd\mathbf{h}\cdot\dd\mathbf{h}$. Because of this quadratic relation, one can infer that \textbf{s} is a unit vector, and this reduces the number of independent constants of integration from six to five. Together, \textbf{h} and \textbf{s} provide the description of the evolution of the separation vector as well as the direction of the ray.

For convenience, we introduce the new following quantity
\begin{equation}
\label{eq:S}
\bm\nabla\mathscr{S}\equiv n\textbf{s}
\end{equation}
where $\mathscr{S}(\textbf{h})$ is a real scalar function of the position and is usually defined as the \emph{optical path} \cite{1999prop.book.....B}. Here, the gradient is computed with respect to $\textbf{h}$, and $n(\textbf{h})$ is an arbitrary scalar field called the index of refraction of the medium. Remembering that $\textbf{s}$ is a unit vector, the square of the gradient of the optical path leads to the well-known \emph{eikonal equation}
\begin{equation}
\label{eq:eikonal_eq}
\bm\nabla\mathscr{S}\cdot\bm\nabla\mathscr{S}=n^2\text{,}
\end{equation}
which is the fundamental equation of geometrical optics. The function $\mathscr{S}$ is also referred to as the \emph{eikonal}.

The height of the ray, its direction of propagation or the eikonal, changes in space according to the spatial variation of the index of refraction of the medium in which the light ray propagates. Thus, it is possible to directly specify the ray with the index of refraction by differentiating Eq. \eqref{eq:S} with respect to $s$, which leads to \cite{1999prop.book.....B}
\begin{equation}
\label{eq:optics}
\frac{\dd}{\dd s}\big(\bm\nabla\mathscr{S}\big)=\bm\nabla n\text{.}
\end{equation}
This expression is given in a reference frame at rest with respect to the refractive medium. Eq. \eqref{eq:optics} can only be integrated after having introduced some hypotheses about the dependencies of $n$. In the following, we will refer to Eq. \eqref{eq:optics} as the equation of geometrical optics.

In order to assess a time description along the path, we introduce an additional well-known formula describing the rate of change of the light time with respect to the geometrical length
\begin{equation}
\label{eq:dtds}
\frac{\dd t}{\dd s}=\frac{n}{c}\text{.}
\end{equation}
The introduction of this additional equation, leads the total number of independent constants of integration from five to six.

Eqs.~\eqref{eq:s}, \eqref{eq:optics}, and \eqref{eq:dtds} are the equations that we have to deal with in order to solve simultaneously the photon path and the light time which in turn can be used to determine the observables, as shown in the next section.

\subsection{Observables given by the geometry}
\label{subsec:obs}

We recall how the observables like the range, the range-rate (Doppler shift), or also the defocussing can be determined once $\bm\nabla\mathscr{S}$ and $t$ are known. Therefore, this discussion highlights the great benefit of having analytical expressions describing the evolution of the tangent to the ray or the light time.

The range is directly given by integration of Eq. \eqref{eq:dtds}. From it, we deduce the group delay expression which is the time-delay due to atmospheric effects
\begin{equation}
\label{eq:time_delay_atm}
\tau_{\text{atm}}=\frac{1}{c}\int_{s_1}^{s_2}n\dd s-t_{\text{vac}}\text{.}
\end{equation}
Here, the integral represents the total light time accounting for the atmospheric effects and $t_{\text{vac}}$ is the light time in vacuum. The integration is performed along the light path, from the transmission point (labeled 1) to the receiving point (labeled 2).

The instantaneous relativistic Doppler shift derives from the fact that the ratio between the received and the transmitted frequencies can be expressed as \cite{SyngeBookGR,2008ASSL..349..153T}
\begin{equation}
\label{eq:doppler_instant}
\frac{\nu_2}{\nu_1}=\frac{(u^{\alpha}k_{\alpha})_2}{(u^{\alpha}k_{\alpha})_1}=\frac{(u^0k_0)_2}{(u^0k_0)_1}\frac{(1+\hat{u}^i\hat{k}_{i})_2}{(1+\hat{u}^i\hat{k}_i)_1}\text{,}
\end{equation}
where $\nu_1$ and $\nu_2$ are the transmitted and the received frequencies respectively, $u^{\alpha}$ denotes the 4-velocity vector, $k_{\alpha}$ is the covariant form of the 4-wave vector which is tangent to light ray, and $\hat{u}^i$ and $\hat{k}_i$ are respectively given by $\hat{u}^i=u^i/u^0$, and $\hat{k}_i=k_i/k_0$ (see \cite{2008ASSL..349..153T,2014PhRvD..89f4045H} for definitions and notations). In this expression, the subscripts 1 and 2 specify that the terms between parenthesis must be evaluated at the position of the transmitter and the receiver respectively.

We can specify the frequency ratio in term of $\bm\nabla\mathscr{S}$ by following e.g. \cite{1999PhRvA..60.4301L}. Indeed, for a static medium, it can be shown that
\begin{equation}
\label{eq:wave_vector}
%\hat{\textbf{k}}=\bm\nabla\mathscr{S}\text{.}
\hat{k}^i=\partial^i\mathscr{S}\text{.}
\end{equation} 
Using the flat Minkowski metric\footnote{Without loss of generality for the discussion, we neglect general relativity effects and focus our attention on the refraction only. If needed, a more generic treatment of the Doppler shift in general relativity can be found in \cite{SyngeBookGR}.} with the signature $(+,-,-,-)$, we have $k_i=-k_0\hat{k}^i$ and $u^0=\Gamma$, where $\Gamma$ is the Lorentz factor of special relativity which is defined as $\Gamma\equiv (1-\hat{\textbf{u}}\cdot\hat{\textbf{u}})^{-1/2}$. In addition, it turns out that $k_0$ is constant along null geodesic for that space-time metric, then Eq. \eqref{eq:doppler_instant} becomes
\begin{equation}
\label{eq:doppler_instant_sv}
\frac{\nu_2}{\nu_1}=\frac{\Gamma_2}{\Gamma_1}\frac{\Big(1-\hat{\textbf{u}}\cdot\bm\nabla\mathscr{S}\Big)_2}{\Big(1-\hat{\textbf{u}}\cdot\bm\nabla\mathscr{S}\Big)_1}\text{.}
\end{equation}

Finally, one can introduce another observable, namely the attenuation of the signal. The expression for the change in intensity, in dB (also called the attenuation), due to the refractive defocussing, is given by \cite{Kliore2004} as
\begin{equation}
\label{eq:defocus}
A_{\text{def}}=10\log\left(1-S\frac{\dd\epsilon}{\dd K}\right)\text{,}
\end{equation}
where $S$ is the distance from the transmission till the closest approach, $K$ is the impact parameter of the light beam trajectory, and $\epsilon$ is the refractive bending experienced by the ray between the transmission and the reception\footnote{If $\epsilon$ can be derived from the cross product between $\bm{\nabla}\mathscr S_2$ and $\bm{\nabla}\mathscr S_1$ as $\epsilon=\arcsin(|\bm{\nabla}\mathscr{S}_2\times\bm{\nabla}\mathscr{S}_1|/n_1n_2)$, in the following, we will use an alternative way for computing it.}.	

Those three quantities [cf. Eqs. \eqref{eq:time_delay_atm}, \eqref{eq:doppler_instant_sv}, and \eqref{eq:defocus}] are three examples of observables that can directly be determined knowing the evolution of $\bm\nabla\mathscr{S}$ and $t$ between the transmission and the reception points. However, as seen in Eqs.~\eqref{eq:s}--\eqref{eq:dtds}, they can only be determined after specifying the index of refraction's dependencies.

\subsection{Refractive profile dependency}
\label{subsec:ref_prof_dep}

Except in Sec. \ref{sec:varconst} where we remain as general as possible, in this work we will consider that the index of refraction coincides with the surface of constant $\Phi$, where $\Phi$ is a generalized potential which will be defined later in Sec.~\ref{sec:applications}.

This hypothesis is motivated by the hydrostatic equilibrium assumption applied to a body which is made up with fluid elements. Indeed, assuming a non-rotating fluid body, it can be demonstrated (cf. p.~91--92 of \cite{2014grav.book.....P}) that the surfaces of constant density, pressure, and gravitational potential all coincide. This result can even be extended to a steady rotating body as shown in \cite{2014grav.book.....P}. However in that case, if surfaces of constant density and pressure still coincide together they do not coincide anymore with surfaces of constant gravitational potential, $U$. Instead, they coincide with the level surfaces of a more generalized potential $\Phi=U-\Phi_C$, where $\Phi_C$ is known as the centrifugal potential.

So, if we now assume that the index of refraction is related to the spatial distribution of the fluid elements (e.g. in the context of elementary theory of dispersion \cite{1999prop.book.....B}), it becomes obvious that the surface of constant $n$ also coincide with the surfaces of constant $\Phi$. Such a dependence between $n$ and $\Phi$ has also been used recently in \cite{2015RaSc...50..712S} to analyze the radioscience data of Cassini in the context of occultations by Saturn's atmosphere.

In order to simplify future notations, we introduce
\begin{equation}
\label{eq:dn/dPhi}
\alpha\equiv-\frac{\dd n}{\dd \Phi}
\end{equation}
which possesses the same dimensions as the inverse of a potential ($L^{-2}T^2$).

In the context of GB observations, $\alpha$ must be seen as an input parameter describing the variation of the index of refraction in different layers of the atmosphere. Conversely, in the context of AO experiments it is seen as an output parameter which must be determined following an accurate modeling of the observable. In general, $\alpha$ is not a constant parameter and several measurements are needed through the atmosphere in order to integrate Eq. \eqref{eq:dn/dPhi}, as it is discussed in \cite{2015RaSc...50..712S}.

\subsection{Presentation of the work}

In this work, we introduce a mathematical framework which provides a comprehensive description of the light ray trajectory inside planetary atmospheres using geometrical optics.

In Chap. \ref{sec:ref}, we derive a simple solution to Eq.~\eqref{eq:optics} assuming that the index of refraction of a planetary atmosphere is a linear function of the monopole term of the gravitational potential of the central planet. In other words, we assume that $\alpha$ [see Eq.~\eqref{eq:dn/dPhi}] is constant, which is a reasonable assumption between each layer of a spherically symmetric atmosphere. This solution is then referred to as the \emph{reference solution}. It is given in terms of constant parameters called \emph{hyperbolic elements} (in reference to \emph{elliptic elements} of celestial mechanics) which describe together the shape and the spatial orientation of the photon path as well as the time-delay.

In Chap.~\ref{sec:varconst}, we derive a comprehensive mathematical framework which allows to study generic refractive profiles. It is obtained using the reference solution derived in Sec.~\ref{sec:ref} as an osculating solution in order to turn Eqs.~\eqref{eq:s}, \eqref{eq:optics}, and \eqref{eq:dtds} into a set of \emph{oscultating equations} describing the rate of change of the hyperbolic elements following a change in the index of refraction profile. The osculating equations (also called the perturbation equations) are perfectly equivalent to the equation of optics, therefore, in that sense, they can used in place of Eqs.~\eqref{eq:s}, \eqref{eq:optics}, and \eqref{eq:dtds}. The main interest is that they are well suited for analytical study of light propagation inside planetary atmospheres.

In Chap.~\ref{sec:applications}, we find approximate solutions to the perturbation equations for a large class of additional contributions in the refractive profile of the central planet's atmosphere. First, we study the effects on the hyperbolic elements due to non-linear dependencies of the refractive profile to the generalized potential. We focus on the limiting case where the generalized potential reduces to the monopole term of the gravitational potential of the central planet (this reduces to the spherically symmetric assumption). Then, we assume that besides the monopole term there exists an additional contribution to the refractive profile, which is simultaneously the centrifugal potential, the axisymmetric part of the gravitational potential of the central planet, and finally an external gravitational potential raised by a perturbing body. 
We also spotlights the versatility offered by the formalism by studying the light dragging effect caused by a rotating medium. All those perturbations are mainly related to the shape of the atmosphere. The last application concerns the effect due to horizontal gradients\footnote{Physically, such gradients may be generated e.g. by horizontal changes in temperature.} inside a spherically symmetric atmosphere for a specific application to GB observations. 

Finally in Chap. \ref{sec:num_sim}, we compare the analytical results obtained for the axisymmetric part of the gravitational potential with a solution obtained from a numerical integration of the equations of geometrical optics (numerical ray-tracing).

\section{Reference solution}
\label{sec:ref}

The purpose of this chapter is to build a reference solution, for an idealized problem. As we will see in the next chapter, the main advantage of this reference solution lays in the fact that it is exact and simple, providing a good understanding of the effect of a central refractive profile on the photon path, and also a good starting point to acquire insights in more complex problems.

The reference solution is found by means of successive approximations made on the refractive profile dependencies. The least stringent is the monopole approximation which implies spherical symmetry. The most stringent is the constancy of $\alpha$ which imposes a particular type of spherical symmetry. Each are discussed in turn.

\subsection{Spherical symmetry}
\label{subsec:shape}

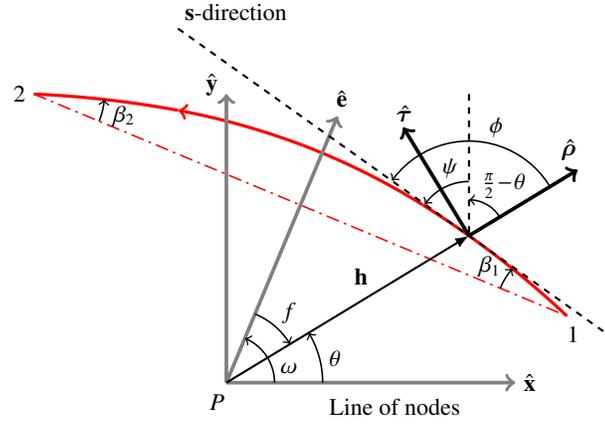
\begin{figure}
\begin{center}
\begin{tikzpicture}[scale=0.64]

\definecolor{ffqqqq}{rgb}{1.,0.,0.}

% Light path
\draw [shift={(-3.8948351751093573,-14.394928229200339)},line width=1.2pt,color=ffqqqq]  plot[domain=1.387892454382943:1.5289960489932601,variable=\t]({1.*21.41363313226521*cos(\t r)+0.*21.41363313226521*sin(\t r)},{0.*21.41363313226521*cos(\t r)+1.*21.41363313226521*sin(\t r)});
\draw [->,shift={(-2.8054474430183967,-8.310378307943488)},line width=1.2pt,color=ffqqqq]  plot[domain=1.2092840940755671:1.3855634362440608,variable=\t]({1.*15.232472718928593*cos(\t r)+0.*15.232472718928593*sin(\t r)},{0.*15.232472718928593*cos(\t r)+1.*15.232472718928593*sin(\t r)});
\draw [shift={(-2.601218558734624,-7.631534237245412)},line width=1.2pt,color=ffqqqq]  plot[domain=1.0138449443840962:1.2059057460496254,variable=\t]({1.*14.525356425389612*cos(\t r)+0.*14.525356425389612*sin(\t r)},{0.*14.525356425389612*cos(\t r)+1.*14.525356425389612*sin(\t r)});
\draw [shift={(-5.24282493332735,-11.78230823906336)},line width=1.2pt,color=ffqqqq]  plot[domain=0.816209733390795:1.0113620974719384,variable=\t]({1.*19.445241372009686*cos(\t r)+0.*19.445241372009686*sin(\t r)},{0.*19.445241372009686*cos(\t r)+1.*19.445241372009686*sin(\t r)});

% Inertial frame
\draw [line width=1.4pt,->,color=gray] (1.,1.)-- (7.,1.);
\draw [line width=1.4pt,->,color=gray] (1.,1.)-- (1.,7.);
\draw [line width=1.4pt,<-,color=gray] (3.3076923076923075,6.538461538461538)-- (1.,1.);
%\draw [line width=1.4pt,<-,color=gray] (-4.538461538461539,3.3076923076923075)-- (1.,1.);

% Light path projection
\draw [line width=0.6pt,dash pattern=on 5pt off 2pt on 1pt off 2pt,color=ffqqqq] (-3.,7.)-- (8.076923076923077,2.384615384615385);

% Radius vector
\draw [line width=0.8pt,>=latex,->] (1.,1.)-- (6.039580219129844,4.055139345428419);

% Moving frame
\draw [line width=1.4pt,<-] (4.673755748465789,6.308123974578563)-- (6.039580219129844,4.055139345428419);
\draw [line width=1.4pt,->] (6.039580219129844,4.055139345428419)-- (8.292564848279984,5.420963816092471);

% Angles
\draw [shift={(1.,1.)},line width=0.6pt,->]  plot[domain=0.:1.176005207095135,variable=\t]({1.*1.*cos(\t r)+0.*1.*sin(\t r)},{0.*1.*cos(\t r)+1.*1.*sin(\t r)});
\draw [shift={(1.,1.)},line width=0.6pt,->]  plot[domain=0.:0.5449870153486758,variable=\t]({1.*2.*cos(\t r)+0.*2.*sin(\t r)},{0.*2.*cos(\t r)+1.*2.*sin(\t r)});
\draw [shift={(1.,1.)},line width=0.6pt,<-]  plot[domain=0.5449870153486759:1.1760052070951352,variable=\t]({1.*1.547337568653714*cos(\t r)+0.*1.547337568653714*sin(\t r)},{0.*1.547337568653714*cos(\t r)+1.*1.547337568653714*sin(\t r)});
\draw [shift={(6.039580219129844,4.055139345428419)},line width=0.6pt,->]  plot[domain=0.5449870153486756:1.5707963267948966,variable=\t]({1.*0.7465039820534816*cos(\t r)+0.*0.7465039820534816*sin(\t r)},{0.*0.7465039820534816*cos(\t r)+1.*0.7465039820534816*sin(\t r)});
\draw [shift={(6.039580219129844,4.055139345428419)},line width=0.6pt,->]  plot[domain=1.5707963267948966:2.5226009933568645,variable=\t]({1.*1.126902998218986*cos(\t r)+0.*1.126902998218986*sin(\t r)},{0.*1.126902998218986*cos(\t r)+1.*1.126902998218986*sin(\t r)});
\draw [shift={(6.039580219129844,4.055139345428419)},line width=0.6pt,->]  plot[domain=0.5449870153486761:2.522600993356864,variable=\t]({1.*1.968134246062641*cos(\t r)+0.*1.968134246062641*sin(\t r)},{0.*1.968134246062641*cos(\t r)+1.*1.968134246062641*sin(\t r)});
\draw [shift={(8.076923076923077,2.384615384615385)},line width=0.6pt,<-]  plot[domain=2.425684159912168:2.7468015338900322,variable=\t]({1.*1.5038349495043406*cos(\t r)+0.*1.5038349495043406*sin(\t r)},{0.*1.5038349495043406*cos(\t r)+1.*1.5038349495043406*sin(\t r)});
\draw [shift={(-3.,7.)},line width=0.6pt,->]  plot[domain=5.8883941874798245:6.208772264813741,variable=\t]({1.*1.467940934940436*cos(\t r)+0.*1.467940934940436*sin(\t r)},{0.*1.467940934940436*cos(\t r)+1.*1.467940934940436*sin(\t r)});

% Pointilles
\draw [line width=0.8pt,dash pattern=on 2.5pt off 2.5pt] (-0.004720920883299229,8.361026008116657)-- (8.97456298146244,1.9642933007324515);
\draw [line width=0.8pt,dash pattern=on 2.5pt off 2.5pt] (6.039580219129844,7.)-- (6.039580219129844,4.);
%\draw [line width=0.8pt,dash pattern=on 2.5pt off 2.5pt] (1.9682849342200273,6.955480619146055)-- (8.97456298146244,1.9642933007324515);
%\draw [line width=0.8pt,dash pattern=on 2.5pt off 2.5pt] (3.,4.055139345428419)-- (9.,4.055139345428419);

% Verbose
\draw (3.4,6.9) node[]{$\hat{\textbf{e}}$};
%\draw (-4.3,3.8) node[]{$\hat{\mathbf{q}}_2$};
\draw (8.1,5.85) node[]{$\hat{\bm{\rho}}$};
\draw (4.7,6.6) node[]{$\hat{\bm{\tau}}$};
\draw (3.8,3.2) node[]{$\mathbf{h}$};
\draw (3.25,1.5) node[]{$\theta$};
\draw (6.6,6.3) node[]{$\phi$};
\draw (5.65,5.45) node[]{$\psi$};
\draw (6.75,5.1) node[]{$\frac{\pi}{2}\!-\!\theta$};
\draw (2.3,1.3) node[]{$\omega$};
\draw (6.48,3.42) node[]{$\beta_1$};
\draw (-1.12,6.55) node[]{$\beta_2$};
\draw (2.3,2.45) node[]{$f$};
\draw (8.2,2.0) node[]{$1$};
\draw (0.8,0.6) node[]{$P$};
\draw (-3.3,7.0) node[]{$2$};
\draw (7.0,1.0) node[right]{$\hat{\textbf{x}}$};
\draw (0.7,6.8) node[above]{$\hat{\textbf{y}}$};
\draw (4.5,0.5) node[]{Line of nodes};
\draw (1.2,8.7) node[]{\textbf{s}-direction};

\end{tikzpicture}

%%% Local Variables:
%%% mode: latex
%%% TeX-master: "bourgoin"
%%% End: 
\end{center}
\caption{Trajectory of the light ray as seen in the propagation plane $(\hat{\mathbf{x}},\hat{\mathbf{y}})$. $P$ is the position of the central planet, $1$ and $2$ are respectively the positions of a transmitter and a receiver along the light trajectory. The unit vector $\hat{\textbf{e}}$ points toward the direction of the closest approach.}
\label{fig:path}
\end{figure}

In this section, we remind some general results which arise assuming a spherically symmetric refractive profile. The spherical symmetry implies that the index of refraction is only function of the magnitude of the separation vector, i.e. $n=n(h)$ with $h=|\mathbf{h}|$. Therefore, it is shown from Eqs. \eqref{eq:s} and \eqref{eq:optics} (see also \cite{1999prop.book.....B}) that the quantity which is defined as
\begin{equation}
\label{eq:K}
\mathbf{K}=\mathbf{h}\times\bm\nabla\mathscr{S}
\end{equation}
is constant all along the path of the light ray. Regarding the magnitude, we deduce the following relation which is known as the Bouguer's rule
\begin{equation}
\label{eq:Bouger}
K=n h\sin\phi\text{,}
\end{equation}
in which $\phi$ is the angle between the separation vector \textbf{h} and the unit tangent vector to the ray \textbf{s} (see Fig. \ref{fig:path}). $K=|\mathbf{K}|$ is a constant value called the impact parameter, hence, in the following $\mathbf{K}$ will be referred to as the impact parameter vector.

When the index of refraction is a function of the height only, the impact parameter vector is constant in direction as well as in magnitude and the light ray propagates into a plane which remains orthogonal to \textbf{K}. In such a case, it is helpful to describe the light path with the help of the polar coordinates $h$ and $\theta$ which are defined such that the components of the separation vector are given by
\begin{equation}
\label{eq:h_cp}
\mathbf{h}=h\cos\theta\ \hat{\mathbf{x}}+h\sin\theta\ \hat{\mathbf{y}}\text{.}
\end{equation}
We have imposed that the $(\hat{\mathbf{x}},\hat{\mathbf{y}})$ plane coincides with the propagation plane. The $\hat{\mathbf{z}}$-direction is collinear to \textbf{K}, the $\hat{\mathbf{x}}$-axis is directed along the line of nodes (defined in Sec.~\ref{subsec:hyp_path}), and the $\hat{\mathbf{y}}$-axis is orthogonal to $\hat{\mathbf{x}}$, and $\hat{\mathbf{z}}$. We take the triad of vector $(\hat{\mathbf{x}},\hat{\mathbf{y}},\hat{\mathbf{z}})$ to form a right-handed vectorial basis that we will refer to as the propagation basis since $(\hat{\mathbf{x}},\hat{\mathbf{y}})$ coincides with the propagation plane of the ray. The geometry is depicted in Fig.~\ref{fig:path}.

Let us introduce a new rotating vectorial basis which is well suited for the polar description of the light propagation. Later, this frame will be referred to as the polar basis. The first member of the basis is $\hat{\bm \rho}\equiv\mathbf{h}/h$. The third member is $\hat{\bm \sigma}\equiv\mathbf{K}/K$, which is necessarily orthogonal to $\hat{\bm \rho}$ and collinear to $\hat{\mathbf{z}}$. We notice that in the specific case where the index of refraction is a function of the height only, $\hat{\bm \sigma}$ is a constant vector. The second member of the basis is $\hat{\bm \tau}$ which is defined to be orthogonal to $\hat{\bm \rho}$ and $\hat{\bm \sigma}$. We take the triad of vectors $(\hat{\bm \rho},\hat{\bm \tau},\hat{\bm \sigma})$ to form a right-handed vectorial basis. In the propagation plane, we have the relations
\begin{align}
\hat{\bm \rho}&=\cos\theta\ \hat{\mathbf{x}}+\sin\theta\ \hat{\mathbf{y}}\text{,} & \hat{\bm \tau}&=-\sin\theta\ \hat{\mathbf{x}}+\cos\theta\ \hat{\mathbf{y}}\text{.}\label{eq:rhotau}
\end{align}

\subsection{Monopole approximation and constancy of $\alpha$}
\label{subsec:monop}

In Sec.~\ref{subsec:ref_prof_dep}, we have discussed that assuming hydrostatic equilibrium together with elementary theory of dispersion, one can infer that $n=n(\Phi)$ where $\Phi$ is a generalized potential containing the gravitational potential $U$ of the central planet. The first approximation of the gravitational potential is the so-called monopole term (point-mass limit), which is given by the well-known Newtonian potential
\begin{equation}
\label{eq:pot_mono}
U_0=-\frac{\mu}{h}\text{,}
\end{equation}
in which $\mu=GM$. Here, $G$ is the Newtonian gravitational constant and $M$ is the mass of the central planet.

Therefore, assuming that the generalized potential contains only the monopole term ($\Phi=U_0$), we deduce that the refractivity profile becomes spherically symmetric (see Sec.~\ref{subsec:shape} for the implications) since it reduces to $n=n(h)$. 

We derive the reference solution, for a very peculiar type of spherical symmetry. Indeed, we are to consider that the refractive profile evolves linearly with the monopole term of the gravitational potential. This is the case when $\alpha$ is constant. Therefore, the expression of the index of refraction is explicitly inferred from Eqs.~\eqref{eq:dn/dPhi} and \eqref{eq:pot_mono} assuming $\alpha$ constant
\begin{equation}
\label{eq:n0}
n(h)=\eta+\frac{\alpha\mu}{h}\text{.}
\end{equation}
In this expression, $\eta$ is a constant representing the value of the index of refraction at infinity\footnote{If the value of the index of refraction is known for a given height, $H$ [let us call $n_H=n(H)$ this value], we have the relation $\eta=n_H-\alpha\mu/H$.}. The quantity which is defined as $N(h)\equiv n(h)-\eta$ is usually referred to as the refractivity of the medium. From Eq.~\eqref{eq:n0}, it is seen that the refractivity profile is spherically symmetric and evolves as $1/h$. 

We can now determine the quantities related to the description of the geometrical path of photons and the light time of the ray in the propagation plane. The complete derivation is given in appendix~\ref{sec:hyp}, we just summarize the results here.

From the conservation of $K$ and from Eq. \eqref{eq:n0}, we obtain the evolution of the height of the ray as well as the tangent to the ray, both expressed in the propagation plane [making use of Eqs.~\eqref{eq:rhotau}]
\begin{subequations}\label{eq:h_ns_sol}
\begin{align}
\textbf{h}&=\frac{p}{1+e\cos\kappa f}\hat{\bm{\rho}}\text{,}\label{eq:h_sol}\\
\bm\nabla\mathscr{S}&=\frac{\sqrt{\eta}}{\kappa}\sqrt{\frac{\alpha\mu}{p}}\left[e\kappa\sin\kappa f\hat{\bm{\rho}}+(1+e\cos\kappa f)\hat{\bm{\tau}}\right]\text{.}\label{eq:s_sol}
\end{align}
\end{subequations}
The light path, which is described by $h$, is an hyperbola. The parameter $\kappa$ is a constant which is a simple function of $K$ [cf. Eq.~\eqref{eq:kappa}] and $p$ and $e$ are two constants of integration being respectively the semi-latus rectum and the eccentricity of the conic-section. The constancy of $p$ and $e$ [cf. Eqs. \eqref{eq:p_mono} and \eqref{eq:ecc}] are both linked to the conservation of the magnitude of the impact parameter. Simultaneously, $e$ is also linked to the conservation of $E$ which is the dimensionless energy [cf. Eq.~\eqref{eq:E_n2}]. In Eqs.~$\eqref{eq:h_ns_sol}$, $f$ is the true anomaly and is defined in Eq. \eqref{eq:f_mono} by means of an other constant of integration, $\omega$, known as the argument of the closest approach. The constancy of $\omega$ is linked to the conservation of the eccentricity vector defined in Eq.~\eqref{eq:ecc_vec}.

The planar description is completed by using the fact that the index of refraction does not depend explicitly on the length or the time. This gives rise to an additional constant of integration, $S$, being the geometrical length till the closest approach. With that constant we now assess a length description of the trajectory given the true anomaly by means of a Kepler-like equation [cf. Eq.~\eqref{eq:Kepler_s}]. In order to directly assess a time description instead of a geometrical length, we derived a second Kepler-like equation for the time [cf. Eq.~\eqref{eq:Kepler_t}]. For that last case, the arbitrary constant of integration is, $T$, the elapsed time till the closest approach.

The solution is consequently described in the propagation plane by means of five constants of integration $(p,e,\omega,S,T)$. Its validity has been assessed by comparison with a numerical integration of Eqs.~\eqref{eq:s}, \eqref{eq:optics}, and \eqref{eq:dtds} for the index of refraction given by Eq. \eqref{eq:n0}. The hyperbolic elements remain constants at the level of the numerical noise.

\subsection{Hyperbolic path in space}
\label{subsec:hyp_path}

%Until now, we have demonstrated that considering the point-mass limit of the gravitational potential of the central planet, let one to deduce that the light path is described by an hyperbola which is parametrized in the propagation plane thanks to five constants of integration namely $(p,e,\omega,S,T)$. Two of them, $p$ and $e$, let to determine the shape of the hyperbola, one, $\omega$, lets to determine the location of the closest approach in the plane, and the two last ones, $S$ and $T$, provide a unique transformation between a location along the hyperbola and the traveled geometrical length, and the light time respectively.

The solution which is described in Sec.~\ref{subsec:monop} and appendix~\ref{sec:hyp} achieves a remarkable degree of simplicity. By assuming that the index of refraction is a linear function of the Newtonian potential of the planet, we have shown that the light path is totally contained inside a plane which remains orthogonal to the parameter vector. We have taken advantage of this particularity by choosing for convenience the polar coordinates $(h,\theta)$ attached to the propagation plane. Then, we have demonstrated that the equation of light propagation possesses an exact solution which is an hyperbola contained inside that plane. However, a more general description involving a more generic frame can be preferred. Indeed, when the reference solution is perturbed, the propagation plane is no longer fixed in space, and the introduction of a fixed frame is required to describe the changes. The projection of the reference solution within a more generic frame is the purpose of the present section.

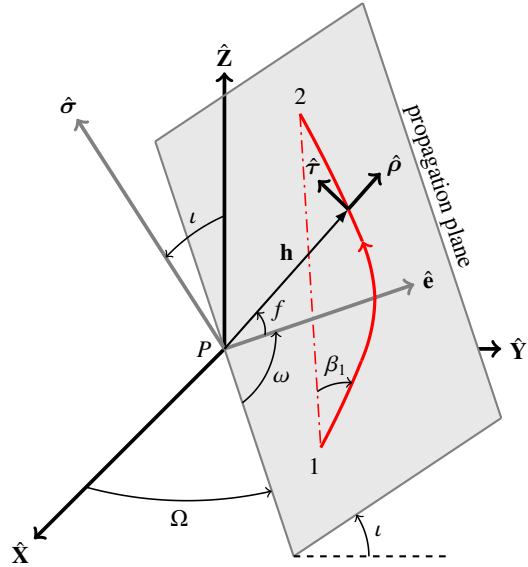
\begin{figure}
\begin{center}
\begin{tikzpicture}[scale=0.46]

\definecolor{zzttqq}{rgb}{0.1,0.1,0.1}
\definecolor{ffqqqq}{rgb}{1.,0.,0.}

% Inertial frame
\draw [line width=1.4pt,->] (2.,2.)-- (2.,10.);
\draw [line width=1.4pt,->] (2.,2.)-- (-3.5,-3.5);
\draw [line width=1.4pt,->] (9.333333333333334,2.)-- (10.,2.);

% Plan
\fill[line width=1.pt,color=zzttqq,fill=zzttqq,fill opacity=0.10,] (0.,8.) -- (4.,-4.) -- (10.,0.) -- (6.,12.) -- cycle;
\draw [line width=0.8pt,color=gray] (0.,8.)-- (4.,-4.);
\draw [line width=0.8pt,color=gray] (4.,-4.)-- (10.,0.);
\draw [line width=0.8pt,color=gray] (10.,0.)-- (6.,12.);
\draw [line width=0.8pt,color=gray] (6.,12.)-- (0.,8.);
\draw [line width=0.8pt,dash pattern=on 2.5pt off 2.5pt] (4.,-4.)-- (8.5,-4.);

% Light path projection
\draw [line width=0.6pt,dash pattern=on 5pt off 2pt on 1pt off 2pt,color=ffqqqq] (4.182089666343105,8.799423515439834)-- (4.78401348848078,-0.8575282397254609);

% Other frames
\draw [line width=1.4pt,->,color=gray] (2.,2.)-- (7.488768545763761,3.869316759499026);
\draw [line width=1.4pt,->,color=gray] (2.,2.)-- (-2.265121746472857,8.654159954123115);

% Radius vector
\draw [line width=0.8pt,>=latex,->] (2.,2.)-- (5.579378323565667,6.040103517572572);

% Light path
\draw [shift={(-28.320244884511656,-9.392839310574622)},line width=1.2pt,color=ffqqqq]  plot[domain=0.3999251603393532:0.5102764567970374,variable=\t]({1.*37.24728416927165*cos(\t r)+0.*37.24728416927165*sin(\t r)},{0.*37.24728416927165*cos(\t r)+1.*37.24728416927165*sin(\t r)});
\draw [shift={(1.4560691275127962,3.2532638558354474)},line width=1.2pt,color=ffqqqq,->]  plot[domain=0.046476813183693705:0.38873262253281193,variable=\t]({1.*4.897169005382511*cos(\t r)+0.*4.897169005382511*sin(\t r)},{0.*4.897169005382511*cos(\t r)+1.*4.897169005382511*sin(\t r)});
\draw [shift={(1.6011616969575269,3.531100532860069)},line width=1.2pt,color=ffqqqq]  plot[domain=5.876850437141317:6.27258615801228,variable=\t]({1.*4.747054859102818*cos(\t r)+0.*4.747054859102818*sin(\t r)},{0.*4.747054859102818*cos(\t r)+1.*4.747054859102818*sin(\t r)});
\draw [shift={(-17.74051376638984,11.233050814221965)},line width=1.2pt,color=ffqqqq]  plot[domain=5.790553142111332:5.899144352531924,variable=\t]({1.*25.5643585860307*cos(\t r)+0.*25.5643585860307*sin(\t r)},{0.*25.5643585860307*cos(\t r)+1.*25.5643585860307*sin(\t r)});

% Moving frame
\draw [line width=1.4pt,->] (5.579378323565667,6.040103517572572)-- (6.5093761004329584,7.089807390705859);
\draw [line width=1.4pt,->] (5.579378323565667,6.040103517572572)-- (4.70496593989295,6.878782942464994);

% Angles
\draw [shift={(0.9146820089951168,8.664859513445496)},line width=0.6pt,->]  plot[domain=4.445606034206899:4.936481399207757,variable=\t]({1.*11.055975744143467*cos(\t r)+0.*11.055975744143467*sin(\t r)},{0.*11.055975744143467*cos(\t r)+1.*11.055975744143467*sin(\t r)});
\draw [shift={(4.,-4.)},line width=0.6pt,->]  plot[domain=0.:0.5880026035475676,variable=\t]({1.*2.1746760176662967*cos(\t r)+0.*2.1746760176662967*sin(\t r)},{0.*2.1746760176662967*cos(\t r)+1.*2.1746760176662967*sin(\t r)});
\draw [shift={(4.091232396034398,1.1042045566912595)},line width=0.6pt,->]  plot[domain=1.9858899616266368:2.3896567827668607,variable=\t]({1.*5.185615783042486*cos(\t r)+0.*5.185615783042486*sin(\t r)},{0.*5.185615783042486*cos(\t r)+1.*5.185615783042486*sin(\t r)});
\draw [shift={(0.9744555497820739,2.4252116809350515)},line width=0.6pt,->]  plot[domain=-0.909192464506317:0.03248165872156305,variable=\t]({1.*2.5152427930357835*cos(\t r)+0.*2.5152427930357835*sin(\t r)},{0.*2.5152427930357835*cos(\t r)+1.*2.5152427930357835*sin(\t r)});
\draw [shift={(2.4636425661496473,2.51878596789338)},line width=0.6pt,->]  plot[domain=-0.16921503664920046:0.8499923762277848,variable=\t]({1.*0.7158727746735933*cos(\t r)+0.*0.7158727746735933*sin(\t r)},{0.*0.7158727746735933*cos(\t r)+1.*0.7158727746735933*sin(\t r)});
\draw [shift={(5.5364781872119115,-0.48350495178681147)},line width=0.6pt,<-]  plot[domain=1.4656994551424596:2.170799976383305,variable=\t]({1.*1.5116424102386776*cos(\t r)+0.*1.5116424102386776*sin(\t r)},{0.*1.5116424102386776*cos(\t r)+1.*1.5116424102386776*sin(\t r)});

% Verbose
\draw (-3.9,-3.9) node[]{$\hat{\mathbf{X}}$};
\draw (10.5,2) node[]{$\hat{\mathbf{Y}}$};
\draw (2,10.5) node[]{$\hat{\mathbf{Z}}$};
\draw (7.9,4.0) node[]{$\hat{\textbf{e}}$};
\draw (-2.5,9.05) node[]{$\hat{\bm{\sigma}}$};
\draw (6.9,7.3) node[]{$\hat{\bm{\rho}}$};
\draw (4.56,7.3) node[]{$\hat{\bm{\tau}}$};
\draw (3.55,3.1) node[]{$f$};
\draw (0.7,-2.9) node[]{$\Omega$};
\draw (3.65,1.2) node[]{$\omega$};
\draw (6.45,-3.25) node[]{$\iota$};
\draw (1.1,5.8) node[]{$\iota$};
\draw (5.25,1.5) node[]{$\beta_1$};
\draw (3.8,4.8) node[]{$\mathbf{h}$};
\draw (4.2,9.3) node[]{$2$};
\draw (1.4,2) node[]{$P$};
\draw (4.6,-1.4) node[]{$1$};
\draw (8.2,6.7) node[rotate=-72]{propagation plane};

\end{tikzpicture}

%%% Local Variables:
%%% mode: latex
%%% TeX-master: "bourgoin"
%%% End:
\end{center}
\caption{Trajectory of the light ray as viewed from the fundamental reference frame $(\hat{\mathbf{X}},\hat{\mathbf{Y}},\hat{\mathbf{Z}})$.}
\label{fig:path_plan}
\end{figure}

Now, let us introduce the new fundamental reference frame $(\hat{\mathbf{X}},\hat{\mathbf{Y}},\hat{\mathbf{Z}})$. We adopt the $(\hat{\mathbf{X}},\hat{\mathbf{Y}})$ plane as a reference plane and the $\hat{\mathbf{Z}}$-direction as a reference direction. As before, the $(\hat{\mathbf{x}},\hat{\mathbf{y}})$ plane is the propagation plane while the $\hat{\mathbf{z}}$-direction is the orthogonal direction to the propagation plane. The reference plane could be superimposed with the inertial equator of the central planet at a given time and the reference direction would be the direction of the pole. In that case, we impose that the two frames share the same origin. To achieve a complete description of the light path in the reference frame, we have to introduce two additional angles as it can be seen from Fig.~\ref{fig:path_plan}. The first one, labeled $\Omega$, is called the \emph{longitude of the node} and is the angle between the $\hat{\mathbf{X}}$-direction and the intersection between the propagation and the reference planes. The second one, labeled $\iota$, is called the \emph{inclination} and represents the angle between the $\hat{\mathbf{z}}$-direction orthogonal to the propagation plane and the $\hat{\mathbf{Z}}$-direction orthogonal to the reference plane. The line forming the intersection between the two planes is called the \emph{line of nodes}. From Fig.~\ref{fig:path_plan}, we can easily determine the components of the unit vectors $\hat{\bm \rho}$, $\hat{\bm \tau}$, and $\hat{\bm \sigma}$ into the reference frame. The radial, tangent, and normal unit vectors are given by
\begin{subequations}\label{eq:rhotausigma_XYZ}
\begin{align}
\hat{\bm \rho}=&+\big[\cos\Omega\cos(f+\omega)-\cos\iota\sin\Omega\sin(f+\omega)\big]\hat{\mathbf{X}}+\big[\sin\Omega\cos(f+\omega)+\cos\iota\cos\Omega\sin(f+\omega)\big]\hat{\mathbf{Y}}+\sin\iota\sin(f+\omega)\hat{\mathbf{Z}}\text{,}\label{eq:rho_XYZ}\\
\hat{\bm \tau}=&-\big[\cos\Omega\sin(f+\omega)+\cos\iota\sin\Omega\cos(f+\omega)\big]\hat{\mathbf{X}}-\big[\sin\Omega\sin(f+\omega)-\cos\iota\cos\Omega\cos(f+\omega)\big]\hat{\mathbf{Y}}+\sin\iota\cos(f+\omega)\hat{\mathbf{Z}}\text{,}\label{eq:tau_XYZ}\\
\hat{\bm \sigma}=&+\sin\iota(\sin\Omega\hat{\mathbf{X}}-\cos\Omega\hat{\mathbf{Y}})+\cos\iota\hat{\mathbf{Z}}\label{eq:sigma_XYZ}\text{.}
\end{align}
\end{subequations}
Regarding the direction of the closest approach, we have
\begin{align}
\hat{\mathbf{e}}=&+\big[\cos\Omega\cos\omega-\cos\iota\sin\Omega\sin\omega\big]\hat{\mathbf{X}}+\big[\sin\Omega\cos\omega+\cos\iota\cos\Omega\sin\omega\big]\hat{\mathbf{Y}}+\sin\iota\sin\omega\hat{\mathbf{Z}}\text{.}\label{eq:x_XYZ}
\end{align}
Thanks to these expressions, the height and the direction of the ray [cf. Eqs. \eqref{eq:h_ns_sol}] can now be expressed into the reference frame. We can also provide some simple definitions for the angles of the problem using the solution in the reference frame. For instance, from the previous equations, we deduce the following useful relationships
\begin{subequations}\label{eq:sincos}
\begin{align}
K\cos\iota&\equiv \mathbf{K}\cdot\hat{\mathbf{Z}}\text{,}\label{eq:cosi}\\
-K\sin\iota\cos\Omega&\equiv \mathbf{K}\cdot\hat{\mathbf{Y}}\text{,}\label{eq:cosOmega}\\
K\sin\iota\sin\Omega&\equiv \mathbf{K}\cdot\hat{\mathbf{X}}\text{,}\label{eq:sinOmega}\\
K\sin\iota\cos\omega&\equiv (\mathbf{K}\times\hat{\mathbf{e}})\cdot\hat{\mathbf{Z}}\text{,}\label{eq:cosomega}\\
\sin\iota\sin\omega&\equiv \hat{\mathbf{e}}\cdot\hat{\mathbf{Z}}\text{.}\label{eq:sinomega}%\\
%Kh\sin\iota\cos(f+\omega)&\equiv (\mathbf{K}\times\mathbf{h})\cdot\hat{\mathbf{Z}}\text{,}\label{eq:cosf+omega}\\
%h\sin\iota\sin(f+\omega)&\equiv \mathbf{h}\cdot\hat{\mathbf{Z}}\text{,}\label{eq:sinf+omega}\\
%h\cos f&\equiv \mathbf{h}\cdot\hat{\mathbf{e}}\text{,}\label{eq:cosf}\\
%Kh\sin f&\equiv \mathbf{h}\cdot(\mathbf{K}\times\hat{\mathbf{e}})\text{.}\label{eq:sinf}
\end{align}
\end{subequations}
We have used the previous expressions of $\hat{\bm \rho}$, $\hat{\bm \tau}$, $\hat{\bm \sigma}$, and $\hat{\mathbf{e}}$. These definitions are fundamental since they are expressed with vectors of the problem. So, an alternative description of the light path consists in using the set of the \emph{hyperbolic elements}\footnote{If we do not specify that $\textbf{s}$ is a unit vector (like in appendix \ref{sec:hyp}), we count seven hyperbolic elements. Otherwise, one of the hyperbolic elements is not independent anymore and we end up with six hyperbolic elements. This point is further developed in Sec.~\ref{subsec:data}.} $(p,e,\iota,\Omega,\omega,S,T)$ instead of the \textbf{h}, \textbf{s}, and $t$.

The complete solution described here and in the previous section is the analogous to the solution of the two body problem in celestial mechanics \cite{1950clme.book.....G,1961mcm..book.....B,2000ssd..book.....M,2014grav.book.....P}. However, as in celestial mechanics, the usefulness of this solution is not in its direct application, but rather in the fact that it is exact and simple, and can therefore be used as a reference solution (or \emph{osculating solution}) for studying much more complex problems, as discussed in Sec.~\ref{sec:varconst}.

\subsection{Limiting case}
\label{subsec:data}

In this section and the next one, we close the topic related to the hyperbolic solution i) by showing that the solution can be further simplified by making use of the unit vector property of $\textbf{s}$, and ii) by discussing its direct application to realistic cases.

In the derivation proposed so far, we never made use of the fact that $\textbf{s}$ is a unit vector. This property of $\textbf{s}$ was left aside on purpose, and the reason will become obvious in Sec. \ref{sec:var_arb_const}. Here, we are about to further simplify the previous expressions derived in appendix \ref{sec:hyp} by using $\dd s=(\dd\mathbf{h}\cdot\dd\mathbf{h})^{1/2}$, which implies that $\textbf{s}$ is a unit vector. Thus, all formulas which are presented in this section are determined under this assumption and should only be applied when the index of refraction is exactly given by Eq.~\eqref{eq:n0}. Although they should not be used in general, they provide a simplified and compact version of the formulas derived in appendix~\ref{sec:hyp}.

If the tangent to the ray is a unit vector, by definition [cf. Eq.~\eqref{eq:S}] we end up with the eikonal equation  [cf. Eq.~\eqref{eq:eikonal_eq}], and the total number of independent constants of integration becomes six. After inserting the eikonal equation into Eq.~\eqref{eq:E_n2}, we deduce $E=\eta^2/2$, and thus, from Eq.~\eqref{eq:ecc}, one infers
\begin{equation}
\label{eq:val_K/e}
K=\alpha\mu e\text{,}
\end{equation}
which shows that the eccentricity reduces to the inverse of the refractivity at $K$. This expression allows to express $\kappa$ [cf. Eq.~\eqref{eq:kappa}] as a unique function of the eccentricity
\begin{equation}
\label{eq:kappa1}
\kappa^2=\frac{e^2-1}{e^2}\text{.}
\end{equation}
From these two last equations, it is straightforward to show that the semi-latus rectum [cf. \eqref{eq:p_mono}] is expressed as
\begin{equation}
\label{eq:p1}
p=\frac{\alpha\mu}{\eta}(e^2-1)\text{,}
\end{equation}
and is therefore a function of the eccentricity only. The meaning is that $p$ and $e$ are not independent constants anymore when $\textbf{s}$ is a unit vector.

Inserting these simplifications into Eqs.~\eqref{eq:h_ns_sol} provides the evolution of the separation vector as well as the direction of the tangent to the ray
\begin{subequations}\label{eq:eq:h_ns_sol1}
\begin{align}
\textbf{h}&=\frac{\kappa^2K}{\eta}\left(\frac{\alpha\mu}{K}+\cos\kappa f\right)^{-1}\hat{\bm{\rho}}\text{,}\label{eq:h_sol1}\\
\bm\nabla\mathscr{S}&=\frac{\eta}{\kappa}\sin\kappa f\hat{\bm{\rho}}+\frac{\eta}{\kappa^2}\left(\frac{\alpha\mu}{K}+\cos\kappa f\right)\hat{\bm{\tau}}\text{.}\label{eq:s_sol1}
\end{align}
\end{subequations}

From Eqs.~\eqref{eq:val_K/e}, \eqref{eq:p1}, and \eqref{eq:a}, it is easily shown that the semi-major axis, $a$, is now given by
\begin{equation}
\label{eq:val_a}
-a=\frac{\alpha\mu}{\eta}\text{.}
\end{equation}
Insertion of Eqs. \eqref{eq:val_K/e} and \eqref{eq:val_a} into the two Kepler-like equations [cf. Eqs.~\eqref{eq:Kepler_s} and \eqref{eq:Kepler_t}], let one to deduce
\begin{equation}
\label{eq:s_F_hyp}
s(f)=S+\frac{K}{\eta}\sinh F(f)\text{,}
\end{equation}
as well as
\begin{equation}
t(f)=T+\frac{K}{c}\sinh F(f)+\frac{\alpha\mu}{c}F(f)+\frac{(\alpha\mu)^2}{cK}f\text{.}
\label{eq:t_F_hyp}
\end{equation}
$F$ is the hyperbolic anomaly [cf. Eqs.~\eqref{eq:F_fetf_F}]. From these two relationships, we deduce that the two events $(ct,s)$ and $(cT,S)$ are linked by the expression
\begin{equation}
c(t-T)-\eta(s-S)=\alpha\mu F(f)+\frac{(\alpha\mu)^2}{K}f\text{.}
\label{eq:ct_nus_hyp}
\end{equation}
This last expression can be useful to convert directly the length in time or vice-versa.

In addition, after having inserted the simplifications into the expression of $\bar e$ [cf. Eq. \eqref{eq:epsilon}], we obtain $\bar e=0$, and we deduce the evolution of the argument of the bending angle
\begin{equation}
\label{eq:psi_F_hyp}
\psi(f)=\omega+f-2\arctan\left[\sqrt{\frac{e-1}{e+1}}\tan\left(\frac{\kappa f}{2}\right)\right]\text{.}
\end{equation}

We can now examine the vacuum limiting case. As already mentioned, the parameter $\alpha$ characterizes the coupling between the geometrical propagation of the light ray and the gravitational attraction of the central planet through the refractive properties of the medium. Thus, it is understood that if $\alpha\rightarrow0$, the light ray should become insensitive to the presence of the medium and should therefore propagate along a straight line trajectory. This can be easily checked by inserting the expression of $p$ [cf. Eq.~\eqref{eq:p1}] into Eq.~\eqref{eq:h_sol_mono}, and by taking the limit when $\alpha\rightarrow0$. In this case, it is seen from Eqs.~\eqref{eq:val_K/e} and \eqref{eq:val_a} that
\begin{align}
\displaystyle\lim\limits_{\alpha\rightarrow 0} e&=+\infty\text{,} & \lim\limits_{\alpha\rightarrow 0} a&=0^-\text{,}\label{eq:lim_e_a}
\end{align}
%\begin{equation}
%\label{eq:lim_e_a}
%\!\!\!\begin{array}{l l}
%\displaystyle\lim\limits_{\alpha\rightarrow 0} e=+\infty\text{,} & \lim\limits_{\alpha\rightarrow 0} a=0^-\text{,}
%\end{array}
%\end{equation}
also, we end up with
\begin{equation}
\label{eq:straightLine}
\lim\limits_{\alpha\rightarrow 0} h=\frac{K}{\eta\cos f}\text{,}
\end{equation}
which represents, indeed, the equation of a straight line in polar coordinates (let us note that the vacuum limit is obtained for $\eta=1$). In addition, from Eq.~\eqref{eq:ct_nus_hyp}, it is seen that $\alpha\rightarrow0$ implies $c(t-T)-\eta(s-S)=0$, which just reflects the fact that the two events $(ct,s)$ and $(cT,S)$ are linked by a signal traveling along a straight line at the phase velocity $c/\eta$ (which reduces to $c$ when $\eta=1$ in vacuum). The straight line propagation behavior can also be inferred from Eqs.~\eqref{eq:psi_F_hyp} and \eqref{eq:lim_e_a}
\begin{equation}
\label{eq:psi1}
\lim\limits_{\alpha\rightarrow 0} \psi=\omega\text{,}
\end{equation}
which shows that $\psi$ remains constant whatever is the value of the true anomaly, meaning that the bending angle is null and thus that the trajectory is a straight line.

\subsection{Applications}
\label{subsec:ev_obs}

At this level, the formulas presented in Sec.~\ref{subsec:data} can already be used either to model atmospheric delays or to process AO data. In this section, we describe succinctly the procedure to simulate an atmospheric time-delay or to analyze Doppler data using the hyperbolic solution described so far.

The solutions in Eqs. \eqref{eq:eq:h_ns_sol1}, \eqref{eq:t_F_hyp}, and \eqref{eq:psi_F_hyp} allow us to compute explicit expressions of the observables [cf. Eqs.~\eqref{eq:time_delay_atm}, \eqref{eq:doppler_instant_sv}, and \eqref{eq:defocus}] using directly the reference solution.

The expression of the atmospheric time-delay is immediate, since we only have to subtract the light time in a vacuum from the total light time which is explicitly given in Eq.~\eqref{eq:t_F_hyp}. The light time in vacuum is simply given by taking the limit of Eq.~\eqref{eq:t_F_hyp} when $\alpha\rightarrow0$ and by using Eq.~\eqref{eq:F_f}. We end up with
\begin{align}
\tau_{\text{atm}}&=\frac{K}{c}\left(\sinh F_2-\sinh F_1\right)-\frac{K}{c}\left(\tan f_2-\tan f_1\right)+\frac{\alpha\mu}{c}\left(F_2-F_1\right)+\frac{(\alpha\mu)^2}{cK}(f_2-f_1)\text{,}\label{eq:atm_time_delay}
\end{align}
where $F_1=F(f_1)$ and $F_2=F(f_2)$.

The expression for the Doppler shift [cf. Eq.~\eqref{eq:doppler_instant_sv}] is also immediate since it is given by the scalar product between Eq.~\eqref{eq:s_sol1} and $\hat{\textbf{u}}$, where $\hat{\textbf{u}}$ is the coordinate velocity vector of the transmitter (subscript 1) and the receiver (subscript 2), divided by the speed of light in vacuum. % Also, all we need is to project $\hat{\textbf{u}}$ on the vectorial basis $(\hat{\bm{\rho}},\hat{\bm{\tau}})$.
The substitution leads to
\begin{equation}
\label{eq:DopplerShift}
\frac{\nu_2}{\nu_1}=\frac{\Gamma_2}{\Gamma_1}\frac{\left[1-\frac{\eta}{\kappa}(\hat{\bm{\rho}}\cdot\hat{\mathbf{u}})_2\sin\kappa f_2-\frac{\eta}{\kappa^2}(\frac{\alpha\mu}{K}+\cos\kappa f_2)(\hat{\bm{\tau}}\cdot\hat{\mathbf{u}})_2\right]}{\left[1-\frac{\eta}{\kappa}(\hat{\bm{\rho}}\cdot\hat{\mathbf{u}})_1\sin\kappa f_1-\frac{\eta}{\kappa^2}(\frac{\alpha\mu}{K}+\cos\kappa f_1)(\hat{\bm{\tau}}\cdot\hat{\mathbf{u}})_1\right]}\text{,}
\end{equation}
where $\hat{\bm{\rho}}_1=\hat{\bm{\rho}}(f_1)$ and $\hat{\bm{\rho}}_2=\hat{\bm{\rho}}(f_2)$, likewise for $\hat{\bm{\tau}}$. If Eq.~\eqref{eq:DopplerShift} is the most general form that can be obtained for the hyperbolic solution, it is also interesting to express the Doppler frequency shift (see appendix~\ref{app:Doppler_bend} for the complete derivation) in term of the total refractive bending angle $\epsilon\equiv\psi(f_2)-\psi(f_1)$, which is easily inferred from Eq.~\eqref{eq:psi_F_hyp}.
%As we will see in the following, the equation which is obtained after inserting Eq.~\eqref{eq:s_sol} into \eqref{eq:doppler_instant_sv}, is actually really general and remains the same in any situation involving atmospheric refraction, even beyond the spherical symmetry. This is a consequence of a requirement that we impose in the incoming Sec.~\ref{sec:var_arb_const} by applying the method of variation of arbitrary constants in order to save the actual expression of $\bm\nabla\mathscr{S}$, and this holds whatever the index of refraction's dependencies are. However, even if Eq.~\eqref{eq:doppler_instant_sv} is a very general expression, it can be interesting to express the Doppler frequency shift in term of the total refractive bending angle $\epsilon$, which can be inferred from Eq.~\eqref{eq:psi(F)} (see appendix~\ref{app:Doppler_bend} for the complete derivation).

Finally, we can also find an expression for the refractive defocussing effect [cf. Eq. \eqref{eq:defocus}] by performing directly the differentiation of Eq.~\eqref{eq:psi_F_hyp} with respect to $K$, remembering that the constants of integration, e.g. the eccentricity, are function of the impact parameter. Such an expression may be tedious to derive, thus a numerical evaluation of the derivative $\dd\epsilon/\dd K$ from finite differences may be preferred.

From the evolution of the observables, we can now think of an easy procedure to apply to simulate delays and to analyze range-rate measurements. However, we remind once again that all the analytical computations made in Secs.~\ref{subsec:data} and \ref{subsec:ev_obs} have been carried out assuming that the index of refraction is exactly given by Eq.~\eqref{eq:n0}. Furthermore, we have seen in Sec.~\ref{subsec:monop} that this refractive profile corresponds to a very peculiar type of spherical symmetry ($\alpha=\text{cst}$). So, in order to extend the reference solution to any kind of spherical symmetries ($\alpha\neq\text{cst}$), we are to assume (only for this discussion) that the atmosphere under study is made with $m$ concentric layers, where each layer possesses its own value of $\alpha_k$ with $k=1,\ldots,m$. In addition, inside each layer the index of refraction follows Eq.~\eqref{eq:n0} where the value of $\eta_k$ must be chosen in such a way that it connects the values of the index of refraction at the interfaces between two layers.

For GB observations, values of $\alpha_k$ can be assumed or deduced from different sources. For instance, we can use the profile of temperature and pressure of the International Standard Atmosphere \cite{2010_GRSAM_VW}, or we can apply the same procedure as the one described in \cite{1973_Mar_Mur} or \cite{2004GeoRL..3114602M}. For a given geometry, we can thus determine the value of the hyperbolic elements inside each layer. The atmospheric time-delay inside each layer is simply given by Eq.~\eqref{eq:atm_time_delay}, and finally the total delay is the sum of all the contributions.

Conversely, the purpose of AO experiments is to deduce the value of $\alpha_k$ at different heights inside the atmosphere. The observables are the Doppler frequency shift measurements, for which the analytical counterpart is given in Eq.~\eqref{eq:DopplerShift}. It depends on the evolution of $\bm\nabla\mathscr{S}$ which relies on $\alpha_k$. Therefore, the difference between the observed and the computed Doppler shift can be minimized by varying the value of $\alpha_k$. At the end, each layer provides its own determination of $\alpha_k$. Therefore, $n$ can be retrieved overall the vertical profile by integrating Eq.~\eqref{eq:dn/dPhi}.

\section{Perturbation equations}
\label{sec:varconst}

In the case of a spherically symmetric refractivity, we saw that the direction of \textbf{K} [see Eq. \eqref{eq:K}] remains the same while the beam ray propagates through the atmosphere. However, if the gradient of the index of refraction have non-null transverse or normal components, the vector \textbf{K} does not remain constant anymore and the previous light path which used to occur in a fixed plane is no longer a solution.

In this chapter, we address the problem of finding a new solution to a more complex situation where the index of refraction is as much general as possible.

\subsection{Perturbing gradient of the index of refraction}

Let us introduce an additional variable contribution besides the hyperbolic term of the index of refraction. We will use the following notation
\begin{equation}
\label{eq:n}
n=n_0+\delta n
\end{equation}
with $n_0$ being the index of refraction given in Eq. \eqref{eq:n0} and $\delta n$ being a new non-constant contribution\footnote{\label{foot:1}If $\delta n$ is a pure constant, it must be treated as a new limit value at infinity, such $\eta'=\eta+\delta n$.}. Therefore, this new contribution will modify the second member of Eq.~\eqref{eq:optics} which is now given by a much more general expression than in Eq. \eqref{eq:nablan}
\begin{equation}
\label{eq:gradn}
\bm\nabla n=-\alpha g\hat{\bm\rho}+\mathbf{f}_{\text{pert}}\text{.}
\end{equation}
Here, $g$ still refers to the magnitude of the local spherical acceleration [cf. Eq. \eqref{eq:g}], and the first term on the right-hand side is the gradient of the index of refraction which produces the hyperbolic solution for the light path. The additional term, $\mathbf{f}_{\text{pert}}$, represents a supplementary spatial variation of the index of refraction and is given by $\mathbf{f}_{\text{pert}}=\bm{\nabla}\delta n$. We do not make any assumption about its magnitude with respect to the hyperbolic part, and we do not specify its dependency. In addition, we release the isopotential surfaces hypothesis. We only assume that the perturbing gradient can be decomposed in a certain rotating basis such that its radial, transverse, and normal components are given by
\begin{equation}
\label{eq:f}
\mathbf{f}_{\text{pert}}=\mathcal{R}\hat{\bm \rho}+\mathcal{T}\hat{\bm \tau}+\mathcal{S}\hat{\bm \sigma}\text{.}
\end{equation}
Here it must be pointed out that the basis $(\hat{\bm \rho},\hat{\bm \tau},\hat{\bm \sigma})$ has a different meaning with respect to the one introduced in Chap. \ref{sec:ref}. Indeed, as discussed above, if $\mathbf{f}_{\text{pert}}=\bm{0}$, only the monopole approximation remains and thus the direction of \textbf{K} remains constant. However, when the perturbing gradient is no longer null, \textbf{K} is changing in direction (also in norm) and then $\hat{\bm \sigma}\equiv \mathbf{K}/K$ does not remain fixed in space as before. Good insight can be obtained on the effects of the perturbing gradient on the impact parameter vector by differentiating Eq. \eqref{eq:K} with respect to $s$. Making use of Eqs.~\eqref{eq:optics}, \eqref{eq:s}, and \eqref{eq:gradn}, we quickly arrive to $\dd\mathbf{K}/\dd s=\mathbf{h}\times\mathbf{f}_{\text{pert}}$, which can be expressed in terms of the perturbing gradient's components
\begin{equation}
\label{eq:dK/ds1}
\frac{\dd\mathbf{K}}{\dd s}=h(\mathcal{T}\hat{\bm \sigma}-\mathcal{S}\hat{\bm \tau})\text{.}
\end{equation}
Then, a comparison between Eq. \eqref{eq:dK/ds1} and the derivative of $\hat{\bm \sigma}$ with respect to $s$ reveals that the magnitude of \textbf{K} changes according to
\begin{equation}
\label{eq:dmK/ds}
\frac{\dd K}{\dd s}=h\mathcal{T}\text{.}
\end{equation}
From these two last equations, we predict that a normal perturbation will only affect the direction of the impact parameter vector while a transverse one will also change its magnitude. These equations are perfectly equivalents to the ones of celestial mechanics where only the non-radial components of the perturbing acceleration change the angular momentum vector \cite{1976AmJPh..44..944B,2014grav.book.....P}.

We can also demonstrate that \textbf{K} is not the only first integral being affected by the perturbing gradient. Indeed, as seen from the form of the relationships in Eqs. \eqref{eq:E_SS} and \eqref{eq:ecc_vec}, we can expect that changes in the index of refraction are going to impact the dimensionless energy parameter and the eccentricity vector. Then, since all the hyperbolic elements $(p,e,\iota,\Omega,\omega,S,T)$ are determined from \textbf{K}, $E$, and $\textbf{e}$ (see appendix. \ref{sec:hyp}), they are also expected for not being constant anymore. The determination of their evolution is the subject of the next two sections.

\subsection{Method of variation of arbitrary constants}
\label{sec:var_arb_const}

In this section, we will refer to the general solution of the unperturbed problem as
\begin{align}
\mathbf{h}&=\mathbf{h}_0\big(s,\mathbf C\big)\text{,} & \bm\nabla\mathscr{S}&=\bm\nabla\mathscr{S}_0\big(s,\mathbf C\big)\text{,} & t&=t_0\big(s,\mathbf C\big)\text{,}\label{eq:sol_var_const0}
\end{align}
where $\mathbf{h}_0$, $\bm\nabla\mathscr{S}_0$, and $t_0$ are the solutions respectively for the separation vector, the direction, and the light time along the ray for an hyperbolic path, i.e. for $n_0$. The $\mathbf C$-vector represents the hyperbolic elements which are constants for the unperturbed trajectory. These solutions have been fully described in Chap.~\ref{sec:ref}.

We are now interested in studying the very general case where the perturbing gradient is non-null. In this context, the method of variation of arbitrary constants is very well adapted (see e.g. \cite{2014grav.book.....P} for application of the method of variation of arbitrary constants in the context of celestial mechanics). The core of the method is to consider that $\mathbf{h}_0$ and $\bm\nabla\mathscr{S}_0$ [cf. Eqs.~\eqref{eq:h_ns_sol}] are still solutions of the perturbed problem, avoiding the apparent contradiction by allowing the constants of integration (hyperbolic elements) to change as $s$ evolves. The physical meaning is that at any length $s$ along the ray, the trajectory is taken to be hyperbolic with the parameters $\mathbf C(s)$. The general solution must be expressed as
\begin{align}
\mathbf{h}&=\mathbf{h}_0\big(s,\mathbf C(s)\big)\text{,} & \bm\nabla\mathscr{S}&=\bm\nabla\mathscr{S}_0\big(s,\mathbf C(s)\big)\text{.}\label{eq:sol_var_const}
\end{align}

A subtle consequence of this choice, is that $\textbf{s}$ and $\textbf{s}_0$ cannot be two unit vectors at the same time. Indeed, from the definition of $\bm\nabla\mathscr{S}$ in Eq.~\eqref{eq:S} and from \eqref{eq:sol_var_const}, we notice that $\textbf{s}_0\cdot\textbf{s}_0=(n/n_0)^2\ \textbf{s}\cdot\textbf{s}$. Moreover, considering that the eikonal equation \eqref{eq:eikonal_eq} must hold true for the index of refraction under study (in this case $n$), we infer that $\mathbf{s}$ is a unit vector, thus $\textbf{s}_0$ is not. Consequently, from now, we are not allowed to use the hyperbolic expressions derived in Secs.~\ref{subsec:data} and \ref{subsec:ev_obs}, which are based on the assumption that $\textbf{s}_0$ is a unit vector. However, we remind that the solution which is derived in appendix~\ref{sec:hyp} is free of this assumption, then all the expressions in there can be used.

Another important consequence is that, saving the expression of $\bm\nabla\mathscr S$ [cf. Eq.~\eqref{eq:s_sol}], allows us to fix the expression of the Doppler frequency shift once and for all. Indeed, Eq.~\eqref{eq:doppler_instant_sv} depends only of $\bm\nabla\mathscr S$, and if Eq.~\eqref{eq:s_sol} must hold by virtue of the method of variation of arbitrary constant, then the expression of the Doppler frequency shift [as inferred using Eq.~\eqref{eq:s_sol}] must also hold true for any arbitrary index of refraction. However, since the hyperbolic elements are not constants anymore, their values at the reception point might be different from the ones at the transmission point.

Applying the proposition \eqref{eq:sol_var_const}, we can determine an explicit expression for the variation of the dimensionless hyperbolic energy $E$. Differentiating Eq. \eqref{eq:E_SS} and making use of Eqs.~\eqref{eq:f}, and \eqref{eq:h_ns_sol}, we get the following expression
\begin{align}
\frac{\dd E}{\dd s}&=\sqrt{\eta}\sqrt{\frac{\alpha\mu}{p}}\Big[e\mathcal{R}\sin\kappa f+\frac{1}{\kappa}(1+e\cos\kappa f)\mathcal{T}-e\frac{\alpha\mu}{p}(1+e\cos\kappa f)^2\mathcal{N}\sin\kappa f\Big]\text{,}\label{eq:dEds}
\end{align}
in which
\begin{equation}
\label{eq:N}
\mathcal{N}=\frac{\delta n}{n}p^{-1}
\end{equation}
is a new perturbing component. The term $\delta n$ has been introduced in Eq. \eqref{eq:n} and refers to the difference between the actual and the hyperbolic index of refraction. We recall that $\delta n$ must contain only the variable part of the difference as discussed in footnote~\ref{foot:1}. We see from Eq. \eqref{eq:dEds} that only the components of the perturbing gradient which are contained inside the propagation plane can impact the dimensionless parameter $E$.

Considering that all the hyperbolic elements can be expressed in terms of the first integrals $E$, $\mathbf{K}$, and $\textbf{e}$, we can now infer their length rate of change as functions of the perturbing gradient's components using Eqs. \eqref{eq:dK/ds1}, \eqref{eq:dmK/ds}, and \eqref{eq:dEds}. However, we first need to express the differentials of the hyperbolic elements as functions of $\dd E/\dd s$, $\dd \mathbf{K}/\dd s$, and $\dd\mathbf{e}/\dd s$.

Let us start with the semi-major axis. Differentiation of Eq.~\eqref{eq:E_a} yields
\begin{equation}
\label{eq:dads}
\frac{\dd a}{\dd s}=\frac{\alpha\mu\eta}{2E^2}\frac{\dd E}{\dd s}\text{.}
\end{equation}

The equation for the rate of change of the eccentricity is given by differentiating Eq. \eqref{eq:ecc}. After some little algebra we find
\begin{equation}
\label{eq:deds}
\frac{\dd e}{\dd s}=\frac{K^2E}{e(\alpha\mu\eta)^2}\left(\frac{2}{K}\frac{\dd K}{\dd s}+\frac{\kappa^2}{E}\frac{\dd E}{\dd s}\right)\text{.}
\end{equation}

The variation of the semi-latus rectum is determined either from Eqs. \eqref{eq:dads} and \eqref{eq:deds} making use of Eq. \eqref{eq:a}, or directly by differentiating Eq. \eqref{eq:p_mono}. Both ways lead to
\begin{equation}
\label{eq:dpds}
\frac{\dd p}{\dd s}=\frac{2K}{\alpha\mu\eta}\frac{\dd K}{\dd s}\text{.}
\end{equation}

The variation of the inclination is given after differentiating Eq.~\eqref{eq:cosi}
\begin{equation}
\label{eq:dids}
\frac{\dd\iota}{\dd s}=-\frac{1}{K\sin\iota}\left(\frac{\dd\mathbf{K}}{\dd s}\cdot\hat{\mathbf{Z}}-\cos\iota\frac{\dd K}{\dd s}\right)\text{.}
\end{equation}

Eqs. \eqref{eq:cosOmega} and \eqref{eq:sinOmega} let to determine the tangent of the longitude of the node. After differentiation, we arrive to
\begin{equation}
\label{eq:dOmegads}
\frac{\dd\Omega}{\dd s}=\frac{1}{K\sin\iota}\left(\cos\Omega\frac{\dd\mathbf{K}}{\dd s}\cdot\hat{\mathbf{X}}+\sin\Omega\frac{\dd \mathbf{K}}{\dd s}\cdot\hat{\mathbf{Y}}\right)\text{.}
\end{equation}

\begin{figure}
\begin{center}
\begin{tikzpicture}[scale=0.82]

% repere inertiel
\draw [line width=1.4pt,->] (0.,0.)-- (-3.,-3.);
\draw [line width=1.4pt,->] (0.,0.)-- (0.,3.);
\draw [line width=1.4pt,->] (3.945625594362348,0.)-- (4.40441164547726,0.);

% plan gris 
\fill[line width=1.pt,color=gray,fill=gray,fill opacity=0.1] (4.,3.) -- (0.,0.) -- (4.149842000480156,1.5294927659031805) -- (4.149281255393987,1.7308759460209382) -- (4.140967901570542,1.9610700694915197) -- (4.124951436437847,2.1855216954908343) -- (4.105027095144695,2.3773680156775496) -- (4.083257633504186,2.5423309952412807) -- (4.057513250701998,2.7052002418833996) -- (4.02536003966695,2.878819369585123) -- cycle;
\fill[line width=1.pt,color=gray,fill=gray,fill opacity=0.1] (2.331093417237952,-2.889824800889869) -- (1.5799303982780004,-1.95861822390343) -- (2.088782352504481,-1.5052069676361537) -- (2.303368305587477,-1.291535497865702) -- (2.5435429949642847,-1.0340713408531954) -- (2.7910447734040655,-0.7458081026484278) -- (3.0333341623031753,-0.43746128794588834) -- (3.2381644425812386,-0.1530819708537221) -- (3.460547263130362,0.18509740233231575) -- (3.639880982481154,0.4847810294426065) -- (3.8201135991066,0.8159792582671708) -- (3.9241610217739726,1.0238611708453713) -- (4.0486101491014965,1.2921474277433118) -- (4.149842000480156,1.5294927659031805) -- (4.129209036277456,1.0889005942394032) -- (4.10349868327799,0.8341746361901017) -- (4.066571349318212,0.5734029821056874) -- (4.0085083780133335,0.2636758440125684) -- (3.9586828252512403,0.047702396691649884) -- (3.8788323872868284,-0.24401774674403587) -- (3.803975733076447,-0.4781213235067714) -- (3.718838701725682,-0.7130096926057861) -- (3.583212169037993,-1.0390041928980498) -- (3.459709290201218,-1.298838288050555) -- (3.3086511130309466,-1.5816040327246204) -- (3.132237266238821,-1.8747191933108995) -- (2.9654526827370127,-2.1231309015780226) -- (2.770506725674271,-2.3851988398110198) -- (2.550595264313288,-2.650895581554368) -- cycle;
\draw [shift={(-2.3295400557112815,1.6121435025513304)},line width=0.8pt,color=gray]  plot[domain=-0.7680832874634138:0.21585062056476423,variable=\t]({1.*6.479909179484236*cos(\t r)+0.*6.479909179484236*sin(\t r)},{0.*6.479909179484236*cos(\t r)+1.*6.479909179484236*sin(\t r)});
\draw [line width=0.8pt,color=gray] (4.,3.)-- (0.,0.);
\draw [line width=0.8pt,color=gray] (2.331093417237952,-2.889824800889869)-- (1.5799303982780004,-1.95861822390343);

% plan noir 
\fill[line width=1.pt,fill=black,fill opacity=0.2] (0.,-3.) -- (0.,0.) -- (4.149842000480156,1.5294927659031805) -- (4.0486101491014965,1.2921474277433118) -- (3.9241610217739726,1.0238611708453713) -- (3.8201135991066,0.8159792582671708) -- (3.639880982481154,0.4847810294426065) -- (3.460547263130362,0.18509740233231575) -- (3.2381644425812386,-0.1530819708537221) -- (3.0333341623031753,-0.43746128794588834) -- (2.7910447734040655,-0.7458081026484278) -- (2.5435429949642847,-1.0340713408531954) -- (2.303368305587477,-1.291535497865702) -- (2.088782352504481,-1.5052069676361537) -- (1.5799303982780004,-1.95861822390343) -- (1.8850204799481594,-1.695234400815182) -- (1.7264035808622111,-1.835180980652443) -- (1.4583227488091843,-2.0570922159076073) -- (1.2061177489319288,-2.250406083191721) -- (0.9577053828627395,-2.4273455509652573) -- (0.7181909976189207,-2.586184019331127) -- (0.4967863257426952,-2.7233763754533307) -- (0.2584989194866285,-2.8612643477470465) -- cycle;
\fill[line width=1.pt,fill=black,fill opacity=0.2] (4.066552871219056,2.6480119925143932) -- (4.083257633504186,2.5423309952412807) -- (4.105027095144695,2.3773680156775496) -- (4.124951436437847,2.1855216954908343) -- (4.140967901570542,1.9610700694915197) -- (4.149281255393987,1.7308759460209382) -- (4.149842000480156,1.5294927659031805) -- (4.24354193249965,1.768093622656771) -- (4.3198140988636355,1.978901893050891) -- (4.393106152957533,2.1989758789281804) -- (4.458582500300678,2.4138498913409516) -- (4.519448660764712,2.6335124484964463) -- (4.568727909369517,2.8297009243633573) -- (4.60777884280425,3.000441410944214) -- cycle;
\draw [shift={(-4.071060846692352,4.895587388942938)},line width=0.8pt]  plot[domain=5.18844816528579:6.068196028849973,variable=\t]({1.*8.883346026914984*cos(\t r)+0.*8.883346026914984*sin(\t r)},{0.*8.883346026914984*cos(\t r)+1.*8.883346026914984*sin(\t r)});
\draw [line width=0.8pt] (0.,0.)-- (0.,-3.);
\draw [line width=0.8pt] (4.60777884280425,3.000441410944214)-- (4.066552871219056,2.6480119925143932);

% pointilles
\draw [line width=0.8pt,dash pattern=on 2.5pt off 2.5pt] (0.,-3.)-- (1.2080659041229542,-2.989098478725143);
\draw [line width=0.8pt,dash pattern=on 2.5pt off 2.5pt] (2.331093417237952,-2.889824800889869)-- (3.5387604565183204,-2.6200268453059565);
\draw [line width=0.8pt,dash pattern=on 2.5pt off 2.5pt] (0.,-3.)-- (0.,-3.775877887181333);
\draw [line width=0.8pt,dash pattern=on 2.5pt off 2.5pt] (2.331093417237952,-2.889824800889869)-- (2.8949695493097747,-3.5888543717518373);
\draw [line width=0.8pt,dash pattern=on 2.5pt off 2.5pt] (4.149842000480156,1.5294927659031805)-- (5.029086677564712,1.8535529043142591);

% angles
\draw [shift={(-2.14326324038191,1.5264014182528662)},line width=0.6pt,<->]  plot[domain=-0.7929838835487804:0.04558127940226591,variable=\t]({1.*7.1798072007750395*cos(\t r)+0.*7.1798072007750395*sin(\t r)},{0.*7.1798072007750395*cos(\t r)+1.*7.1798072007750395*sin(\t r)});
\draw [shift={(0.6523861745074707,8.625103499898426)},line width=0.6pt,<->]  plot[domain=4.659829806896537:4.893974714296262,variable=\t]({1.*12.418129773979137*cos(\t r)+0.*12.418129773979137*sin(\t r)},{0.*12.418129773979137*cos(\t r)+1.*12.418129773979137*sin(\t r)});
\draw [shift={(-0.5925321569236781,2.2137161933213663)},line width=0.6pt,<->]  plot[domain=4.390020049449128:4.846163806316202,variable=\t]({1.*4.442563436480408*cos(\t r)+0.*4.442563436480408*sin(\t r)},{0.*4.442563436480408*cos(\t r)+1.*4.442563436480408*sin(\t r)});
\draw [shift={(-3.834148265607907,4.93886820447577)},line width=0.6pt,<->]  plot[domain=5.233989391790912:5.750308623695119,variable=\t]({1.*7.694946151493365*cos(\t r)+0.*7.694946151493365*sin(\t r)},{0.*7.694946151493365*cos(\t r)+1.*7.694946151493365*sin(\t r)});
\draw [shift={(2.331093417237952,-2.889824800889869)},line width=0.6pt,<->]  plot[domain=0.21979507152403263:0.8795724526390418,variable=\t]({1.*1.0603264399127212*cos(\t r)+0.*1.0603264399127212*sin(\t r)},{0.*1.0603264399127212*cos(\t r)+1.*1.0603264399127212*sin(\t r)});
\draw [shift={(0.,-3.)},line width=0.6pt,<->]  plot[domain=0.009023700894691091:0.5227297147081946,variable=\t]({1.*1.0888855795480274*cos(\t r)+0.*1.0888855795480274*sin(\t r)},{0.*1.0888855795480274*cos(\t r)+1.*1.0888855795480274*sin(\t r)});

% vecteur
\draw [line width=0.8pt,>=latex,->] (0.,0.)-- (4.149842000480156,1.5294927659031805);

% Verbose
\draw (-3.2,-3.2) node[]{$\hat{\mathbf{X}}$};
\draw (4.6,0.0) node[]{$\hat{\mathbf{Y}}$};
\draw (0.0,3.3) node[]{$\hat{\mathbf{Z}}$};
\draw (-1.1,-2.5) node[]{$\Omega$};
\draw (1.5,-4.1) node[]{$\delta\Omega$};
\draw (1.8,-0.7) node[]{$\theta$};
\draw (5.2,-1.1) node[]{$\theta\!+\!\delta\theta$};
\draw (1.25,-2.65) node[]{$\iota$};
\draw (3.6,-1.95) node[]{$\iota\!+\!\delta\iota$};
\draw (2.4,1.2) node[]{$\mathbf{h}$};

\end{tikzpicture}

%%% Local Variables:
%%% mode: latex
%%% TeX-master: "bourgoin"
%%% End:
\end{center}
\caption{Illustration of the effect on the $\theta$ angle due a tilt $\delta\Omega$ along the longitude of the node. The \emph{darker} plane corresponds to the propagation plane before the tilt while the \emph{lighter} one is the same plane after the tilt. From that sketch, we infer the relation $(\theta+\delta\theta)\cos\delta\iota+\delta\Omega\cos\iota=\theta$, which reduces to $\delta\theta=-\delta\Omega\cos\iota$ for infinitesimal tilts.}
\label{fig:Omega_var}
\end{figure}
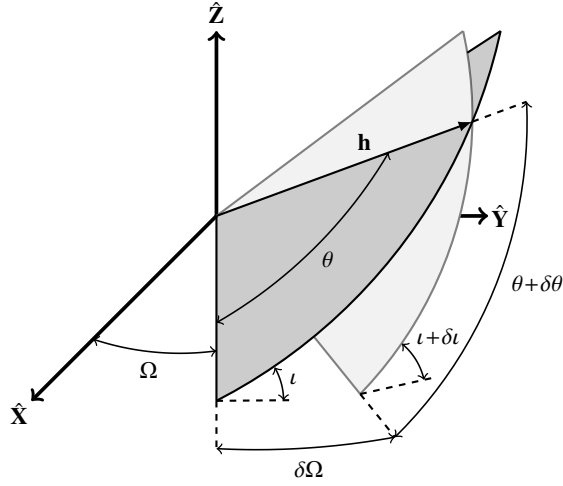

For the derivation of the rate of change of the argument of the closest approach, we do not differentiate the eccentricity vector as it is usually done in celestial mechanics \cite{2014grav.book.....P}. Instead, we are to simplify the algebra by using a geometrical relation \cite{1976AmJPh..44..944B}. The computation starts by differentiating Eq.~\eqref{eq:h_sol_mono} and by making use of Eqs. \eqref{eq:const_sol_mono}, \eqref{eq:kappa}, \eqref{eq:dpds}, and \eqref{eq:deds}, which leads to
\begin{align}
\frac{\dd\omega}{\dd s}&=-\cos\iota\frac{\dd\Omega}{\dd s}-\frac{\kappa}{e^2}\frac{K^2}{(\alpha\mu\eta)^2}\cot\kappa f\frac{\dd E}{\dd s}+\frac{2}{e\kappa}\frac{K}{\alpha\mu\eta}\csc\kappa f\bigg(h^{-1}+\frac{e\eta}{2\kappa}\frac{(\alpha\mu)^3}{K^4}f\sin\kappa f-\frac{E}{e\alpha\mu\eta}\cos\kappa f\bigg)\frac{\dd K}{\dd s}\label{eq:domegads}\text{.}
\end{align}
On the right-hand side we have substituted the following relationship
\begin{equation}
\label{eq:dthetads}
\frac{\dd\theta}{\dd s}-\frac{\hat{\bm{\tau}}\cdot\bm\nabla\mathscr{S}}{nh}=-\cos\iota\frac{\dd\Omega}{\dd s}
\end{equation}
which has been determined geometrically from Fig.~\ref{fig:Omega_var}. The left-hand side is obtained by making use of the method of variation of arbitrary constants [cf. Eqs. \eqref{eq:sol_var_const}]. The right-hand side contribution is determined remembering that for a certain location along the ray, \textbf{h} remains unchanged, so we conclude that a tilt along $\Omega$ will decrease $\theta$ by an amount equal to the projection of the magnitude of the tilt on the propagation plane. So, the geometrical contribution to $\theta$ changes is exactly given by the term $-\delta\Omega\cos\iota$. This point is illustrated in Fig. \ref{fig:Omega_var} and is also discussed in \cite{1976AmJPh..44..944B}. With this geometrical relation, the expression of the change in $\omega$ has been determined without using the change in the eccentricity vector.

Finally, we consider the rate of change of the true anomaly for closing the system, even if it is not constant along the photon path. The relation is obtained by differentiating Eq.~\eqref{eq:f_mono} and by substituting it into Eq.~\eqref{eq:dthetads}
\begin{equation}
\label{eq:dfds_oO}
\frac{\dd f}{\dd s}-\frac{\hat{\bm{\tau}}\cdot\bm\nabla\mathscr{S}}{nh}=-\left(\frac{\dd\omega}{\dd s}+\cos\iota\frac{\dd\Omega}{\dd s}\right)\text{.}
\end{equation}
The left-hand side represents the hyperbolic contribution to the change in the true anomaly. The right-hand side represents the change in the direction of the closest approach relative to the $\hat{\textbf{X}}$-direction and measured inside the propagation plane (see Fig.~\ref{fig:Omega_var}). The same is also true in celestial mechanics as mentioned by \cite{2014grav.book.....P}. Substituting Eq.~\eqref{eq:domegads} into Eq.~\eqref{eq:dfds_oO} provides the following expression
\begin{widetext}
\begin{align}
\frac{\dd f}{\dd s}&=\frac{\sqrt{\eta}}{\kappa n}\sqrt{\frac{\alpha\mu}{p^3}}(1+e\cos\kappa f)^2+\frac{\kappa}{e^2}\frac{K^2}{(\alpha\mu\eta)^2}\cot\kappa f\frac{\dd E}{\dd s}-\frac{2}{e\kappa}\frac{K}{\alpha\mu\eta}\csc\kappa f\bigg(h^{-1}+\frac{e\eta}{2\kappa}\frac{(\alpha\mu)^3}{K^4}f\sin\kappa f-\frac{E}{e\alpha\mu\eta}\cos\kappa f\bigg)\frac{\dd K}{\dd s}\label{eq:dfds}\text{.}
\end{align}
\end{widetext}
The right-hand side reduces to $\hat{\bm{\tau}}\cdot\bm\nabla\mathscr{S}/nh$ when the perturbation is null (i.e. when $\dd E/\dd s$ and $\dd K/\dd s$ vanish).

\subsection{Osculating equations}
\label{subsec:eq_osc}

We are now ready to derive the final expressions for the evolution of the hyperbolic elements as functions of the components of the perturbing gradient of the index of refraction. However, for a matter of compactness, we do not provide the differentials of $(S,T)$, but rather the differentials of $(f,t)$, even if $f$ or $t$ are not constants along the light path for the hyperbolic trajectory.

Substituting Eqs. \eqref{eq:dK/ds1}, \eqref{eq:dmK/ds}, and \eqref{eq:dEds} into all the previous equations describing the rate of change of the hyperbolic elements with $\dd E/\dd s$, $\dd \textbf{K}/\dd s$, and $\dd K/\dd s$, it is possible to infer the following relationships
\begin{widetext}
\begin{subequations}\label{eq:guauss}
\begin{align}
\frac{\dd p}{\dd s}&=\frac{2}{\kappa\sqrt{\eta}}\sqrt{\frac{p^3}{\alpha\mu}}\frac{1}{1+e\cos\kappa f}\mathcal{T}\text{,}\label{eq:dpds1}\\
\frac{\dd e}{\dd s}&=\frac{1}{\sqrt{\eta}}\sqrt{\frac{p}{\alpha\mu}}\Bigg\{\mathcal{R}\sin\kappa f+\frac{1}{\kappa}\Bigg[\frac{2\cos\kappa f+e(1+\cos^2\kappa f)}{1+e\cos\kappa f}\Bigg]\mathcal{T}-\frac{\alpha\mu}{p}(1+e\cos\kappa f)^2\mathcal{N}\sin\kappa f\Bigg\}\text{,}\label{eq:deds1}\\
\frac{\dd\iota}{\dd s}&=\frac{\kappa}{\sqrt{\eta}}\sqrt{\frac{p}{\alpha\mu}}\frac{\cos(f+\omega)}{1+e\cos\kappa f}\mathcal{S}\text{,}\label{eq:dids1}\\
\frac{\dd\Omega}{\dd s}&=\frac{\kappa}{\sqrt{\eta}}\sqrt{\frac{p}{\alpha\mu}}\frac{\sin(f+\omega)}{1+e\cos\kappa f}\mathcal{S}\csc\iota\text{,}\label{eq:dOmegads1}\\
\frac{\dd \omega}{\dd s}&=-\frac{1}{e\kappa\sqrt{\eta}}\sqrt{\frac{p}{\alpha\mu}}\Bigg\{\mathcal{R}\cos\kappa f-\frac{1}{\kappa}\Bigg[\frac{2+e\cos\kappa f}{1+e\cos\kappa f}\sin\kappa f+\frac{e\kappa(1-\kappa^2) f}{1+e\cos\kappa f}\Bigg]\mathcal{T}\nonumber\\
&+e\kappa^2\frac{\sin(f+\omega)}{1+e\cos\kappa f}\mathcal{S}\cot\iota-\frac{\alpha\mu}{p}(1+e\cos\kappa f)^2\mathcal{N}\cos\kappa f\Bigg\}\text{,}\label{eq:domegads1}\\
\frac{\dd f}{\dd s}&=\frac{\sqrt{\eta}}{\kappa n}\sqrt{\frac{\alpha\mu}{p^3}}(1+e\cos\kappa f)^2\nonumber\\
&+\frac{1}{e\kappa\sqrt{\eta}}\sqrt{\frac{p}{\alpha\mu}}\Bigg\{\mathcal{R}\cos\kappa f-\frac{1}{\kappa}\Bigg[\frac{2+e\cos\kappa f}{1+e\cos\kappa f}\sin\kappa f+\frac{e\kappa(1-\kappa^2) f}{1+e\cos\kappa f}\Bigg]\mathcal{T}-\frac{\alpha\mu}{p}(1+e\cos\kappa f)^2\mathcal{N}\cos\kappa f\Bigg\}\text{,}\label{eq:dfds1}\\
\frac{\dd t}{\dd s}&=\frac{n}{c}\text{.}\label{eq:dtds1}
\end{align}
\end{subequations}
\end{widetext}
These equations provide a good understanding of how the components of the perturbing gradient govern the length rate of change of the hyperbolic elements. For instance, it can be seen that $p$ is only affected by the transverse component of the perturbing gradient, where $\iota$ and $\Omega$ only change because of the normal component. We also see that $e$ and $f$ are affected by the radial and the transverse components, therefore any perturbation which is contained inside the propagation plane will induce variations of these two elements. In addition, they are also expected to vary because of a term in $\delta n$ (through $\mathcal{N}$), which represents a non-constant change in the expression of the index of refraction. Finally, we see that $\omega$ is the only one being affected by all the components of the perturbing gradient.

This set of equations is similar to the \emph{osculating equations} of celestial mechanics which are also called \emph{perturbation equations}. However, one important difference stands in the terms in $\mathcal{N}$ which does not possess any equivalent in celestial mechanics. The reason is that the motion of planets is not directly sensitive to the gravitational potential itself but rather to its gradient. In the case of geometrical optics, the equations of light propagation not only contain the gradient of the index of refraction, but also the value of that index, see e.g. Eq.~\eqref{eq:dtds}. This can also be inferred once Eq.~\eqref{eq:optics} is written as $\dd\textbf{s}/\dd s=n^{-1}[\bm{\nabla}n-\textbf{s}(\textbf{s}\cdot\bm{\nabla})n]$, which represents the rate of change of the curvature of the ray \cite{1999prop.book.....B}.

As in celestial mechanics, when the inclination is null, $\omega$ and $\Omega$ are not defined since the direction of the line of nodes is not determined as seen in Fig \ref{fig:path_plan}. These singularities are related to the choice we have made in the definition of angles. They can be removed \cite{1961mcm..book.....B,2000ssd..book.....M,2002mcma.book.....M,2014grav.book.....P} by introducing new angles, such the longitude of the periapsis which is defined as $\varpi=\omega+\Omega$. In the next, we will continue with $\omega$ and $\Omega$, keeping in mind that a null inclination induces coordinate singularities.

The perturbation equations provide an alternative description to the fundamental equations of optics [cf. Eq.~\eqref{eq:optics}], also in that sense, they are generic and no approximations have been introduced in the transcription. Their validity has been assessed by comparing numerical integration of Eqs.~\eqref{eq:guauss} with numerical integration of Eqs.~\eqref{eq:s}, \eqref{eq:optics}, and \eqref{eq:dtds} showing that the differences remain at the level of the numerical noise. The main advantage of the perturbation equations is to provide an easy geometrical picture to tackle the effects of a perturbing gradient with respect to an hyperbolic path.

It is worth noticing that any exact solution to the fundamental equations of optics could have been used as a reference solution, or osculating solution. For instance, we could have chosen the straight line as the osculating solution instead of the hyperbolic path. Actually, if we insert Eqs.~\eqref{eq:val_K/e}--\eqref{eq:p1} into Eqs.~\eqref{eq:guauss}, and if we take the limit $\alpha\rightarrow0$ (which implies $e\rightarrow\infty$), the perturbation equations reduce to a set of equations describing the evolution of an osculating straight line.

In the following, we will see how to use the set of perturbation equations when the perturbing gradient can be considered small with respect to the spherical contribution of the gravitational potential of the central planet.\\

\subsection{First-order approximation}

The great benefit of having written the equation of fundamental optics in terms of the components of the perturbing gradient will become obvious in this section.

Indeed, in the case where the perturbing gradient is small with respect to the hyperbolic contribution (i.e. $\alpha g\gg |\mathbf{f}_{\text{pert}}|$), the changes in the hyperbolic elements are expected to be small, too. Thus, a good understanding can be reached by inserting the constant values of the hyperbolic elements in the right-hand side of Eqs.~\eqref{eq:guauss} in order to get the first-order approximation. For this, it is convenient to use the true anomaly as independent variable instead of the geometrical path length. This requires to invert Eq. \eqref{eq:dfds1} keeping only the linear order in the components of the perturbing gradient. We find the following relations
\begin{subequations}\label{eq:guauss_pert}
\begin{widetext}
\begin{align}
\frac{\dd p}{\dd f}&\simeq 2\frac{n_0}{\eta}\frac{p^3}{\alpha\mu}\frac{1}{(1+e\cos\kappa f)^3}\mathcal{T}\text{,}\label{eq:dpdf}\\
\frac{\dd e}{\dd f}&\simeq\frac{\kappa n_0}{\eta}\frac{p^2}{\alpha\mu}\Bigg\{\frac{\sin\kappa f}{(1+e\cos\kappa f)^2}\mathcal{R}+\frac{1}{\kappa}\Bigg[\frac{2\cos\kappa f+e(1+\cos^2\kappa f)}{(1+e\cos\kappa f)^3}\Bigg]\mathcal{T}-\frac{\alpha\mu}{p}\mathcal{N}\sin\kappa f\Bigg\}\text{,}\label{eq:dedf}\\
\frac{\dd\iota}{\dd f}&\simeq\frac{\kappa^2n_0}{\eta}\frac{p^2}{\alpha\mu}\frac{\cos(f+\omega)}{(1+e\cos\kappa f)^3}\mathcal{S}\text{,}\label{eq:didf}\\
\frac{\dd\Omega}{\dd f}&\simeq\frac{\kappa^2n_0}{\eta}\frac{p^2}{\alpha\mu}\frac{\sin(f+\omega)}{(1+e\cos\kappa f)^3}\mathcal{S}\csc\iota\text{,}\label{eq:dOmegadf}\\
\frac{\dd \omega}{\dd f}&\simeq-\frac{n_0}{\eta e}\frac{p^2}{\alpha\mu}\Bigg\{\frac{\cos\kappa f}{(1+e\cos\kappa f)^2}\mathcal{R}-\frac{1}{\kappa}\Bigg[\frac{2+e\cos\kappa f}{(1+e\cos\kappa f)^3}\sin\kappa f+\frac{e\kappa(1-\kappa^2) f}{(1+e\cos\kappa f)^3}\Bigg]\mathcal{T}\nonumber\\
&+e\kappa^2\frac{\sin(f+\omega)}{(1+e\cos\kappa f)^3}\mathcal{S}\cot\iota-\frac{\alpha\mu}{p}\mathcal{N}\cos\kappa f\Bigg\}\text{,}\label{eq:domegadf}\\
\frac{\dd\delta s}{\dd f}&\simeq-\frac{\kappa}{e}\frac{n_0^2}{\sqrt{\eta^3}}\sqrt{\frac{p^7}{(\alpha\mu)^3}}\Bigg\{\frac{\cos\kappa f}{(1+e\cos\kappa f)^4}\mathcal{R}-\frac{1}{\kappa}\Bigg[\frac{2+e\cos\kappa f}{(1+e\cos\kappa f)^5}\sin\kappa f+\frac{e\kappa(1-\kappa^2) f}{(1+e\cos\kappa f)^5}\Bigg]\mathcal{T}\nonumber\\
&-\frac{\alpha\mu}{p}\frac{(e\eta/n_0+\cos\kappa f)}{(1+e\cos\kappa f)^2}\mathcal{N}\Bigg\}\text{,}\label{eq:dsdf}\\
\frac{\dd\delta t}{\dd f}&\simeq -\frac{\kappa}{ec}\frac{n_0^3}{\sqrt{\eta^3}}\sqrt{\frac{p^7}{(\alpha\mu)^3}}\Bigg\{\frac{\cos\kappa f}{(1+e\cos\kappa f)^4}\mathcal{R}-\frac{1}{\kappa}\Bigg[\frac{2+e\cos\kappa f}{(1+e\cos\kappa f)^5}\sin\kappa f+\frac{e\kappa(1-\kappa^2) f}{(1+e\cos\kappa f)^5}\Bigg]\mathcal{T}\nonumber\\
&-\frac{\alpha\mu}{p}\frac{(2e\eta/n_0+\cos\kappa f)}{(1+e\cos\kappa f)^2}\mathcal{N}\Bigg\}\text{,}\label{eq:dtdf}
\end{align}
\end{widetext}
In the two last equations, we have introduced the non-hyperbolic contributions to the geometrical length and the light time, such that $\delta s=s-s_0$ and $\delta t=t-t_0$, where $s_0$ and $t_0$ are the hyperbolic contributions which are given explicitly in Eqs.~\eqref{eq:Kepler_s} and \eqref{eq:Kepler_t}, respectively.
    
We can also add an expression for the evolution of the argument of the refractive bending which is needed to determine the defocussing and which can also be used to approximate the Doppler frequency shift (cf. discussion in Sec. \ref{subsec:ev_obs}). To do so, one can differentiate Eqs.~\eqref{eq:Bouger} and \eqref{eq:psi}, and then makes use of Eqs.~\eqref{eq:trig_phi} to deduce
\begin{equation*}
\label{eq:dpsi_df}
\dd\psi=-\frac{h}{n}\frac{\dd n}{\dd h}\left(1+\frac{\dd\omega}{\dd f}-\frac{1}{h^2}\frac{\dd K}{\dd n}\frac{\dd s}{\dd f}\right)\dd f\text{,}
\end{equation*}
which is more general than Eq. \eqref{eq:dpsidf} since it includes, besides the hyperbolic term, the variations of $K$ and $\omega$. Substituting the differentials into that equation and keeping only the first order in the components of the perturbation, we end up with
\begin{widetext}
\begin{align}
\frac{\dd\delta\psi}{\dd f}&\simeq-\frac{p}{\eta e}\Bigg\{\frac{(e\eta/n_0+\cos\kappa f)}{(1+e\cos\kappa f)}\mathcal{R}-\frac{1}{\kappa}\Bigg[\frac{2+e\cos\kappa f}{(1+e\cos\kappa f)^2}\sin\kappa f+\frac{e\kappa(1-\kappa^2) f}{(1+e\cos\kappa f)^2}+e^2\kappa^2\frac{\eta}{n_0}\frac{\sin\kappa f}{(1+e\cos\kappa f)^2}\Bigg]\mathcal{T}\nonumber\\
&+e\kappa^2\frac{\sin(f+\omega)}{(1+e\cos\kappa f)^2}\mathcal{S}\cot\iota+\frac{\alpha\mu}{p}(e\eta/n_0-\cos\kappa f)(1+e\cos\kappa f)\mathcal{N}\Bigg\}\text{,}\label{eq:dpsidf_pert}
\end{align}
\end{widetext}
\end{subequations}
where we have removed the hyperbolic contribution, $\psi_0$, which is given exactly in Eq.~\eqref{eq:psi(F)}. We have defined $\delta\psi=\psi-\psi_0$ as the non-hyperbolic contribution to the argument of the refractive bending. Once integrated, that last equation will give us the net change in the refractive bending due to a perturbation and then, Eq. \eqref{eq:DopplerShift_bis} gives us directly the first order effect on the Doppler frequency shift.

\subsection{Method of integration}
\label{subsec:meth_int}

Eqs. \eqref{eq:guauss_pert} are general enough to be applied to different geometries and different kind of problems as AO experiments or GB observations. In both cases, the integration can be performed along the hyperbolic light path to get the first order effects on the hyperbolic elements. Calling $\mathbf{C}$ the vector of the following elements, $\mathbf{C}=(p,e,\iota,\Omega,\omega,\delta s,\delta t,\delta\psi)$, we have
\begin{equation}
\label{eq:int_hyp_elmts}
\Delta\textbf{C}\equiv \mathbf{C}(f_2)-\mathbf{C}(f_1)=\int_{f_1}^{f_2}\frac{\dd\mathbf{C}}{\dd f'}\dd f'\text{,}
\end{equation}
where $f_1$ and $f_2$ are respectively the true anomaly at the transmission and at the reception point, with $f_2\geq f_1$. The term $\dd\mathbf{C}/\dd f'$ represents the rate of change of the elements with respect to the true anomaly and is obviously given by Eqs. \eqref{eq:guauss_pert}.

In general, direct integration, as in Eq. \eqref{eq:int_hyp_elmts}, may lead to complex solutions. So, if small quantities show up in Eqs. \eqref{eq:guauss_pert}, a Taylor expansion in term of these quantities can first of all let to simplify the integration and secondly lead to a more user-friendly solution.

For instance, in the context of AO experiments and GB observations, we can think of the inverse of the eccentricity as a small quantity. Indeed, in most experiments, we have to deal with a small refractivity which implies $\alpha\ll 1$ and then, from Eq.~\eqref{eq:val_K/e}, we deduce $e\gg 1$ (as long as $K\gg\alpha\mu$). The meaning is that the shape of the hyperbola departs slightly from a straight-line trajectory. However, if this approximation is largely relevant for AO, it can become somehow inaccurate for GB experiments, in particular at very high elevations (close to the zenith-direction of the site). Indeed, $K$ might become the same order as $\alpha\mu$, which implies $e\sim 1$. Hence, the solutions at first order in $1/e$ are not valid for GB experiments at very high elevations (i.e. when $K\sim\alpha\mu$). For instance in the Earth's atmosphere \footnote{Following \cite{doi1010292010JD015214}, for an average parcel of air at sea level, the refractivity is approximately $(3\pm1)\times10^{-4}$, which represents a value of $\alpha=(5\pm2)\times10^{-6}\,\mathrm{km^{-2}\cdot s^2}$. So, the region around the zenith-direction, in which the eccentricity is $e\lesssim10$, is enclosed inside a cone with a half top angle which remains $\lesssim0.3^{\circ}$. All the light path trajectories with $K\lesssim 30\,\mathrm{km}$, are enclosed inside that region.}, if we consider that the expansion in $1/e$ does not hold for $e\lesssim10$, we must consider only trajectories with $K\gtrsim 30\,\mathrm{km}$, which corresponds to elevations angles $\lesssim89.7^{\circ}$.

For GB observations at very high elevations, the light ray trajectory is close to be radial, that is to say that the change in the true anomaly between the transmitter and the receiver $(f_2-f_1)$ is a small quantity. For that particular case, we do not need to integrate Eqs.~\eqref{eq:int_hyp_elmts}, instead an evaluation of Eqs.~\eqref{eq:guauss_pert} at the level of the transmitter is sufficient to determine the change in the hyperbolic elements at first order in $(f_2-f_1)$
\begin{equation}
\Delta\mathbf{C}=(f_2-f_1)\frac{\dd\mathbf{C}}{\dd f}(f_1)\text{.}
\label{eq:int_eq_guauss}
\end{equation}
Therefore, for all the following applications (see Chap.~\ref{sec:applications}), we consider the leading order in $1/e$ since: i) it encompasses almost all experimental cases (usually, at zenith angles less than few degrees, source tracking can be difficult), ii) the transposition to GB observations at very high elevations (zenith angles less than few degrees) reduces to an evaluation of Eqs.~\eqref{eq:guauss_pert} at the level of the transmitter.

\section{Applications}
\label{sec:applications}

In this chapter, different types of perturbations are analyzed within the osculating equations formalism. In each application, we aim at determining the variations of the photon path and the light time due to these perturbations. Then, the change in the observables can be easily inferred as discussed in Sec.~\ref{subsec:obs}. 

We first consider that the index of refraction possesses non-linear dependences to the monopole term of the gravitational potential. Then, we explore the impact of the centrifugal potential as well as the drag effect on the light propagation due to the moving medium. We also apply the perturbation equations considering the non-spherical part of the self-gravitational potential of the central planet. Then, we study the perturbation on the light beam trajectory caused by a perturbing body raising tides on the central planet's atmosphere. Finally, we close the section by studying the effects due to the presence of horizontal gradients in a spherically shaped atmosphere.

\subsection{Definitions and hypothesis}
\label{subsec:app_def}

In this section, we introduce the basic ideas which will help us to formulate the internal problem\footnote{In the Solar System, the internal problem (study of the internal structure of a self-gravitating bodies) can be mainly decoupled from the external one (inter-body dynamics). This is true as long as the inter-body distance is much more higher than a characteristic length scale within the extended bodies. Consequently, an external stress can be considered as a perturbation in the internal problem and conversely, an internal stress may be seen as a perturbation in the external problem.}. The idea is to determine a generic refractive profile assuming the influence of different sources of stress.

Lately, we have made use of three reference frames, all centered at the planet's center of mass and referred to as the propagation frame $(\hat{\mathbf{x}},\hat{\mathbf{y}},\hat{\mathbf{z}})$, the polar basis $(\hat{\bm{\rho}},\hat{\bm{\tau}},\hat{\bm{\sigma}})$, and the reference frame $(\hat{\mathbf{X}},\hat{\mathbf{Y}},\hat{\mathbf{Z}})$. The first one was helpful to derive the solution of reference (hyperbolic path) by making use of the second one to introduce polar coordinates. The last one was introduced to infer the perturbation equations in an arbitrary oriented frame. It was assumed that the medium was at rest simultaneously in the propagation frame and in the frame of reference, also the equations of geometrical optics [cf. Eq.~\eqref{eq:optics}] were consequently available inside these two frames. 

When considering the internal motion of the medium with respect to the reference frame, the equation of geometrical optics [cf. Eq. \eqref{eq:optics}] no longer stands into the propagation or the reference frames since it is expressed in a frame comoving with the medium. Therefore, in order to maintain the coherence with the reference solution which was expressed in the reference frame, we deal with the theory of light propagation in moving medium \cite{1818...AF}. Following \cite{1999PhRvA..60.4301L} and \cite{Rozanov2005}, the equation of geometrical optics expressed in the medium comoving frame is
\begin{subequations}\label{eq:optics_acc}
\begin{align}
\frac{\dd}{\dd s}\big(\bm\nabla\mathscr{S}\big)&=\bm{\nabla}n+\mathbf{f}_{\text{drag}}\text{,}\label{eq:optics_drag}
\end{align}
where $n$ still refers to the value of the index of refraction as measured by an observer at rest in the refractive medium. The additional term refers to a dragging contribution which is given by
\begin{align}
\mathbf{f}_{\text{drag}}&=\frac{2}{c}\Big[(\bm\nabla\mathscr{S}\cdot\bm{\nabla}n)\mathbf{v}-(\mathbf{v}\cdot\bm\nabla\mathscr{S})\bm{\nabla}n\Big]+\frac{\gamma n}{c}\bm\nabla\mathscr{S}\times(\bm{\nabla}\times\textbf{v})+\mathcal{O}\left(v^2/c^2\right)\text{,}\label{eq:f_drag}
\end{align}
\end{subequations}
where $\textbf{v}$ is the velocity vector of the medium, with $v=|\textbf{v}|$. $\gamma$ is the Fresnel's dragging coefficient introduced in Eq.~\eqref{eq:drag_coeff}. It might be seen from this expression, that the corrections to Eq.~\eqref{eq:optics} are of the order of $v/c$. For gas giant planets in the Solar System (e.g. Jupiter), the typical upper bound value of the velocity for the solid rotation at the equator is $5\times10^4\,\mathrm{km/h}$ with zonal winds asymmetric velocity ranging between $\pm 550\,\mathrm{km/h}$ \cite{2018Natur.555..223K}, also $v/c\lesssim 10^{-4}$. Therefore, we will consider the effect of the moving medium as a perturbation to the hyperbolic path,
\begin{equation}
\label{eq:fpert_moving}
\mathbf{f}_{\text{pert}}=\mathbf{f}_{\text{drag}}\text{.}
\end{equation}
%It is obvious that the perturbation is turned on only when the height of the ray is below the radii of the top of the atmosphere.
This contribution will be considered later in the context of a steady rotating atmosphere.

To handle the internal motion let us introduce a new reference frame $(\hat{\mathbf{X}}',\hat{\mathbf{Y}}',\hat{\mathbf{Z}}')$ centered at the planet's center of mass and rotating with it (we call $P$ the central planet). The $\hat{\mathbf{Z}}'$-axis is chosen to be orthogonal to the equatorial plane of $P$. The angular velocity vector is considered to be constant and given by $\textbf{w}=w\hat{\mathbf{Z}}'$. Since $\textbf{w}$ is independent of time, the $\hat{\mathbf{Z}}'$-axis is spatially fixed. Therefore, we choose the $\hat{\mathbf{Z}}$-axis of the reference frame to be collinear with the $\hat{\mathbf{Z}}'$-axis. The frame $(\hat{\mathbf{X}}',\hat{\mathbf{Y}}',\hat{\mathbf{Z}}')$ is well suited for the study of internal motions of $P$ and will be referred to as the fluid rotating frame.

It is shown in \cite{2014grav.book.....P} (cf. p.~106) that Euler's equation describing the time evolution of the velocity of a fluid element makes appear the following generalized potential\footnote{We took the convention that the Newtonian gravitational potential is a negative quantity, also the signs in that expression differs from \cite{2014grav.book.....P}.}
\begin{equation}
\label{eq:gen_pot}
\Phi=U_0+\sum_{l\geq 2} U_l-\Phi_C+U_{\text{tidal}}(t)\text{,}
\end{equation} 
once written in the fluid rotating frame. In the following, we consider that the dominant term of that expression is the monopole term $U_0$ of the gravitational potential of the fluid planet. The other terms will be considered as perturbations before $U_0$. The $U_l$ terms are the non-spherical parts of the central planet self gravitational potential, $\Phi_C$ is the centrifugal potential due to planet's proper rotation, and $U_\text{tidal}(t)$ is the external potential due to the presence of other massive bodies in the surrounding of the central planet. This last term is dynamic since it evolves with the positions of the perturbing bodies. Each contribution will be studied in turn.

We now have to define an expression for the index of refraction. If we assume that the refractive index is directly related to the density of the fluid, we can consider that Eq.~\eqref{eq:gen_pot} constitutes the generalized potential in Eq.~\eqref{eq:dn/dPhi}. In addition, successive integrations by part of Eq.~\eqref{eq:dn/dPhi} leads to the following infinite series
\begin{equation}
\label{eq:n(Phi)}
n=\eta-\alpha\Phi+\sum_{k=2}^{+\infty}\frac{(-1)^k}{k!}\alpha_{(k)}\Phi^k\text{,}
\end{equation}
where $\Phi$ is given by Eq. \eqref{eq:gen_pot}, and $\alpha_{(k)}\equiv \dd^{k}n/\dd\Phi^{k}$ for $k\geq 2$. We keep the same notation as in previous chapters, i.e. $\alpha\equiv \alpha_{(1)}$ is still the linear term with the generalized potential. This series can represent any function of the gravitational potential, and the choice is governed by the numerical values which are assigned to the $\alpha_k$ coefficients. 

Eqs. \eqref{eq:optics_acc}--\eqref{eq:n(Phi)} are all we need to formulate the internal problem and study first order effect on the light propagation due to the different pieces in Eqs.~\eqref{eq:gen_pot} and \eqref{eq:n(Phi)}.

\subsection{Non-linearity}
\label{subsec:nonlinear}

Previously in Chap. \ref{sec:ref}, we have assumed $\dd n/\dd\Phi=\text{cst}$ [cf. Eq.~\eqref{eq:dn/dPhi}] in order to simplify the integration of Eq.~\eqref{eq:optics}. However, this could be too restrictive in some applications where it is not possible to build a multi-layers modeling of the atmosphere. If a multi-layers modeling can be achieved as in \cite{2015RaSc...50..712S} or in \cite{doi:10.1029/2006JB004834}, a constant value of $\alpha$ can be associated to each layer and the total profile can be recovered by integrating Eq. \eqref{eq:dn/dPhi}.

In this section, we propose to explore the first order effect of non-linearity with isopotentials in the refractive profile [see Eq.~\eqref{eq:n(Phi)}]. We focus on the simple case where the generalized potential is given by the monopole term of the Newtonian potential, $\Phi=U_0$. All the mathematical details are exposed in Sec.~\ref{subsec:detail_nonlin} and the change in hyperbolic elements for any degree $k$ is given at leading order in $1/e$ in Eq.~\eqref{eq:Delta_guauss_nonlin}. The application to the quadratic contribution $(k=2)$ is given in Eq.~\eqref{eq:Delta_guauss_nonlin1}.

After integration, it is shown that whatever function of the Newtonian potential the index of refraction is, $p$, $\iota$, and $\Omega$ remain always constants regardless degree $k$. This is due to the fact that the index of refraction acts like a central field with no transverse or normal components. %Another interesting fact concerns AO experiments for which it might be seen from Eq. \eqref{eq:Delta_guauss_nonlin_e}, that the variation in eccentricity is expected to be small. Indeed, as stated before the major part of the variation is expected to be contained in between $f_i$ and $f_e$ where $f_e\simeq-f_i$ for a nearly spherical atmosphere. Then, $e(f_e)$ is expected to have the same value as $e(f_i)$, also $\Delta e\simeq0$. Therefore, we might expect that the determination of the eccentricity as provided by the only hyperbolic contribution is sufficiently good, meaning that the eccentricity remains almost constant along the all light path. This is no longer true for GB observations for which $|f_e|\neq |f_i|$.

We copy here the expressions of the non-hyperbolic contribution to the light time and the refractive bending for $f_2\geq f_1$ and $k=2$
\begin{align*}
\Delta\delta t&=2\eta\frac{\alpha_{(2)}\mu^2}{cK}(f_2-f_1)\text{,}\\
\Delta\delta\epsilon&=\frac{\eta}{2}\frac{\alpha_{(2)}\mu^2}{K^2}\left(f_2-f_1+\cos f_2\sin f_2-\cos f_1\sin f_1\right)\text{.}
\end{align*}
Obviously, from an observational point of view, these expressions are of a great interest. The first one provide directly the additional time-delay with respect to the hyperbolic path, and the second one provides the supplementary contribution to the hyperbolic refractive bending. Both are function of the geometry of the problem only. %From Eq. \eqref{eq:Delta_guauss_nonlin_psi}, the argument of the refractive bending at the reception is given by $\psi(f_2)=\psi_0(f_2)+\delta\psi(f_2)$, where $\psi_0(f_2)$ is the hyperbolic contribution which is given by $\psi_0(f_2)=\psi(f_1)+\epsilon_0(f_2)$ with $\epsilon_0$ the hyperbolic refractive bending. Also using the definition of the total refractive bending, namely $\epsilon(f_2)\equiv \psi(f_2)-\psi(f_1)$, we deduce that $\epsilon(f_2)=\epsilon_0(f_2)+\delta\psi(f_2)$.
In addition, from Eq.~\eqref{eq:DopplerShift_bis}, it can be seen that the later provides directly the change in the Doppler measurement due to $\alpha_{(2)}$.

\subsection{Steady rotating atmosphere}
\label{subsec:centrifuge}

The second application concerns the rotation of the atmosphere. All bodies in the Solar System are rotating. The rotation motion can be decomposed into two parts. The first concerns the proper rotation of the body around a certain axis of rotation, while the second concerns the spatial orientation of that axis. The generalized potential in Eq. \eqref{eq:gen_pot} neglects the second part and considers a steady rotating planet.

In the frame rotating with the fluid, the media experiences the effect of the proper rotation via the simplified following generalized potential $\Phi=U_0-\Phi_C$, where $\Phi_C$ represents the centrifugal potential. In addition, the dragging effect due to the rotational motion of the medium is described by the additional contribution $\textbf{f}_{\text{drag}}$ in Eqs.~\eqref{eq:optics_acc}. The solutions describing the change in hyperbolic elements due to a steady rotating atmosphere are given in Eqs.~\eqref{eq:Delta_guauss_PhiC} and \eqref{eq:Delta_guauss_drag}. All the mathematical details are presented in Sec.~\ref{subsec:detail_steadrot}.

On one hand, Eqs.~\eqref{eq:Delta_guauss_PhiC} describe the change in hyperbolic elements due to a centrifugal potential induced by the proper rotation of the planet. The expressions of the non-hyperbolic contribution to the light time and the refractive bending are
\begin{align*}
\Delta\delta t&=-\frac{\alpha w^2K^3}{24c\eta^3}\sin^2\iota\Big[8\cos 2\omega\left(\tan^3 f_2-\tan^3 f_1\right)+3\sin 2\omega\left(\sec^4 f_2\big[1+2\cos 2 f_2\big]-\sec^4 f_1\big[1+2\cos 2 f_1\big]\right)\Big]\text{,}\\
\Delta\delta\epsilon&=\frac{\alpha w^2K^2}{4\eta^3}\Big[\sin 2\omega\sin^2\iota\left(\sec^2f_2-\sec^2f_1\right)-\big(3+\cos 2\iota+2\cos 2\omega\sin^2\iota\big)\big(\tan f_2-\tan f_1\big)\Big]\text{,}
\end{align*}
with $f_2\geq f_1$.

On the other hand, Eqs.~\eqref{eq:Delta_guauss_drag} describe the change in hyperbolic elements induced by the light dragging effect due to the rotation of the fluid material which composes the atmosphere of the central planet. The expressions of the non-hyperbolic contribution to the light time and the refractive bending are
\begin{align*}
\Delta\delta t&=-\frac{2\gamma}{3}\frac{K^2w}{c^2}\Big[(2+\cos 2f_2)\sec^2 f_2\tan f_2-(2+\cos 2f_1)\sec^2 f_1\tan f_1\Big]\cos\iota\text{,}\\
\Delta\delta\epsilon&=-2\gamma \frac{K w}{c}\big(\tan f_2-\tan f_1\big)\cos\iota\text{,}
\end{align*}
with $f_2\geq f_1$.

%An important feature of these two set of equations is that they cannot be employed when $h\rightarrow\infty$. Indeed, when the magnitude of the separation vector grows to infinity, the true anomaly goes to values similar to $\pm\pi/2$ (in the limit of small refractivity\footnote{Actually, $f\rightarrow\pm\pi/2$ when the refractivity is null.}). Also, because of terms in $\sec f$ or in $\tan f$, the two set of solutions grow to infinity. Such terms are due to the high velocity of the steady rotating medium ($w h$) that is observed when $h$ goes to infinity. However, because the atmosphere of the central planet does not extend to infinity, the true anomaly can never reach values close to $\pm\pi/2$, especially in the limit of small refractivity. 
These relationships show that the first set is quadratic with the magnitude of the angular velocity vector while the latter evolves linearly with it. Therefore, the change in hyperbolic elements due to the centrifugal potential is independent of the orientation of $\textbf w$ and remains the same for either a direct or an indirect rotation. On the contrary, the change in hyperbolic elements due to the dragging of light is dependent of the direction of the medium's rotation. %To spotlight this point, let us consider e.g. a light ray entering the atmosphere at $f_i=-f_{\text{ini}}$ and exiting the same atmosphere at $f_e=f_{\text{ini}}$ (e.g. like in AO experiments). In that case, the sens of propagation of the ray is the same as the velocity of the moving medium. From Eq.~\eqref{eq:Deltat_drag}, we see that the dragging effect due to the direct rotation of the atmosphere decreases the light time by the following amount
%\begin{equation*}
%\Delta\delta t=-\frac{4\gamma}{3}\frac{K^2w}{c^2}\cos\iota(2+\cos 2f_{\text{ini}})\sec^2f_{\text{ini}}\tan f_{\text{ini}}\text{.}
%\end{equation*}
%Similarly, from Eq.~\eqref{eq:Deltapsi_drag}, we deduce that the refractive bending is also decreased by
%\begin{equation*}
%\Delta\delta\epsilon=-4\gamma\frac{Kw}{c^2}\cos\iota\tan f_{\text{ini}}\text{.}
%\end{equation*}
%However, if the rotation is indirect the sign of $\textbf w$ is reversed, then both the light time and the refractive bending are increased (respectively by the amounts $|\Delta\delta t|$ and $|\Delta\delta\epsilon|$). Also, one can see that for AO experiments [see Eq.~\eqref{eq:DopplerShift_bis_plan_occ}], the received frequency is either smaller or greater than the received frequency computed without dragging effect, for a direct or an indirect rotation of the atmosphere respectively.
Because of a cosine of the inclination, we can also see that the effect is maximum in the planetary equator, where the distance with respect to the axis of rotation is maximum.

Another important difference between the effects due to the centrifugal potential and those due to the dragging of light can be inferred by comparing the expressions of the non-hyperbolic contribution to the light time and the path length. In the case of light dragging, it is seen in Eq.~\eqref{eq:Deltat_drag} that the non-hyperbolic contribution to the light time reduces to the non-hyperbolic contribution to the geometrical length divided by the speed of light in vacuum. This comes from the fact that the $\mathcal{N}$ perturbing component is null for the light dragging effect while it is not for the centrifugal potential contribution.

\subsection{Axisymmetric gravitational potential}
\label{subsec:nonspherical}

Because of their proper rotation, non-rigid bodies tend to be flattened due to centrifugal forces. Then, the mass repartition becomes slightly different from what would be expected in spherical symmetry, and consequently, the gravitational potential also changes. This mass redistribution due to centrifugal forces is usually the most important departure from the monopole contribution to the total self-gravitational potential of the planet.

The non-spheric components of the gravitational potential of the central planet can be modeled using the simplified following generalized potential $\Phi=U_0+\sum_{l}U_l$ with $l\geq 2$. The solutions for the change in the hyperbolic elements are derived in Sec.~\ref{subsec:detail_axisym} and the results for $l=2$ are given in Eqs.~\eqref{eq:Delta_guauss_J2}. We copy here the expressions of the non-hyperbolic contribution to the light time and the refractive bending for $f_2\geq f_1$
\begin{align*}
\Delta\delta t&=\frac{\eta^2 J_2}{8}\frac{\alpha\mu R^2}{cK^2}\Big[\sin^2\iota\Big(9\sin 2\omega\big[\cos f_2-\cos f_1\big]+7\big[\sin(3f_2+2\omega)-\sin(3f_1+2\omega)\big]\Big)\nonumber\\
&+\Big(20-\sin^2\iota\big[30-9\cos 2\omega\big]\Big)\big(\sin f_2-\sin f_1\big)\Big]\text{,}\\
\Delta\delta\epsilon&=\frac{\eta^2 J_2}{32}\frac{\alpha\mu R^2}{K^3}\Big[\big(1+3\cos 2\iota\big)\big(\sin 3f_2-\sin 3f_1\big)+6\Big(6-\sin^2\iota\big[9-4\cos 2\omega\big]\Big)\big(\sin f_2-\sin f_1\big)\nonumber\\
&+2\sin^2\iota\Big(\big[1+3\cos 2f_2\big]\big[3\sin(f_2+2\omega)+\sin(3f_2+2\omega)\big]-\big[1+3\cos 2f_1\big]\big[3\sin(f_1+2\omega)+\sin(3f_1+2\omega)\big]\Big)\Big]\text{.}
\end{align*}

Considering that the centrifugal potential and the quadrupole moment of the gravitational potential of the central planet are both axisymmetric fields, one might expect similar signatures in the changes induced on the hyperbolic elements. However, because they evolve differently with $h$ (the centrifugal effect tends to grow with $h$, where the $J_2$ effect decreases with $h$), a comparison shows that the signatures produced on the hyperbolic elements differ. However, one can see that the ratio between Eqs.~\eqref{eq:Delta_guauss_J2} and \eqref{eq:Delta_guauss_PhiC} is always proportional to
\begin{equation*}
\frac{\textbf{C}_{J_2}}{\textbf{C}_C}\propto\frac{\mu J_2R^2\eta^5}{w^2K^5}\text{,}
\end{equation*}
where $\textbf{C}$ represents the set of hyperbolic elements. The subscript $J_2$ refers to the effect due to the quadrupolar moment of the gravitational potential, where the subscript $C$ refers to the contribution due to the centrifugal potential.

For a planet like Jupiter, considering that the impact parameter of the ray remains at the level of the equatorial radius, the ratio is $0.16$. For Saturn, the Earth, and Titan we find respectively $0.11$, $0.31$, and $0.75$. In each case the centrifugal contribution is the most important. However, it must be noticed that this ratio can grow while the light ray goes deeper inside the atmosphere, since the impact parameter decreases. This can make the $J_2$ effect being more important than the centrifugal one, especially for Titan for which the rotation rate is slow. The ratio can also exceed unity for GB observations carried out at high elevation because $K$ might become much smaller than the equatorial radius.

\subsection{Tidal potential}
\label{subsec:tide}

In systems containing close orbiting bodies, e.g. a planet and its satellites, the question of knowing whether the tidal effect due to a perturbing body on the planetary atmosphere can affect the light propagation or not can be raised.

To study this possibility we focus on the tidal contribution in the generalized potential and ignore the body's rotational deformation and the non-spherical gravitational potential contribution which were treated previously. Hence, we consider the simplified following generalized potential $\Phi=U_0+U_{\text{tidal}}(t)$. In addition, we will assume that the characteristic time of variation of the external tidal field is so slow that it never takes the central planet's atmosphere out of hydrostatic equilibrium. This assumption is known as the static tides approximation.

All the computations are detailed in Sec.~\ref{subsec:detail_tide}, and the evolution of the hyperbolic elements are given in Eqs.~\eqref{eq:delta_gauss_sat}. We remind that these equations are derived considering a tide raising body moving on an equatorial circular orbit around the central planet, and a photon path lying inside the equatorial plane (i.e. $\iota=0$). We copy here the expressions of the non-hyperbolic contribution to the light time and the refractive bending for $f_2\geq f_1$
\begin{align*}
\Delta\delta t&=-\frac{1}{320\eta^4}\frac{\alpha\mu K^4}{cr^4}\frac{\mu_A}{\mu}\Big\{\sec^5 f_2\big[15\sin(5f_2+3Nt-3\varpi)+15\sin(5f_2-3Nt+3\varpi)\nonumber\\
&+7\sin(5f_2+Nt-\varpi)+7\sin(5f_2-Nt+\varpi)+75\sin(3f_2+3Nt-3\varpi)+75\sin(3f_2-3Nt+3\varpi)\nonumber\\
&+35\sin(3f_2+Nt-\varpi)+35\sin(3f_2-Nt+\varpi)+150\sin(f_2+3Nt-3\varpi)-150\sin(f_2-3Nt+3\varpi)\nonumber\\
&+70\sin(f_2+Nt-\varpi)+10\sin(f_2-Nt+\varpi)\big]\nonumber\\
&-\sec^5 f_1\big[15\sin(5f_1+3Nt-3\varpi)+15\sin(5f_1-3Nt+3\varpi)\nonumber\\
&+7\sin(5f_1+Nt-\varpi)+7\sin(5f_1-Nt+\varpi)+75\sin(3f_1+3Nt-3\varpi)+75\sin(3f_1-3Nt+3\varpi)\nonumber\\
&+35\sin(3f_1+Nt-\varpi)+35\sin(3f_1-Nt+\varpi)+150\sin(f_1+3Nt-3\varpi)-150\sin(f_1-3Nt+3\varpi)\nonumber\\
&+70\sin(f_1+Nt-\varpi)+10\sin(f_1-Nt+\varpi)\big]\Big\}\text{,}\\
\Delta\delta\epsilon&=-\frac{1}{16\eta^4}\frac{\alpha\mu K^3}{r^4}\frac{\mu_A}{\mu}\Big\{\sec^3 f_2\big[5\sin(3f_2+3Nt-3\varpi)+5\sin(3f_2-3Nt+3\varpi)+6\sin(f_2+Nt-\varpi)\nonumber\\
&+2\sin(3f_2+Nt-\varpi)+2\sin(3f_2-Nt+\varpi)+15\sin(f_2+3Nt-3\varpi)-15\sin(f_2-3Nt+3\varpi)\big]\nonumber\\
&-\sec^3 f_1\big[5\sin(3f_1+3Nt-3\varpi)+5\sin(3f_1-3Nt+3\varpi)+6\sin(f_1+Nt-\varpi)\nonumber\\
&+2\sin(3f_1+Nt-\varpi)+2\sin(3f_1-Nt+\varpi)+15\sin(f_1+3Nt-3\varpi)-15\sin(f_1-3Nt+3\varpi)\big]\Big\}\text{.}
\end{align*}

%Let us emphasis that the method of integration employed here is different from what has been done in the previous sections (see discussion in Sec.~\ref{subsec:meth_int}). Because we expended the integrand around a certain value $f_x$ before integration, the solutions which are discussed here are defined at first orders in $f_e-f_i$ (with $f_e$ the true anomaly at the entrance of the refractive region of influence and $f_i$ the true anomaly at the exit of the refractive region of influence), and at first order in $1/e$.

Comparison between the magnitude of the change in hyperbolic elements in Eqs.~\eqref{eq:delta_gauss_sat} and \eqref{eq:Delta_guauss_J2} reveals
\begin{equation*}
\cfrac{\textbf{C}_{\text{tide}}}{\textbf{C}_{J_2}}\propto\cfrac{\mu_AK^6}{\mu J_2r^4R^2\eta^6}\text{,}
\end{equation*}
For the tides raised by the Moon on Earth, the ratio is of the order of $10^{-6}$. For tides raised by Titan on Saturn, the value is $10^{-7}$, and for tides raised by Io on Jupiter the ratio is of the order of $10^{-5}$. For all those cases, the tides effects are at least 5 orders of magnitude smaller than the one of the oblateness. However, they can become important when the light ray passes through the atmosphere of a satellite orbiting a central massive planet. In such a configuration the tidal effects are expected to be more important. For instance, let us consider the case of a light ray crossing Titan's atmosphere; substitution of numerical values (we took the Titan's $J_2$ value from \cite{2012Sci...337..457I}) reveals that the changes due to Saturn's tides are expected to be the same order of magnitude than changes due to Titan's oblateness.   

All the results in Eqs.~\eqref{eq:delta_gauss_sat} are derived under the assumption that the main planet is made with a perfect fluid medium having a non-unity index of refraction. However, such a perfect fluid model might be somewhat inaccurate in the context of tidal dynamics. The main reason is linked to the fact that perfect fluid model does not admit a mechanism to dissipate energy and the fluid's response to a tidal stress is purely elastic. In the context of celestial mechanics, energy dissipation mechanisms are of prime importance in a large number of applications \cite{2000ssd..book.....M}. In Sec.~\ref{subsec:detail_tide}, we compute the first order effect on the hyperbolic elements caused by the non-elastic response of the atmosphere following a tidal stress. Referring to Eq.~\eqref{eq:comp_tide_elnonel}, we see that the non-elastic contribution is a least $w\tau$ times the elastic one, with $w$ the magnitude of the angular velocity of the central planet, and $\tau$ the time-delay which is related to dissipative phenomenon.

For the Earth, a typical value of $\tau$ corresponding to rotational semi-diurnal deformation is $4\,\mathrm{min}$ \cite{2014IPNPR.196C...1F}, so the ratio is of the order $2\times10^{-2}$. So, if the elastic contribution is important in some circumstances, the viscoelastic response of the atmosphere represents a perturbation of a few percent of the amplitude of the elastic one. In the case of Saturn and Jupiter, we can use the fact that $w\tau\approx (2Q)^{-1}$ where $Q$ is the specific dissipation function \cite{2000ssd..book.....M}. The numerical substitution is done by taking value of $k_2/Q$ (with $k_2$ the gravitational Love number of degree 2) from \cite{2009Natur.459..957L} together with the value of $k_2$ provided by \cite{2018Natur.555..220I} for Jupiter and the one provided by \cite{2017Icar..281..286L} for Saturn. We find the ratios to be respectively of the order of $8\times10^{-6}$ and $7\times10^{-5}$, which is well beyond the elastic response.

\subsection{Horizontal gradients}
\label{subsec:horiz_grad}

The last study is a direct application for GB observations operating for the realization of IERS reference systems e.g. SLR, VLBI, or the Global Positioning System (GPS). In \cite{1997JGR...10220489C} or in \cite{doi:10.1029/2006JB004834} it is shown that the group delay due to the horizontal gradients in the Earth atmosphere can overpass the centimetric level while observational techniques like SLR or VLBI currently operate with sub-centimeter accuracy measurements. Therefore, not considering horizontal gradients may lead to systematic errors in the estimation of the station coordinates, which can have repercussions on the determination of the International Terrestrial Reference Frame (ITRF). Consequently, delays due to horizontal gradients must be taken into account, especially at low elevations where the effect is maximum.

The computations are carried out in Sec.~\ref{subsec:detail_HG}, and the equations describing the change in the hyperbolic elements following horizontal gradients are given in Eqs.~\eqref{eq:delta_gauss_HG}. We remind that these equations are derived for a simplified geometry. Indeed, we have assumed that the transmitting source is observed at its highest elevation as seen from the receiving site on Earth. This situation is encountered when the azimuth of the transmitter equals the azimuth of the observer, that is to say when the source is at the meridian of the receiving site. In such a case, the propagation plane is aligned with the meridian and intersects Earth's equator for $\iota=\pi/2$ and $\lambda=\Omega$. For this type of inclination, it is seen from Eqs.~\eqref{eq:delta_gauss_HG} that only a West-East horizontal variation of the refractivity can change the inclination or the longitude of the node. Conversely, only a North-South horizontal gradient is expected to change the other hyperbolic elements.

We copy here the results for the non-hyperbolic contribution to the light time
\begin{align}
\delta t(f)&=\text{sign}(f_1-f)\frac{Kn_{\text{NS}}}{2\eta c}\Big[\sec^2f-\sec^2f_1+4\ln|\cos f|-4\ln|\cos f_1|-4(f_2-f)\tan f+4(f_2-f_1)\tan f_1\Big]\text{.}\label{eq:deltat_horGrad}
%\delta\epsilon(f)&=\frac{n_{\text{NS}}}{\eta}\Big(\ln|\cos f|-\ln|\cos f_1|\Big)\text{.}
\end{align}
For applications, the true anomaly is not a common way of expressing the results. Instead, the colatitude (referred from the direction of the North pole) may be preferred. In order to pass from true anomalies to colatitudes, we make use of hyperbolic relations defined in appendix~\ref{sec:hyp}, in particular Eqs.~\eqref{eq:f_mono} and \eqref{eq:omega}. Because, this transformation is only a matter of rewriting the boundary conditions of Eq.~\eqref{eq:deltat_horGrad}, we can safely approximate the photon path assuming straight line between the source and the receiver. In this case, the hyperbolic relations in appendix~\ref{sec:hyp} can be further simplified taking the limit $\alpha\rightarrow 0$, which corresponds to light propagation in vacuum.

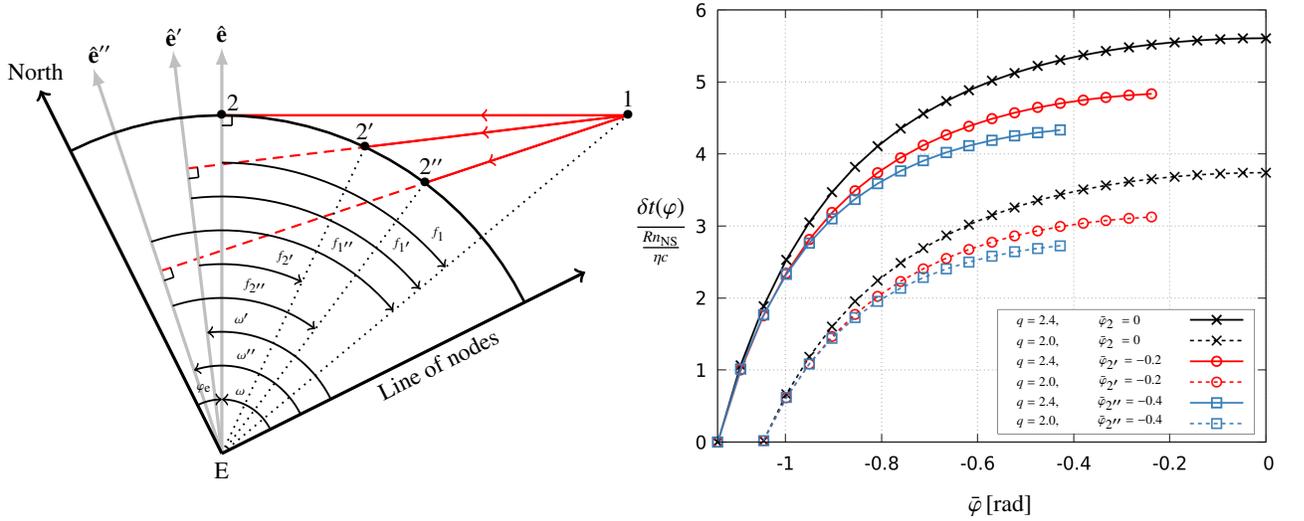
\begin{figure*}
%  \begin{center}
%    \includegraphics[scale=2.22]{horiz_grad}
%  \end{center}  
\begin{minipage}{0.46\textwidth}
\begin{tikzpicture}[scale=0.9]

% light ray
\draw [line width=.8pt,color=red] (0.,5.)-- (6.,5.);
\draw [line width=.8pt,color=red] (2.1152405075533625,4.530536126685824)-- (6.,5.);
\draw [line width=.8pt,color=red] (3.,4.)-- (6.,5.);
\draw [line width=.8pt,color=red,->] (6.,5.)-- (3.8181818181818183,5.);
\draw [line width=.8pt,color=red] (3.8181818181818183,5.)-- (0.,5.);
\draw [line width=.8pt,color=red,->] (6.,5.)-- (3.833941205035555,4.738236986437014);
\draw [line width=.8pt,color=red] (2.1152405075533625,4.530536126685824)-- (3.833941205035555,4.738236986437014);
\draw [line width=.8pt,color=red] (3.,4.)-- (3.930145531526153,4.310048510508717);
\draw [line width=.8pt,color=red,<-] (3.930145531526153,4.310048510508717)-- (6.,5.);
\draw [line width=.8pt,dash pattern=on 4.pt off 2.5pt,color=red] (-0.5091770617666863,4.213381553878333)-- (2.1152405075533625,4.530536126685824);
\draw [line width=.8pt,dash pattern=on 4.pt off 2.5pt,color=red] (-0.9,2.7)-- (3.,4.);

% Positions
\draw [line width=.7pt,dotted] (0.,0.)-- (6.,5.);
\draw [line width=.7pt,dotted] (0.,0.)-- (2.1152405075533625,4.530536126685824);
\draw [line width=.7pt,dotted] (0.,0.)-- (3.,4.);

% Eccentricity vectors
\draw [>=latex,->,line width=1.2pt,color=gray!50] (0.,0.) -- (0.,6.);
\draw [>=latex,->,line width=1.2pt,color=gray!50] (0.,0.) -- (-0.7198482872982089,5.956661686152222);
\draw [>=latex,->,line width=1.2pt,color=gray!50] (0.,0.) -- (-1.8973665961010284,5.692099788303082);

% Vectors of the frame
\draw [->,line width=1.2pt] (0.,0.) -- (-2.6823338911537853,5.367036882337198);
\draw [->,line width=1.2pt] (0.,0.) -- (5.3670368823371986,2.6823338911537857);

% Earth curvature
\draw [shift={(0.,0.)},line width=1.pt]  plot[domain=0.463471026709497:2.0342673535043936,variable=\t]({1.*5.*cos(\t r)+0.*5.*sin(\t r)},{0.*5.*cos(\t r)+1.*5.*sin(\t r)});

% Angles
\draw [shift={(0.,0.)},line width=.7pt,<-]  plot[domain=1.5707963267948966:2.0342673535043936,variable=\t]({1.*0.8*cos(\t r)+0.*0.8*sin(\t r)},{0.*0.8*cos(\t r)+1.*0.8*sin(\t r)});
\draw [shift={(0.,0.)},line width=.7pt,<-]  plot[domain=0.6947382761967031:1.8925468811915391,variable=\t]({1.*3.3*cos(\t r)+0.*3.3*sin(\t r)},{0.*3.3*cos(\t r)+1.*3.3*sin(\t r)});
\draw [shift={(0.,0.)},line width=.7pt,<-]  plot[domain=0.6947382761967033:1.6910607397332542,variable=\t]({1.*3.8*cos(\t r)+0.*3.8*sin(\t r)},{0.*3.8*cos(\t r)+1.*3.8*sin(\t r)});
\draw [shift={(0.,0.)},line width=.7pt,<-]  plot[domain=0.6947382761967031:1.5707963267948966,variable=\t]({1.*4.3*cos(\t r)+0.*4.3*sin(\t r)},{0.*4.3*cos(\t r)+1.*4.3*sin(\t r)});
\draw [shift={(0.,0.)},line width=.7pt,<-]  plot[domain=0.9272952180016121:1.8925468811915391,variable=\t]({1.*2.3*cos(\t r)+0.*2.3*sin(\t r)},{0.*2.3*cos(\t r)+1.*2.3*sin(\t r)});
\draw [shift={(0.,0.)},line width=.7pt,<-]  plot[domain=1.1339896856110152:1.6910607397332542,variable=\t]({1.*2.8*cos(\t r)+0.*2.8*sin(\t r)},{0.*2.8*cos(\t r)+1.*2.8*sin(\t r)});
\draw [shift={(0.,0.)},line width=.7pt,->]  plot[domain=0.463471026709497:1.8925468811915391,variable=\t]({1.*1.3*cos(\t r)+0.*1.3*sin(\t r)},{0.*1.3*cos(\t r)+1.*1.3*sin(\t r)});
\draw [shift={(0.,0.)},line width=.7pt,->]  plot[domain=0.463471026709497:1.6910607397332542,variable=\t]({1.*1.8*cos(\t r)+0.*1.8*sin(\t r)},{0.*1.8*cos(\t r)+1.*1.8*sin(\t r)});
\draw [shift={(0.,0.)},line width=.7pt,->]  plot[domain=0.463471026709497:1.5707963267948966,variable=\t]({1.*0.8*cos(\t r)+0.*0.8*sin(\t r)},{0.*0.8*cos(\t r)+1.*0.8*sin(\t r)});

% Angles droits
\draw [line width=.6pt] (0.,4.85)-- (0.15,4.85);
\draw [line width=.6pt] (0.15,5.)-- (0.15,4.85);
\draw [line width=.6pt] (-0.7576975052924292,2.747434164902521)-- (-0.7102633403899057,2.6051316701949507);
\draw [line width=.6pt] (-0.7102633403899057,2.6051316701949507)-- (-0.8525658350974761,2.5576975052924267);
\draw [line width=.6pt] (-0.4911808545842335,4.064465011724547)-- (-0.3422643124304364,4.082461218907003);
\draw [line width=.6pt] (-0.3422643124304364,4.082461218907003)-- (-0.360260519612889,4.231377761060787);

% Bullets
\draw (0.,5.) node {$\bullet$};
\draw (6.,5.) node {$\bullet$};
\draw (2.1152405075533625,4.530536126685824) node {$\bullet$};
\draw (3.,4.) node {$\bullet$};

% Verbose
\draw (-2.75,5.65) node[]{$\text{North}$};
\draw (0.0,6.2) node[]{$\hat{\textbf{e}}$};
\draw (-0.7,5.9) node[above]{$\hat{\textbf{e}}'$};
\draw (-1.8,5.9) node[]{$\hat{\textbf{e}}''$};
\draw (-.26,0.97) node[]{\tiny{$\varphi_{\text{e}}$}};
\draw (0.27,0.91) node[]{\tiny{$\omega$}};
\draw (0.37,1.47) node[]{\tiny{$\omega''$}};
\draw (0.3,2.0) node[]{\tiny{$\omega'$}};
\draw (0.95,2.9) node[]{\tiny{$f_{2'}$}};
\draw (0.5,2.48) node[]{\tiny{$f_{2''}$}};
\draw (1.8,3.09) node[]{\tiny{$f_{1''}$}};
\draw (2.67,3.09) node[]{\tiny{$f_{1'}$}};
\draw (3.2,3.25) node[]{\tiny{$f_{1}$}};
\draw (0,0) node[below]{$\text{E}$};
\draw (-0.05,4.95) node[above right] {$2$};
\draw (6.,5.0) node[above] {$1$};
\draw (2.12,4.52) node[above] {$2'$};
\draw (3.1,4.) node[above] {$2''$};
\draw (3.2,1.32) node[rotate=26.5]{Line of nodes};

\end{tikzpicture}
\end{minipage}
\begin{minipage}{0.45\textwidth}
\include{plot}
\end{minipage}
\caption{Evolution of the North-South horizontal gradient contribution to the atmospheric time-delay for observations at meridian. The \emph{left panel} is a sketch representing a transmitting source (labeled $1$) being ground tracked simultaneously by three stations on Earth (labeled $2$, $2'$, and $2''$). The source is passing in the meridian of the three stations, i.e. $\iota=\pi/2$ and $\lambda=\Omega$, thus the direction of the line of nodes is the intersection between the propagation plane of photons and the Earth's equator. The \emph{right panel} is a graph of Eq.~\eqref{eq:deltat_horGrad} [making use of \eqref{eq:transfo}], for the three different paths which are depicted on the \emph{left panel}. The plain and dotted curves are computed for $q=2.4$ and $q=2.0$ respectively. The computation has been carried out assuming $\bar\varphi_2=0\,\mathrm{rad}$, $\bar\varphi_{2'}=-0.2\,\mathrm{rad}$, $\bar\varphi_{2''}=-0.4\,\mathrm{rad}$, and $\bar\varphi_1=\bar\varphi_{1'}=\bar\varphi_{1''}=-\arccos[1/q]$.}
\label{fig:plot}
\end{figure*} 

As an example, we give the transformation rule which allows us to pass from true anomaly to colatitude. For the photon path $1\rightarrow2'$, which is depicted on the left panel of Fig.~\ref{fig:plot}, we get
\begin{subequations}\label{eq:transfo}
\begin{align}
f&=\bar{\varphi}-(\omega'-\omega)\text{,} &\bar{\varphi}=\varphi-\varphi_{\text{e}}\text{.}
\label{eq:transfo_f}
\end{align}
Here, for a matter of convenience for the next, we have introduced the colatitude $\bar{\varphi}$ which is referred from $\hat{\textbf{e}}$'s direction instead of North's (see Fig.~\ref{fig:plot}). The difference $\omega'-\omega$ appearing on the right-hand side of Eq.~\eqref{eq:transfo_f}, is function of the location of the receiving antenna ($\bar\varphi_{2'}$) and is also function of the ratio between the magnitude of the separation vectors of the source ($h_1$) and the receiver ($R$) as $q\equiv h_1/R$. It is given by the following expression
\begin{equation}
\omega'-\omega=\arctan\left[\frac{1-\cos\bar\varphi_{2'}}{-\text{sign}(\bar\varphi_{2'})\sqrt{q^2-1}+\sin\bar\varphi_{2'}}\right]\text{.}
\label{eq:transfo_omo}
\end{equation}
\end{subequations}
Finally, insertion of Eqs.~\eqref{eq:transfo} into \eqref{eq:deltat_horGrad} allows one to infer the atmospheric time-delay along the photon path $[\delta t(\bar\varphi)]$ and for the following boundary conditions $(q,\bar\varphi_1,\bar\varphi_2)$. 

In order to emphasize the behavior of $\delta t(\bar\varphi;q,\bar\varphi_1,\bar\varphi_2)$, we have represented in Fig.~\ref{fig:plot}, the evolution of the North-South horizontal gradient contribution to the atmospheric time-delay. The computation has been carried out assuming three stations (labeled $2$, $2'$, and $2''$) which are simultaneously ground tracking a unique source (labeled $1$). We have assumed that the stations are located at different colatitudes along the same meridian.

For a given ratio $q$, it is seen that the largest effect is reached for station $2$, which is ground tracking the source at the minimal elevation. This is due to three cumulative effects. Firstly, the projection of the horizontal gradient along the direction of the ray is maximum for the path $1\rightarrow2$. Secondly, the geometrical length inside the atmosphere is the most important for the path $1\rightarrow2$. Finally, the horizontal contribution in $(\varphi-\varphi_2)n_{\text{NS}}$ into the expression of the index of refraction [see Eq.~\eqref{eq:n_hor_gad}] is the most important for $\varphi=\varphi_1$ and for the path $1\rightarrow2$ which presents the most important value of $\varphi_1-\varphi_2$. Consequently, for this path, the phase velocity is minimum and consequently the time-delay is maximum.

\section{Comparison with numerical integration}
\label{sec:num_sim}

The validity of the approximated solutions which are derived in previous section can be assessed by a direct comparison with the output of a numerical integration of the light path.

As discussed so far, the solutions can indifferently be applied to AO experiments or astro-geodetic GB observations. The differences remain in the role of $\alpha$, which is either an output or an input, and also in the geometry through the values of the true anomaly at the transmission and at the reception (or at the entrance and the exit of the region of refractive influence). However, we have seen that the method of computation for AO experiments may differ for GB observations, especially at very high elevations for which an expansion in $1/e$ might not be accurate enough.

In this chapter, we test the validity of the derived solutions in the context of AO experiments. %Also, we expect a good agreement between the analytical solutions and the numerical ones, because, as discussed in Sec.~\ref{subsec:meth_int}, in the context of AO experiments, considering the leading order in $1/e$ is a very good approximation since the impact parameter is usually $K\gg\alpha\mu$.
We focus on Eqs.~\eqref{eq:Delta_guauss_J2} which solve, in a non-ubiquitous way, the problem of determining analytically the light path in the presence of the atmosphere's oblateness. As discussed in Chap.~\ref{sec:intro}, this problem has never been solved analytically in a complete and satisfactory way.

\begin{figure*}
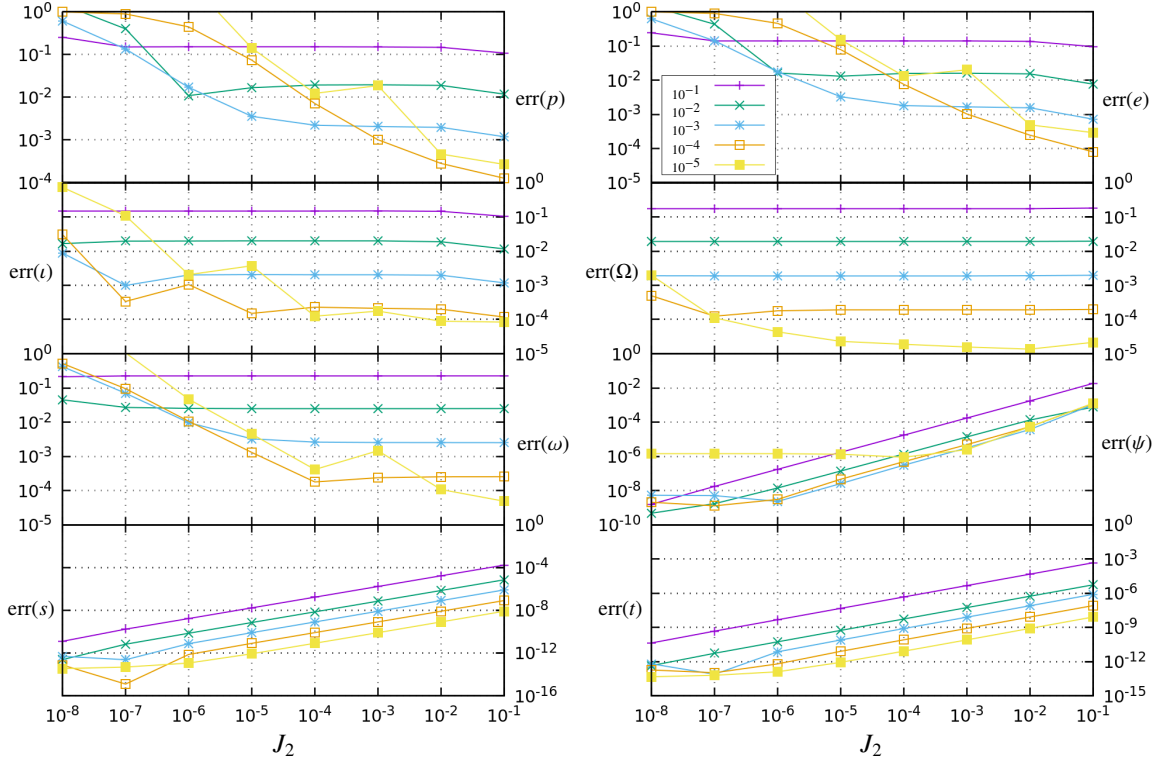

\begin{center}
\include{plot_err}
\end{center}
\caption{Evolution of the relative error on the change in elements $\mathbf{C}=(p,e,\iota,\Omega,\omega,s,t,\psi)$, for different values of the refractivity at the surface (colored curves) and for different values of $J_2$ ($x$-axis). The relative error on the change in elements is computed as $\text{err}(\textbf{C})=|\Delta\mathbf{C}_{\text{num}}-\Delta\mathbf{C}_{\text{ana}}|/|\Delta\mathbf{C}_{\text{num}}|$, with $\Delta\mathbf{C}$ being the total change in the value of the elements between the transmission and the reception of the signal. The subscripts ``num'' refers to the numerical predictions while the subscript ``ana'' refers to the analytical solutions [cf. Eqs.~\eqref{eq:Delta_guauss_J2}]. The change in $s$, $t$, and $\psi$ is the total change including the hyperbolic and the non-hyperbolic contributions.}
\label{fig:plot_err}
\end{figure*}

In order to assess the validity of Eqs.~\eqref{eq:Delta_guauss_J2}, we have performed a numerical integration of Eqs.~\eqref{eq:s}, \eqref{eq:optics}, and \eqref{eq:dtds} across the refractive profile given in Eqs.~\eqref{eq:n_nonsphe} and \eqref{eq:fj1} for $l=2$. The numerical integration has been carried out in double precision, with a numerical error tolerance of $10^{-11}$, for different values of $\alpha$ and $J_2$. [Actually, instead of $\alpha$, we work in the following with the dimensionless parameter $\alpha\mu/R$, which represents the hyperbolic refractivity evaluated at the level of the radius of the central planet, $N_0(R)=n_0(R)-\eta$]. The tested values of the refractivity and $J_2$, range from $10^{-1}$ to $10^{-5}$, and from $10^{-1}$ to $10^{-8}$ respectively. For each numerical integration, we have compared the total change in the hyperbolic elements, between the transmission and the reception, with the analytical predictions given in Eqs.~\eqref{eq:Delta_guauss_J2}. Let us emphasize that $\delta s$, $\delta t$, and $\delta\psi$ cannot be determined easily from the numerical integration of Eqs.~\eqref{eq:s}, \eqref{eq:optics}, and \eqref{eq:dtds}, thus, instead of working with the non-hyperbolic contributions alone, we are considering the total change in $s$, $t$, and $\psi$, which is easily inferred from results of the numerical integration. From an analytical point of view, the total change in $s$, $t$, and $\psi$, is simply given by the sum of hyperbolic [see Eqs.~\eqref{eq:Kepler_s}, \eqref{eq:Kepler_t}, and \eqref{eq:psi(F)}] and non-hyperbolic contributions [see Eqs.~\eqref{eq:Deltas_J2}--\eqref{eq:Deltapsi_J2}]. In Fig.~\ref{fig:plot_err}, we show the evolution of the relative error on the total change in the hyperbolic elements and $s$, $t$, and $\psi$.

Firstly, it is seen that the solutions, for low refractivity [$N_0(R)$ below $10^{-4}$] and for small values of $J_2$, seem to suffer from a lack of accuracy (c.f. errors on $p$, $e$, $\iota$, $\Omega$, and $\omega$ in Fig.~\ref{fig:plot_err}). However, this loss of accuracy is only due to numerical noise, since for small values of $J_2$, the perturbing effect is so small that the hyperbolic elements tend to be constants. The difference between their values at the reception and at the transmission remains at the level of the numerical noise. For instance, for $N_0(R)=10^{-5}$, we have $p(f_1)\sim 10^8$, moreover the analytical solution predicts $\Delta p_{\text{ana}}\sim 10^{-5}$ for $J_2=10^{-6}$, which represents a change in $\Delta p_{\text{ana}}/p(f_1)\sim 10^{-13}$, i.e. one order of magnitude beyond the numerical double precision.  

Secondly, we also notice an important feature linked to the accuracy of the solutions. Indeed, by changing the order of magnitude of the refractivity we also change the order of magnitude of the relative error on the total change in the value of all parameters. This feature reveals the approximation at leading order in $1/e$, which has been assumed in order to simplify the integration of the perturbation equations. In fact, while changing the order of magnitude of the refractivity value, which corresponds to a change in $\alpha$, we also modify the order of magnitude of the eccentricity. From Eq.~\eqref{eq:val_K/e}, we notice that increasing $\alpha$ makes the eccentricity smaller, which decreases the accuracy of the solutions which are derived at leading order in $1/e$. For very high refractivity (e.g. $10^{-1}$ which corresponds here to a value of eccentricity around $10$), it is seen (cf. purple curves in Fig.~\ref{fig:plot_err}) that the change in the hyperbolic elements, as given by the analytical expressions, is accurate at the level of $\sim10\%$. For usual typical values of the refractivity ($\sim10^{-4}$), the analytical solutions in Eqs.~\eqref{eq:Deltap_J2}--\eqref{eq:Deltaomega_J2} are found to be accurate at the level of $\sim0.01\%$.

For the geometrical length, the total light time and the refractive bending, the relative error evolves linearly with $J_2$ in the log-log plot. This behavior is due to the fact that the hyperbolic contribution is included within the analytical computation of $s$, $t$, and $\psi$. Indeed, for hyperbolic elements, as discussed above, when $J_2$ tends to be null, the computation of $\Delta\mathbf{C}_{\text{num}}$ generates a lot of numerical noise since the hyperbolic elements are constants when the perturbation vanishes. In the case of $s$, $t$, and $\psi$, when $J_2$ goes to zero, the computation of  $\Delta\mathbf{C}_{\text{num}}$ does not generates any numerical noise since a non-null hyperbolic contribution remains when the non-hyperbolic one vanishes. In order to prove that the non-hyperbolic evolution of $s$, $t$, and $\psi$ is well described by Eqs.~\eqref{eq:Deltas_J2}--\eqref{eq:Deltapsi_J2}, one can compute the relative error considering only the hyperbolic contribution in $\Delta\mathbf{C}_{\text{ana}}$. The computation reveals that, the relative error computed without the non-hyperbolic contributions is between one and two orders of magnitude larger than the relative error computed with the non-hyperbolic contribution.

The internal accuracy of our solutions describing the evolution of the geometrical length, the light time and the refractive bending can be assessed by looking at Fig.~\ref{fig:plot_err}. For very high refractivity (e.g. $10^{-1}$), the maximum relative error is reached for the maximum value of the oblateness ($J_2=10^{-1}$). For instance, considering the hyperbolic and the non-hyperbolic contributions, we are able to provide analytical solutions which are accurate at the level of $\sim 0.01\%$ for both $s$ and $t$, and at the level of $\sim 1\%$ for $\psi$. For a typical value of the refractivity ($\sim10^{-4}$), and for a value of Jupiter or Saturn's $J_2$ ($\sim10^{-2}$), we might expect errors at the level of $\sim10^{-6}\%$ for both $s$ and $t$, and at the level of $\sim0.001\%$ for $\psi$. For the same values of the refractivity and $J_2$, if we only consider the hyperbolic contribution, the relative error grows to $\sim10^{-5}\%$ for both $s$ and $t$, and to $0.1\%$ for $\psi$.

\section{Conclusion and perspectives}

In this paper we have demonstrated, in Sec. \ref{sec:ref}, that the equations of geometrical optics possess an exact solution, referred to as the reference solution, when one assumes the hydrostatic equilibrium together with the constancy of the variation of the index of refraction with the Newtonian potential. Not surprisingly, since the problem reduces to a central field problem, the solution is found to be a conic-section similar to the solution of the well-known two-body problem in celestial mechanics. For the cases of interest, the reference solution is an hyperbola which is completely described with some constants of integration called hyperbolic elements. These elements describe the geometry of the light beam, namely the shape and the spatial orientation of the hyperbola. We have shown that the hyperbolic elements are related to the refractive response of the medium to the gravitational potential which is parametrized by the derivative of the index of refraction with respect to the potential.

The reference solution is found to be applicable in situation involving light propagation in planetary atmospheres, and thus, to be well suited for future applications to GB observations or AO experiments. In Sec. \ref{subsec:data}, we provided short indications on how applying it in the context of real observations, but its practical application to real space missions' data will be discussed in a successive paper. However, the main feature of this solution is not its direct application, but the fact that it provides a comprehensive framework that can be extended to carry out analytical studies in the case where the index of refraction has generic dependencies.

Indeed, based on the method of variation of arbitrary constants, in Chap. \ref{sec:varconst} we have converted the equation of geometrical optics as length rate of change in the hyperbolic elements. These new equations are shown to be the optical counterpart of the perturbation equations in celestial mechanics. When no assumption is made on the order of magnitude of the change in the refractive profile, they are perfectly equivalent to equations of geometrical optics. If they are less compact, they present the advantage of providing a comprehensive alternative description to the path of light rays. They describe quantitatively the departure from the hyperbolic path due to additional terms besides the hyperbolic contribution into the index of refraction.

In the case where the perturbing gradient can be assumed small with respect to the hyperbolic one, we have shown in Sec.~\ref{subsec:meth_int}, how to approach analytically the solutions of the perturbation equations. This procedure seems to be generic enough to easily handle the first order effects due to any kind of perturbation besides the hyperbolic contribution.

To highlight the capabilities of this formalism,  in Chap.~\ref{sec:applications} we have analyzed different sources of perturbing gradients. For instance, we have studied the effects on the light path due to, the non-linear dependencies with the Newtonian potential, the centrifugal potential and the light dragging effect due to the rigid rotation of the atmosphere, the non-spherical gravitational potential of the planet, the tidal atmospheric bulge raised by the presence of a massive perturbing body, and the horizontal gradients of refractivity inside the atmosphere. These examples of utilization are just a non-exhaustive list showing the possibility offered by the perturbation equations formalism applied to geometrical optics.

Finally, in Sec.~\ref{sec:num_sim}, in order to assess the validity of, i) the perturbation equations, and ii) the method of integration, we have compared the analytical solution derived in the context of a quadrupolar axisymmetry in the gravitational potential of the planet with its numerical counterpart determined from a numerical integration of the equations of geometrical optics. We carried out several numerical integrations for different values of the refractivity and the $J_2$ parameter. For each one of them, we have compared the changes in the hyperbolic elements as provided by the numerical integration to those obtained from the analytical solution. By doing this, we have been able to assess the accuracy of the analytical solution. We have shown that for standard refractivity values ($\sim10^{-4}$), the relative error in the total change in the hyperbolic elements is of the order of $0.01\%$ and is independent of the value of $J_2$. For the other elements like the geometrical length, the time-delay and the refractive bending, the relative error evolves with $J_2$ as shown in Fig.~\ref{fig:plot_err}. For Jupiter or Saturn's typical value of $J_2$ ($\sim10^{-2}$), the relative error is of the order of $10^{-6}\%$ for both the length and the time, and $0.001\%$ for the bending. This represents a really good agreement considering the simplicity of the final solution. Indeed, by looking at the complete analytical expression of the light time, we notice that it is just given by the sum of the hyperbolic [cf. Eq.~\eqref{eq:Kepler_s}] and the non-hyperbolic contribution [cf. Eq.~\eqref{eq:Deltat_J2}].

The main original contribution of this paper is the reformulation of the equations of geometrical optics into a set of perturbation equations [cf. Eqs.~\eqref{eq:guauss}], describing the evolution of an osculating hyperbola along the light trajectory. This reformulation is very convenient for a use inside planetary atmospheres since the hyperbolic trajectory has proved to be an exact solution for spherical symmetry. However, an other possibility for the reformulation of the equations of geometrical optics could have been to use the straight line as an osculating solution instead of the hyperbola, offering in this way the possibility of analytically studying other symmetries involving small refractivity (e.g. for application to occultations by the Io plasma torus \cite{1981JGR....86.8447B,2017JGRA..122.1731P,doi:10.1029/2017JA025113}). This is also something which can can be easily achieved from Eqs.~\eqref{eq:guauss} by taking the limit $\alpha\rightarrow0$.

In the context of the realization of the IERS reference systems, the perturbation equations together with the method of integration, could provide an efficient tool for improving existing models of the Earth tropospheric delay. In particular, an accurate modeling of horizontal gradients is of prime importance for the determination of station coordinates which in turn define the scale and the origin of the ITRF. However, existing models of the tropospheric delay always stick to the spherical symmetry assumption, and only few of them consider horizontal gradients (\cite{1997JGR...10220489C} for VLBI, \cite{doi:10.1029/2006JB004834} for SLR). In addition, even for those which consider horizontal gradients, the integration across the atmosphere is always done numerically assuming spherical layers made of constant index of refraction. The perturbation equations could easily tackle these issues by providing e.g. the contribution to the time-delay due to the atmosphere's oblateness and the one due to horizontal gradients, as discussed in Secs.~\ref{subsec:nonspherical} and \ref{subsec:horiz_grad}.

In the context of AO experiments, the perturbation equations allow to assess for the first time a clear description of the photon path, the light time or the refractive bending into an oblate atmosphere. As mentioned in the introduction, this question is usually tackled using numerical ray-tracing techniques \cite{2015RaSc...50..712S}, and only a few studies \cite{1992AJ....103..967L} aimed at exploring analytically this problem. Yet, the main advantage of analytical formulations is twofold. First, they help to acquire a physical understanding of the phenomenon, and secondly, they do not require high computation time while analyzing observational data. If the solution proposed by \cite{1992AJ....103..967L} is a first analytical approach, it fails to provide a clear physical understanding of the photon path or the atmospheric time-delay due to the atmosphere's oblateness. The reason is due to the fact that it consists in a Taylor series expansion around a local point along the photon path, and in addition it is expressed in terms of Cartesian coordinates which might be somehow abstract. Because of this Taylor expansion, the solution cannot be employed for high refractivity and is only defined in the surroundings of the expansion point. With the perturbation equations, we have solved all these difficulties since the solution is available all along the light path trajectory and can also be applied for high refractivity. In addition, the solution provides a simple geometrical interpretation by fitting at any length along the ray an hyperbolic trajectory. Another advantage which is worth mentioning is that the perturbation equations allows to express directly the light time as well as the refractive bending expressions which are needed for computing range and range-rate observables. Finally, let us mention that if analytical studies are not as precise as purely numerical ones, the formalism of the perturbation equations has been shown to be really accurate (errors of one part in $10^8$ and $10^5$ between the numerical and the analytical predictions on the light time and the refractive bending, in the presence of an oblate atmosphere characterized by $J_2=10^{-2}$ and $N_0(R)=10^{-4}$, respectively). In addition, we have demonstrated that the formalism presented in this paper offers a large flexibility since it is able to handle a wide range of perturbations including light dragging effects caused by atmosphere's winds. Thus, a complete and purely analytical method, only based on analytical solutions derived in Chap.~\ref{sec:applications}, could be developed in order to process real AO data in the context of past, current, and future space missions. Our next opportunity, which actually motivated this work, is ESA's L-class mission JUICE (JUpiter Icy moons Explorer) \cite{Grasset20131,Witasse2016}; here a radioscience experiment named 3GM (Gravity and Geophysics of Jupiter and the Galilean Moons) will take advantage of a careful Jupiter system tour design \cite{Boutonnet20161465}, offering both frequent Jupiter moon's flybys \cite{Lari2018,Dirkx201714} and radio occultation opportunities by Jupiter's oblate atmosphere.

\begin{acknowledgements}
The Authors are grateful to the University of Bologna and to Italian Space Agency (ASI) for financial support through the Agreement 2013-056-RO in the context of ESA's JUICE mission. A.B. is grateful to F. Mignard from Observatoire de la C\^ote d'Azur for interesting discussions about atmospheric and astronomic refraction. A.B. is also grateful to A. Hees from Paris Observatory for valuable comments about a preliminary version of this manuscript.    
\end{acknowledgements}

\bibliographystyle{apsrev4-1}
\bibliography{bourgoin}

\appendix

%\newpage
\section{Hyperbolic path}
\label{sec:hyp}

As discussed in Chap. \ref{sec:ref}, when the index of refraction is only function of the height of the ray (spherically symmetric assumption), all the light path is contained in the propagation plane. Using this fact, we derive, in this chapter, a complete solution to Eqs.~\eqref{eq:s}, \eqref{eq:optics}, and \eqref{eq:dtds}. We start by determining the shape of the light trajectory (i.e. the light path), then we focus on the time-delay. 

\subsection{Shape of the light path}

We wish to write the equations of geometrical optics in the propagation frame and in polar coordinates by making use of the polar basis. The vectors \textbf{h} and $\bm\nabla\mathscr{S}$ are decomposed as
\begin{subequations}
\label{eq:hsdnsds_polar}
\begin{align}
\mathbf{h}&=h\hat{\bm \rho}\text{,}\label{eq:h_polar}\\
\bm\nabla\mathscr{S}&=n\dot h\hat{\bm \rho}+nh\dot\theta \hat{\bm \tau}\text{,}\label{eq:s_polar}
\end{align}
\end{subequations}
where a dot denotes the differentiation with respect to $s$. Moreover, the left-hand side of Eq. \eqref{eq:optics} is given by
\begin{equation}
\frac{\dd}{\dd s}\big(\bm\nabla\mathscr{S}\big)=(n\ddot h+\dot n\dot h-n h\dot\theta^2)\hat{\bm \rho}+\frac{1}{h}\frac{\dd}{\dd s}(n h^2\dot\theta)\hat{\bm \tau}\text{.}\label{eq:dnsds_polar}
\end{equation}
The gradient of the index of refraction can also be expressed in the same coordinate system. With the help of Eq. \eqref{eq:dn/dPhi} and the monopole approximation [cf. Eq. \eqref{eq:pot_mono}], the right-hand side of Eq.~\eqref{eq:optics} is given by
\begin{equation}
\label{eq:nablan}
\bm\nabla n=-\alpha g\hat{\bm\rho}\text{.}
\end{equation}
The term $g$ represents the magnitude of the local acceleration (monopole contribution) experienced by the media which makes up the atmosphere of the central planet. It is explicitly given by 
\begin{equation}
\label{eq:g}
g=\frac{\dd U_0}{\dd h}=\frac{\mu}{h^2}\text{.}
\end{equation}
It is obvious from the presence of $\alpha$ in Eq. \eqref{eq:nablan}, that the light ray experiences the gravitational pull via the media in which the ray propagates through.

From Eqs. \eqref{eq:dnsds_polar} and \eqref{eq:nablan}, it is seen that the absence of a transverse component (along $\hat{\bm \tau}$) in the gradient of the index of refraction implies that $n h^2\dot\theta$ is a conserved quantity. From Eqs.~\eqref{eq:K} and \eqref{eq:hsdnsds_polar} we immediately deduce
\begin{equation}
\label{eq:K_dtheta}
K=n h^2\dot\theta\text{.}
\end{equation}
Inserting this expression into the radial part of Eq. \eqref{eq:dnsds_polar}, we find
\begin{equation}
\label{eq:radial_dot}
\ddot h+\frac{\dot n\dot h}{n}-\frac{K^2}{n^2 h^3}=-\frac{\alpha g}{n}\text{.}
\end{equation}
We change the independent variable from $s$ to $\theta$ adopting the convention that a prime denotes the differentiation with respect to $\theta$. We also adopt $u\equiv 1/h$ as a convenient substitute for $h$ and derive a differential equation for it. Making all these substitutions into Eq.~\eqref{eq:radial_dot}, we quickly arrive to 
\begin{equation}
\label{eq:radial_prime}
u''+\kappa^2 u=\frac{\alpha\mu\eta}{K^2}\text{,}
\end{equation}
with
\begin{equation}
\label{eq:kappa}
\kappa^2=1-\frac{(\alpha\mu)^2}{K^2}
\end{equation}
a constant parameter. The general solution to the simple Eq.~\eqref{eq:radial_prime} is a conic section with origin at the focus. Returning to the original radial variable the spatial solution is
\begin{equation}
\label{eq:h_sol_mono}
h=\frac{p}{1+e\cos\kappa f}\text{,}
\end{equation}
in which $e$ is an arbitrary constant of integration called the \emph{eccentricity} of the conic. We will adopt the convention that the eccentricity is a positive quantity. The two other parameters, $p$ and $f$, are defined such that 
\begin{subequations}
\label{eq:const_sol_mono}
\begin{align}
p&\equiv \frac{\kappa^2K^2}{\alpha\mu\eta}\text{,}\label{eq:p_mono}\\
f&\equiv \theta-\omega\text{.}\label{eq:f_mono}
\end{align}
\end{subequations}
They are respectively known as the \emph{semi-latus rectum} and the \emph{true anomaly}. The last parameter, $\omega$, is also a constant of integration and its role is better understood when $h$ is minimal, that is to say when $\theta=\omega$. It is seen that $\omega$ defines the \emph{argument of the closest approach} which is the minimal separation between the path of the ray and the center of the reference frame. This angle $\omega$ is completely determined by the initial conditions of the problem. For instance, if we call $(h_1,\theta_1)$ the components of the position vector of the transmission point, then from the solution in Eq.~\eqref{eq:h_sol_mono}, we end up with
\begin{equation}
\label{eq:omega}
\omega=\theta_1\pm\kappa^{-1}\arccos\left(\frac{p-h_1}{eh_1}\right)\text{,}
\end{equation}
where the sign is taken to be positive when $-\pi/2\leq\theta_1<\pi/2$ and negative elsewhere.

The minimal separation is also determined from Eq.~\eqref{eq:h_sol_mono} evaluated at the closest approach, such that
\begin{equation}
\label{eq:hC_p}
h_C=\frac{p}{1+e}\text{.}
\end{equation}

In the following, we will consider that the signal was emitted or received (or both) from infinity, in a space region well beyond the planetary atmosphere where the index of refraction can be considered to be unity (or more generally, constant). This restriction let to avoid periodic light path. Consequently, we will restrict our attention to non-periodic solutions ($e\geq 1$). Following this, the only possibility for the eccentricity value is $e>1$, which corresponds to hyperbolic trajectories for the light path. Therefore, we can introduce a new constant known as the \emph{semi-major axis}
\begin{equation}
\label{eq:a}
a\equiv -\frac{p}{e^2-1}\text{,}
\end{equation}
which is a negative quantity.

Non-periodic solutions, which are hyperbolic trajectories, also imply the introduction of a new important angle, $\Delta f_{\infty}$, describing the net change between the two asymptotes' directions. Based on Eq. \eqref{eq:h_sol_mono}, it can be seen that $h$ goes to infinity when $\cos\kappa f$ tends to $-1/e$, which gives rise to two asymptotic solutions for $\kappa f$. Let call $f_{\text{in}}^{\infty}$ the negative solution and $f_{\text{out}}^{\infty}$ the positive one. If we introduce $\Delta f_{\infty}= f_{\text{out}}^{\infty}-f_{\text{in}}^{\infty}$, we immediately deduce
\begin{equation}
\label{eq:Deltaf}
\Delta f_{\infty}\equiv \frac{2}{\kappa}\arccos\big(-1/e\big)\text{,}
\end{equation}
Since, we focus on hyperbolic trajectories $(e>1)$, we will always have $\Delta f_{\infty}<2\pi$.

In AO experiments, the geometry is such that $\Delta f_{\infty}$ can be geometrically related to the refractive bending (indeed from Fig.~\ref{fig:path}, we deduce $\Delta f_{\infty}=\pi+\epsilon$) which is itself directly in relation with the changes in the Doppler measurements, at first order (see Eq.~\eqref{eq:DopplerShift_bis} or \cite{1965JGR....70.3217F,1968P&SS...16.1035F}). Thus, Eq.~\eqref{eq:Deltaf} can be used to infer the eccentricity from the Doppler frequency shift measurements.

Once the evolution of the height is totally determined, we can focus on the evolution of the tangent to the ray [cf. Eq.~\eqref{eq:s_polar}]. We thus need to determine the expressions of the radial and the transverse components; the later depends on the angular velocity. Invoking Eqs. \eqref{eq:h_sol_mono} and \eqref{eq:K_dtheta} and differentiating them making use of Eqs. \eqref{eq:const_sol_mono}, reveals that
\begin{subequations}
\label{eq:doththeta_sol_mono}
\begin{align}
n\dot h&=e\sqrt{\eta}\sqrt{\frac{\alpha\mu}{p}}\sin\kappa f\text{,}\label{eq:doth_sol_mono}\\
n\dot \theta&=\frac{\sqrt{\eta}}{\kappa}\sqrt{\frac{\alpha\mu}{p^3}}(1+e\cos\kappa f)^2\text{.}\label{eq:dottheta_sol_mono}
\end{align}
\end{subequations}
The evolution of the tangent to the ray is provided after inserting those relationships together with Eq. \eqref{eq:h_sol_mono} into Eq. \eqref{eq:s_polar}.

\subsection{First integrals of the light path}

Until now, we are missing an expression for the eccentricity and the argument of the closest approach in term of fundamental constants of the problem, such as $\textbf{K}$. We saw e.g. that the semi-latus rectum is linked to the conservation of the magnitude of the impact parameter [cf. Eq. \eqref{eq:p_mono}]. In this section, we look for additional first integrals of the light path, which then could let to express the constant of integration in a fundamental way.

The first fundamental constant can be obtained by multiplying both side of Eq. \eqref{eq:radial_dot} by $n^2\dot{h}$. Then, one can see that each term is a total differential with respect to $s$. In addition, one can recognized the squared components of \eqref{eq:s_polar} in the integrated equation, and we end up with 
\begin{equation}
\label{eq:E_SS}
E=\frac{(\bm\nabla\mathscr{S})^2}{2}-\frac{\alpha\mu}{h}\left(\eta+\frac{\alpha\mu}{2h}\right)\text{,}
\end{equation}
where $E$ is a constant of integration. That constant would be similar to the energy in classical mechanics. Indeed, by computing the square of $\bm\nabla\mathscr{S}$ from Eqs.~\eqref{eq:doththeta_sol_mono} and \eqref{eq:s_polar}, we end up with
\begin{equation}
\label{eq:n2}
(\bm\nabla\mathscr{S})^2=\alpha\mu\left(\frac{2\eta}{h}-\frac{\eta}{a}+\frac{\alpha\mu}{h^2}\right)\text{,}
\end{equation}
where we have used Eq. \eqref{eq:a}. Then, insertion of Eq. \eqref{eq:n2} into \eqref{eq:E_SS}, leads to
\begin{equation}
\label{eq:E_a}
E=-\frac{\alpha\mu\eta}{2a}\text{,}
\end{equation}
which possesses the same form as the energy in the Kepler problem. From the definition of $a$, the right-hand side of the equation is directly seen for being constant and positive. Therefore, $E$ is conserved along the all light ray trajectory.

As a general remark, let us notice that Eq. \eqref{eq:E_SS} can be put under a more fundamental form. Indeed, by making use of the definition of $n$ [cf. Eq.~\eqref{eq:n0}], we deduce
\begin{equation}
\label{eq:E_n2}
E=\frac{1}{2}\left(\bm\nabla\mathscr{S}\cdot\bm\nabla\mathscr{S}-\gamma n^2\right)\text{,}
\end{equation}
where
\begin{equation}
\label{eq:drag_coeff}
\gamma\equiv \left(1-\frac{\eta^2}{n^2}\right)
\end{equation}
is known as the Fresnel coefficient [$\eta$ has been previously defined in Eq. \eqref{eq:n0}]. This expression of $E$ is more fundamental than Eq. \eqref{eq:E_SS} since it does not require a prior knowledge of the index of refraction. Indeed, the constancy of $E$ can be inferred using only Eqs.~\eqref{eq:s} and \eqref{eq:optics}.

From Eq. \eqref{eq:E_a}, it is seen that the semi-major axis is linked to the conservation of the dimensionless energy parameter, $E$. At the same time, the constancy of the eccentricity is assured, because of the definition \eqref{eq:a}. So, in that sense the eccentricity is linked to both the conservation of energy and the conservation of the magnitude of the impact parameter. Eqs.~\eqref{eq:E_a}, \eqref{eq:a}, and \eqref{eq:p_mono} can be used to infer the eccentricity
\begin{equation}
\label{eq:ecc}
e\equiv\sqrt{1+\frac{2\kappa^2EK^2}{(\alpha\mu\eta)^2}}\text{.}
\end{equation}

In addition to $E$, and $\textbf{K}$, we can determine an other fundamental constant of the problem. This last one, is equivalent to the eccentricity vector of celestial mechanics and comes from the very peculiar form of the index of refraction, which is linear with the gravitational potential making it evolving as $1/h$. However, because $\kappa$ is different from unity, the eccentricity vector does not show up as easily as in celestial mechanics. It is given by the following relationship
\begin{equation}
\label{eq:ecc_vec}
\textbf{e}=\frac{p}{K^2}\bm\nabla\mathscr{S}\times\textbf{K}-\hat{\bm{\rho}}+e\textbf{A}\text{,}
\end{equation}
where we have introduced the vector $\textbf{A}$ which possesses the following components in the polar basis
\begin{equation}
\label{eq:ecc_vec_A}
\textbf{A}=(\cos f-\cos\kappa f)\hat{\bm{\rho}}+(\kappa\sin\kappa f-\sin f)\hat{\bm{\tau}}\text{.}
\end{equation}
The vector $\textbf{A}$ is expressed in terms of fundamental quantities after determining $f$ from Eq. \eqref{eq:h_sol_mono}, and expressing $\hat{\bm{\rho}}$ and $\hat{\bm{\tau}}$ in terms of $\textbf{h}$ and $\textbf{K}$. The eccentricity in Eq.~\eqref{eq:ecc_vec}, must be expressed in terms of $E$ and $K$ thanks to Eq. \eqref{eq:ecc}.

The constancy of $\textbf{e}$ can be demonstrated by substituting Eqs.~\eqref{eq:doththeta_sol_mono} into \eqref{eq:s_polar}, then inserting the result together with Eq.~\eqref{eq:K_dtheta} into \eqref{eq:ecc_vec}, which finally gives with Eq.~\eqref{eq:rhotau}
\begin{equation}
\label{eq:ecc_vec_e}
\textbf{e}=e(\hat{\textbf{x}}\cos\omega+\hat{\textbf{y}}\sin\omega)\text{.}
\end{equation}
Since $e$ and $\omega$ are two constants of integration, $\textbf{e}$ is conserved during the propagation of light. Moreover, it is seen that the constancy of $\omega$ is linked to the conservation of the eccentricity vector. It always points towards the closest approach and its magnitude is constant and equal to the eccentricity of the light path.

The important point of this section is summarized recalling that the light path is totally contained inside a fixed plane which remains orthogonal to the impact parameter vector when the refractive profile is purely radial. In addition, we found that the shape of the trajectory is described by an hyperbola when the refractive profile is given by Eq. \eqref{eq:n0}. That shape is parametrized thanks to two geometrical parameters, $p$ and $e$, and the orientation in the propagation plane is located thanks to one angle, $\omega$. Those three parameters can totally be expressed [cf. Eqs.~\eqref{eq:p_mono}, \eqref{eq:ecc}, and \eqref{eq:ecc_vec_e}] in terms of fundamental quantities of the problem, such that $K$, $E$, and $\textbf{e}$.

\subsection{Kepler-like problems}
\label{subsec:Kepler}

In the previous section, we have defined the shape of the trajectory as well as its orientation in the propagation plane in terms of constant parameters. In order to completely solve the problem in the plane, we need to determine a single location along the light path given a geometrical length or a light time.

The Eq. \eqref{eq:dottheta_sol_mono} can be used to complete the description of the light path by providing a unique relationship between $f$ and $s$. This can be accomplished by integrating Eq. \eqref{eq:dottheta_sol_mono} as
\begin{equation}
\label{eq:s_f}
s(f)-S=\frac{\kappa}{\sqrt{\eta}}\sqrt{\frac{p^3}{\alpha\mu}}\int_{0}^f\frac{n(f')\dd f'}{(1+e\cos\kappa f')^2}\text{,}
\end{equation}
where $S$ is an arbitrary constant of integration representing the traveled \emph{geometrical length to the closest approach}. Additionally, we can also derive a unique relationship between $f$ and $t$, by making use of Eq. \eqref{eq:dtds} in order to turn Eq.~\eqref{eq:s_f} into
\begin{equation}
\label{eq:t_f}
t(f)-T=\frac{\kappa}{c\sqrt{\eta}}\sqrt{\frac{p^3}{\alpha\mu}}\int_0^f\frac{n(f')^2\dd f'}{(1+e\cos\kappa f')^2}\text{,}
\end{equation}
where $T$ is an arbitrary constant of integration representing the \emph{light time till the closest approach}.

Eqs. \eqref{eq:s_f} and \eqref{eq:t_f} are the analogous of the Kepler equation of celestial mechanics. They can be exactly solved by introducing a new variable, $F$, known as the \emph{hyperbolic anomaly}. The change from the true to the hyperbolic anomaly, and conversely, is given by
\begin{subequations}\label{eq:F_fetf_F}
\begin{align}
\cosh F&=\cfrac{e+\cos\kappa f}{1+e\cos\kappa f}\text{,} & \sinh F&=\cfrac{\sqrt{e^2-1}\sin\kappa f}{1+e\cos\kappa f}\text{,}\label{eq:F_f}\\
\cos\kappa f&=\cfrac{e-\cosh F}{e\cosh F-1}\text{,} & \sin\kappa f&=\cfrac{\sqrt{e^2-1}\sinh F}{e\cosh F-1}\text{,}\label{eq:f_F}
\end{align}
and is summarized with the half-angles formula\footnote{Attention must be paid when one want to determine the hyperbolic from the true anomaly using \eqref{eq:tanf_F}. Indeed, the inverse function of the hyperbolic tangent is not well defined when the argument is close to $\pm1$, which might generates numerical noise. To avoid this issue, the inverse function of the hyperbolic sine [cf. Eq. \eqref{eq:F_f}] can be preferred.}
\begin{equation}
\label{eq:tanf_F}
\tan\left(\frac{\kappa f}{2}\right)=\sqrt{\frac{e+1}{e-1}}\tanh\left(\frac{F}{2}\right)\text{.}
\end{equation}
\end{subequations}
Eqs. \eqref{eq:F_fetf_F} allow us to introduce the following relations which relate the differentials of $f$ and $F$
\begin{align}
\cfrac{\dd f}{\dd F}&=\cfrac{\kappa^{-1}\sqrt{e^2-1}}{e\cosh F-1}\text{,} & \cfrac{\dd F}{\dd f}&=\cfrac{\kappa\sqrt{e^2-1}}{1+e\cos\kappa f}\text{.}\label{eq:df_dF}  
\end{align}
The hyperbolic anomaly can be employed instead of the true anomaly as a convenient substitute. Making all these substitutions we quickly arrive to
\begin{subequations}\label{eq:path_F}
\begin{align}
h&=-a(e\cosh F-1)\text{,}\label{eq:h_F}\\ 
n\dot h&=e\sqrt{\eta}\sqrt{\frac{\alpha\mu}{-a}}\frac{\sinh F}{e\cosh F-1}\text{,}\label{eq:doth_F}\\
n\dot F&=\sqrt{\eta}\sqrt{\frac{\alpha\mu}{-a^3}}\frac{1}{e\cosh F-1}\text{.}\label{eq:dotF_F}
\end{align}  
\end{subequations}

The last equation is perfectly equivalent to Eq. \eqref{eq:s_f}, however it possesses the main advantage of being easily integrable. Then, making use of Eqs. \eqref{eq:n0} and \eqref{eq:h_F} in order to express the refractive index in terms of $F$, we find the following analogous to the well-known Kepler equation
\begin{equation}
\label{eq:Kepler_s}
\sqrt{\eta}\sqrt{\frac{\alpha\mu}{-a^3}}(s-S)=\eta e\sinh F-\eta F-\frac{\alpha\mu}{a}F\text{,}
\end{equation}
in which the left-hand side is the \emph{geometrical length mean anomaly}.

Similarly, expressing Eq. \eqref{eq:dotF_F} in term of the light time instead of length path with the help of Eq.~\eqref{eq:dtds}, let one to directly integrate Eq.~\eqref{eq:t_f}, such that
\begin{equation}
\label{eq:Kepler_t}
c\sqrt{\eta}\sqrt{\frac{\alpha\mu}{-a^3}}(t-T)=\eta^2e\sinh F-\eta^2F-2\eta\frac{\alpha\mu}{a}F+\frac{2}{\sqrt{e^2-1}}\frac{(\alpha\mu)^2}{a^2}\arctan\left[\sqrt{\frac{e+1}{e-1}}\tanh\left(\frac{F}{2}\right)\right]\text{,}
\end{equation}
in which the left-hand side is the \emph{light time mean anomaly}

From these two Kepler-like equations, it is seen that the light time [cf. Eqs.~\eqref{eq:Kepler_t}] is not just the geometrical length [cf. Eqs.~\eqref{eq:Kepler_s}] multiplied by the inverse of the speed of light in a vacuum. Indeed, it also involves some additional terms which account for the apparent change in the speed of propagation of the electromagnetic waves. Consequently, the geometrical length mean anomaly considers a ray traveling at the speed of light along the bent trajectory while the time mean anomaly is related to the phase path and takes into account the apparent change in the speed of the waves along the actual path of photons.

\subsection{Argument of the refractive bending}
\label{subsec:taandba}

In this section, we take advantage of the introduction of the hyperbolic anomaly to define a very important geometrical relationship  between the true anomaly and the argument of the refractive bending, $\psi$. As seen in Fig. \ref{fig:path}, $\psi$ is the angle between the orthogonal direction to the line of nodes and the tangent to the light ray.

In the context of AO experiments, this angle plays an important role. Indeed, from Fig.~\ref{fig:path}, it might be seen that the net change in $\psi$ between the transmitter (labeled 1) and the receiver (labeled 2) is exactly the refractive bending, $\epsilon\equiv\psi_2-\psi_1$. Moreover, as shown in Eq. \eqref{eq:DopplerShift_bis} and in \cite{1971AJ.....76..123F} or \cite{1973P&SS...21.1521E}, the refractive bending can be use to infer the Doppler frequency shift which is the main observable for AO experiments.

The differential relationship between $\psi$ and $f$ can be inferred from Fig.~\ref{fig:path} which allows us deduce the following relationship
\begin{equation}
\label{eq:psi}
\psi=\theta+\phi-\frac{\pi}{2}\text{.}
\end{equation}
We remind that the angle $\phi$ has been previously defined in Eq.~\eqref{eq:Bouger} as the angle between $\hat{\bm{\rho}}$ and \textbf{s}. Hence, from Eq. \eqref{eq:s_polar} we immediately deduce
\begin{align}
\cos\phi&=\dot h\text{,} & \sin\phi&=h\dot{\theta}\text{.}\label{eq:trig_phi}
\end{align}
Then, differentiating Eqs. \eqref{eq:Bouger} and \eqref{eq:psi}, and making use of Eqs.~\eqref{eq:trig_phi}, we obtain the following relation
\begin{equation}
\label{eq:dpsidf}
\dd\psi=-\frac{h}{n}\frac{\dd n}{\dd h}\dd f\text{,}
\end{equation}
which can be written as
\begin{equation}
\label{eq:dpsidf_int}
\psi(f)-\omega=\frac{\alpha\mu}{p}\int_0^f\frac{1+e\cos\kappa f'}{n(f')}\dd f'\text{.}
\end{equation}
Making use of the relationships in Eqs. \eqref{eq:path_F}, we change the variable of integration from the true to the hyperbolic anomaly, and after some algebra we integrate the previous expression as
\begin{equation}
\label{eq:psi(F)}
\psi-\omega=f(F)-\frac{2}{e\kappa}\sqrt{\frac{e^2-1}{1-\bar e^2}}\arctan\left[\sqrt{\frac{1+\bar e}{1-\bar e}}\tanh\left(\frac{F}{2}\right)\right]\text{,}
\end{equation}
in which we have introduced
\begin{equation}
\label{eq:epsilon}
\bar e=\frac{\kappa^2+(\kappa^2-1)(e^2-1)}{e\kappa^2}\text{.}
\end{equation}

Since Eq.~\eqref{eq:tanf_F} provides a unique relationship between $f$ and $F$, Eq. \eqref{eq:psi(F)} can be used to find the argument of the bending angle knowing the true anomaly.

\section{Mathematical details for applications}
\label{sec:math_details}

In this chapter, we derive the change in hyperbolic elements for different contributions in the index of refraction.

\subsection{Non-linearity with $U_0$}
\label{subsec:detail_nonlin}

We first simplify the problem by assuming that the generalized potential reduces to the spherical part of the Newtonian potential ($\Phi=U_0$). Consequently, Eq.~\eqref{eq:n(Phi)} and the perturbing gradient now read
\begin{subequations}\label{eq:ndn_nonlin}
\begin{align}
n&=n_0+\sum_{k=2}^{+\infty}\frac{\alpha_{(k)}}{k!}\left(\frac{\mu}{h}\right)^k\text{,}\label{eq:n(h)}\\
\textbf{f}_{\text{pert}}&=-\frac{1}{h}\sum_{k=2}^{+\infty}\frac{\alpha_{(k)}}{(k-1)!}\left(\frac{\mu}{h}\right)^k\hat{\bm\rho}\text{.}\label{eq:dn(h)}
\end{align}
\end{subequations}
We can directly see that the non-null components of the perturbing gradient are the radial and the change in the index of refraction
\begin{subequations}\label{eq:RN_nonlin}
\begin{align}
\mathcal{R}&=-\frac{1}{h}\sum_{k=2}^{+\infty}\frac{\alpha_{(k)}}{(k-1)!}\left(\frac{\mu}{h}\right)^k\text{,}\label{eq:R_nonlin}\\
\mathcal{N}&=\frac{1}{np}\sum_{k=2}^{+\infty}\frac{\alpha_{(k)}}{k!}\left(\frac{\mu}{h}\right)^k\text{.}\label{eq:N_nonlin}
\end{align}
\end{subequations}
Eqs. \eqref{eq:RN_nonlin} can then be inserted into Eqs. \eqref{eq:guauss_pert} and the integrand is expanded in power of $1/e$ (see Sec. \ref{subsec:meth_int}). The change in the hyperbolic elements is given for any $k$ (with $k\geq 2$) by the following expressions
\begin{subequations}\label{eq:Delta_guauss_nonlin}
\begin{align}
\Delta e_{(k)}&=-\frac{\eta^{k-1}}{(k-1)!\,e^{k-1}}\frac{\alpha_{(k)}}{\alpha^k}\int_{f_1}^{f_2}\cos^{k-1}f\sin f\ \dd f\text{,}\label{eq:Delta_guauss_nonlin_e}\\
\Delta\omega_{(k)}&=\frac{\eta^{k-1}}{(k-1)!\,e^{k}}\frac{\alpha_{(k)}}{\alpha^k}\int_{f_1}^{f_2}\cos^{k}f\ \dd f\text{,}\label{eq:Delta_guauss_nonlin_omega}\\
\Delta s_{(k)}&=\frac{(k+1)\eta^{k-2}}{k!\,e^{k-1}}\frac{\alpha_{(k)}}{\alpha^{k-1}}\mu\int_{f_1}^{f_2}\cos^{k-2}f\ \dd f\text{,}\label{eq:Delta_guauss_nonlin_s}\\
\Delta t_{(k)}&=\frac{(k+2)\eta^{k-1}}{k!\,e^{k-1}}\frac{\alpha_{(k)}}{\alpha^{k-1}}\frac{\mu}{c}\int_{f_1}^{f_2}\cos^{k-2}f\ \dd f\text{,}\label{eq:Delta_guauss_nonlin_t}\\
\Delta \psi_{(k)}&=\Delta\omega_{(k)}\text{,}\label{eq:Delta_guauss_nonlin_psi}
\end{align}
\end{subequations}
where we have kept the leading order in $1/e$, and where $f_2\geq f_1$.

Considering e.g. the quadratic dependence with the gravitational potential in the index of refraction profile ($k=2$), we see that the evolution of the hyperbolic elements all along the light path is given by
\begin{subequations}\label{eq:Delta_guauss_nonlin1}
\begin{align}
e(f)&=\frac{\eta}{2e}\frac{\alpha_{(2)}}{\alpha^2}\cos^2f\text{,}\label{eq:Delta_guauss_nonlin_e1}\\
\omega(f)&=\frac{\eta}{2e^2}\frac{\alpha_{(2)}}{\alpha^2}(f+\cos f\sin f)\text{,}\label{eq:Delta_guauss_nonlin_omega1}\\
\delta s(f)&=\frac{3}{2e}\frac{\alpha_{(2)}}{\alpha}\mu f\text{,}\label{eq:Delta_guauss_nonlin_s1}\\
\delta t(f)&=\frac{2\eta}{e}\frac{\alpha_{(2)}}{\alpha}\frac{\mu}{c}f\text{,}\label{eq:Delta_guauss_nonlin_t1}\\
\delta\psi(f)&=\omega(f)\text{.}\label{eq:Delta_guauss_nonlin_psi1}
\end{align}
\end{subequations}
All those equations are defined within a constant which is obviously given by the initial condition of the problem at the level of the transmitter, so Eqs.~\eqref{eq:Delta_guauss_nonlin1} are valid for all $f\geq f_1$. 

\subsection{Steady rotating atmosphere}
\label{subsec:detail_steadrot}

The term $\Phi_C$ in Eq.~\eqref{eq:gen_pot} is the centrifugal potential due to proper rotation of the fluid body and is given by \cite{1966tsga.book.....K,2014grav.book.....P}
\begin{equation}
\label{eq:centrifuge_pot}
\Phi_C=\frac{w^2h^2}{2}\left[(\hat{\bm{\rho}}\cdot\hat{\mathbf{X}}')^2+(\hat{\bm{\rho}}\cdot\hat{\mathbf{Y}}')^2\right]\text{.}
\end{equation}
At linear order in the generalized potential, the index of refraction and its gradient are expressed in the frame comoving with the fluid as
\begin{subequations}\label{eq:ndn_centr}
\begin{align}
n&=n_0+\frac{\alpha w^2h^2}{2}\left[(\hat{\bm{\rho}}\cdot\hat{\mathbf{X}}')^2+(\hat{\bm{\rho}}\cdot\hat{\mathbf{Y}}')^2\right]\text{,}\label{eq:n_PhiC}\\
\textbf{f}_{\text{pert}}&=\alpha w^2h\Big[(\hat{\bm{\rho}}\cdot\hat{\mathbf{X}}')\,\hat{\mathbf{X}}'+(\hat{\bm{\rho}}\cdot\hat{\mathbf{Y}}')\,\hat{\mathbf{Y}}'\Big]\text{.}\label{eq:dn_PhiC} 
\end{align}
\end{subequations}
Once $\hat{\textbf{X}}'$ and $\hat{\textbf{Y}}'$ are projected into the polar basis, we obtain the following components 
\begin{subequations}\label{eq:f_RTSN_PhiC}
\begin{align}
\mathcal{R}&=\frac{\alpha w^2p}{4}\frac{\big[3+\cos 2\iota+2\sin^2\iota\cos 2(f+\omega)\big]}{1+e\cos\kappa f}\text{,}\label{eq:f_R_PhiC}\\
\mathcal{T}&=-\frac{\alpha w^2p}{2}\frac{\sin 2(f+\omega)}{1+e\cos\kappa f}\sin^2\iota\text{,}\label{eq:f_T_PhiC}\\
\mathcal{S}&=-\frac{\alpha w^2p}{2}\frac{\sin(f+\omega)}{1+e\cos\kappa f}\sin 2\iota\text{,}\label{eq:f_S_PhiC}\\
\mathcal{N}&=\frac{\alpha w^2p}{8n}\frac{\big[3+\cos 2\iota+2\sin^2\iota\cos 2(f+\omega)\big]}{(1+e\cos\kappa f)^2}\text{.}\label{eq:f_N_PhiC}
\end{align}
\end{subequations}
Inserting Eqs. \eqref{eq:f_RTSN_PhiC} into Eqs. \eqref{eq:guauss_pert}, expending them in power of $1/e$ and performing the integration, allows one to find, at first order in $1/e$, that the centrifugal contribution impacts all the hyperbolic elements, such
\begin{subequations}\label{eq:Delta_guauss_PhiC}
\begin{align}
p(f)&=-\frac{e^4}{3\eta^4}\alpha^4\mu^3w^2\sin^2\iota\big(3\cos 2\omega+\sin 2\omega\sec f\sin 3f\big)\sec^2 f\text{,}\label{eq:Deltap_PhiC}\\
e(f)&=\frac{e^3}{24\eta^3}\alpha^3\mu^2w^2\Big(9+3\cos 2\iota-\sin^2\iota\big[18\cos 2\omega+\sin 2\omega(3\sin f+7\sin 3f)\sec f\big]\Big)\sec^2 f\text{,}\label{eq:Deltae_PhiC}\\
\iota(f)&=-\frac{e^2}{12\eta^3}\alpha^3\mu^2w^2\sin 2\iota\big(3\cos 2\omega+\sin 2\omega\sec f\sin 3f\big)\sec^2 f\text{,}\label{eq:Deltai_PhiC}\\
\Omega(f)&=-\frac{e^2}{12\eta^3}\alpha^3\mu^2w^2\cos\iota\big(3\sin f+6\sin 2\omega\cos f+[1-2\cos 2\omega]\sin 3f\big)\sec^3 f\text{,}\label{eq:DeltaOmega_PhiC}\\
\omega(f)&=\frac{e^2}{12\eta^3}\alpha^3\mu^2w^2\cos\omega\Big(3\sin\omega\big[3+\cos 2\iota\big]-\cos\omega\big[5+7\cos 2f-2\cos 2\iota\sin^2f\big]\tan f\Big)\sec^2 f\text{,}\label{eq:Deltaomega_PhiC}\\
\delta s(f)&=-\frac{e^3}{48\eta^4}\alpha^4\mu^3w^2\Big(\big[3+\cos 2\iota\big]\sec f\big[3\sin f+\sin 3f\big]\nonumber\\
&+6\sin^2\iota\sin 2\omega\cos 2f\sec^2 f+12\sin^2\iota\cos 2\omega\tan f\big]\Big)\sec^2 f\text{,}\label{eq:Deltas_PhiC}\\
\delta t(f)&=-\frac{e^3}{24c\eta^3}\alpha^4\mu^3w^2\sin^2\iota\Big(3\sin 2\omega\sec^4 f\big[1+2\cos 2 f\big]+8\cos 2\omega\tan^3 f\Big)\text{,}\label{eq:Deltat_PhiC}\\
\delta \psi(f)&=\frac{e^2}{4\eta^3}\alpha^3\mu^2w^2\Big(\sin 2\omega\sin^2\iota\sec^2f-\big[3+\cos 2\iota+2\cos 2\omega\sin^2\iota\big]\tan f\Big)\text{.}\label{eq:Deltapsi_PhiC}
\end{align}
\end{subequations}
All these solutions are defined within a constant which is given by the initial condition of the problem at the level of the transmitter, so Eqs. \eqref{eq:Delta_guauss_PhiC} are valid for any $f\geq f_1$.

We remind that the previous results only account for the contribution of the centrifugal potential and does not consider the dragging effect due to the moving medium. To quantify it, let us express Eq. \eqref{eq:f_drag} as
\begin{align}
\mathbf{f}_{\text{drag}}&=2\frac{w h}{c}\Big[(\hat{\bm{\rho}}\cdot\hat{\mathbf{X}})\,\hat{\mathbf{Y}}-(\hat{\bm{\rho}}\cdot\hat{\mathbf{Y}})\,\hat{\mathbf{X}}\Big](\bm\nabla\mathscr{S}\cdot\bm{\nabla}n)\nonumber\\
&+2\frac{w h}{c}\Big[(\hat{\bm{\rho}}\cdot\hat{\mathbf{Y}})(\bm\nabla\mathscr{S}\cdot\hat{\mathbf{X}})-(\hat{\bm{\rho}}\cdot\hat{\mathbf{X}})(\bm\nabla\mathscr{S}\cdot\hat{\mathbf{Y}})\Big]\bm{\nabla}n\nonumber\\
&+2\gamma n\frac{w}{c}\Big[(\bm\nabla\mathscr{S}\cdot\hat{\mathbf{Y}})\,\hat{\mathbf{X}}-(\bm\nabla\mathscr{S}\cdot\hat{\mathbf{X}})\,\hat{\mathbf{Y}}\Big]\text{,}\label{eq:f_drag_i}
\end{align}
in which, we have assumed a steady rotating atmosphere such that $\mathbf{v}=\textbf{w}\times\mathbf{h}$. If we express the dragging contribution in the polar basis, we obtain the following non-null components
\begin{subequations}\label{eq:RTS_drag}
\begin{align}
\mathcal{R}&=2\frac{\sqrt{\eta}}{\kappa}\frac{w}{c}\sqrt{\frac{\alpha\mu}{p}}(n\gamma+N)\cos\iota(1+e\cos\kappa f)\text{,}\label{eq:R_C_drag}\\
\mathcal{T}&=-2e\sqrt{\eta}\ \frac{w}{c}\sqrt{\frac{\alpha\mu}{p}}(n\gamma+N)\cos\iota\sin\kappa f\text{,}\label{eq:T_C_drag}\\
\mathcal{S}&=-2\frac{\sqrt{\eta}}{\kappa}\frac{w}{c}\sqrt{\frac{\alpha\mu}{p}}\sin\iota\Big[n\gamma(1+e\cos\kappa f)\sin(f+\omega)-e\kappa(n\gamma+N)\cos(f+\omega)\sin\kappa f\Big]\text{,}\label{eq:S_C_drag}
\end{align}
\end{subequations}
where $N$ is the refractivity defined as $N=n-\eta$. We notice the absence of the $\mathcal{N}$-component of the perturbation.

The effect of the fluid rotation on the light propagation is assessed by inserting those components into Eqs. \eqref{eq:guauss_pert}, by keeping the first order in $1/e$, and then by performing the integration over the true anomaly
\begin{subequations}\label{eq:Delta_guauss_drag}
\begin{align}
p(f)&=-\frac{2\gamma e^3}{\eta}\frac{(\alpha\mu)^2w}{c}\cos\iota\sec^2 f\text{,}\label{eq:Deltap_drag}\\
e(f)&=-\gamma e^2\frac{\alpha\mu w}{c}\cos\iota\sec^2 f\text{,}\label{eq:Deltae_drag}\\
\iota(f)&=\frac{\gamma e}{2}\frac{\alpha\mu w}{c}\sin^2\iota\sin 2\omega\big(\tan\omega\sec^2f-2\tan f\big)\text{,}\label{eq:Deltai_drag}\\
\Omega(f)&=-\frac{\gamma e}{2}\frac{\alpha\mu w}{c}\sin 2\omega\big(\sec^2f+2\tan\omega\tan f\big)\text{,}\label{eq:DeltaOmega_drag}\\
\omega(f)&=\gamma e\frac{\alpha\mu w}{c}\cos\iota\cos^2\omega\big(\tan\omega\sec^2f-2\tan f\big)\text{,}\label{eq:Deltaomega_drag}\\
\delta s(f)&=-\frac{2\gamma e^2}{3\eta}\frac{(\alpha\mu)^2w}{c}\cos\iota(2+\cos 2f)\sec^2 f\tan f\text{,}\label{eq:Deltas_drag}\\
\delta t(f)&=\frac{\eta}{c}\delta s(f)\text{,}\label{eq:Deltat_drag}\\
\delta \psi(f)&=-2\gamma e\frac{\alpha\mu w}{c}\cos\iota\tan f\text{.}\label{eq:Deltapsi_drag}
\end{align}
\end{subequations}
Once again let us precise that these equations are defined within a constant which is determine with the transmitter's position. Eqs.~\eqref{eq:Delta_guauss_drag} are defined for $f\geq f_1$.

\subsection{Axisymmetric gravitational potential}
\label{subsec:detail_axisym}

A convenient way to handle non-spherical gravitational potential is to deal with either the spherical harmonic expansion \cite{1966tsga.book.....K,2000ssd..book.....M} or the symmetric and trace free (STF) tensors formalism \cite{1994CeMDA..60..139H,2014grav.book.....P}. Using e.g. the later one, it can be shown that the expansion of the non-spheric part of the gravitational potential is given by
\begin{equation}
\label{eq:pot_nonsphe}
U_{l}=-G\frac{(-1)^l}{l!}I^{<L>}\partial_{<L>}h^{-1}\text{,}
\end{equation}
for $l\geq 2$. We recall that $G$ is the gravitational constant, $I^{<L>}$ is the multipole moment STF tensors of the mass distribution, and $\partial_{<L>}$ is the STF tensor made with the partial derivatives with respect to \textbf{h}. We refer to \cite{1994CeMDA..60..139H} and \cite{2014grav.book.....P} for notations and definitions about the use of the STF tensors.

We remind that the unit vector $\hat{\mathbf{Z}}$ has been chosen in Sec.~\ref{subsec:app_def} for being aligned with the rotation axis of the extended body. We focus on the special case where the spatial changes in the rotation axis are supposed to be negligible during the light propagation event. Thus, we consider that the $\hat{\mathbf{Z}}$-direction is spatially fixed. In addition, we consider a fluid body in hydrostatic equilibrium rotating at a constant rate. Then, the resulting gravitational field is considered for being axisymmetric and independent of the longitude \cite{2014grav.book.....P}. This symmetry imposes the multipole moments STF tensors to be proportional to $\hat{Z}^{<L>}$ which is the STF tensor formed with the $\hat{\mathbf{Z}}$ unit vector. Hence, we have the relation
\begin{equation}
\label{eq:IL_STF}
I^{<L>}\equiv -MR^lJ_l\hat{Z}^{<L>}\text{,}
\end{equation}
where $M$ is the total mass of the extended body, $R$ its equatorial radius, and $J_l$ are its unitless multipole moments.

Making use of basic properties about STF tensors \cite{1994CeMDA..60..139H,2014grav.book.....P}, such
\begin{subequations}\label{eq:STF_prop}
\begin{align}
\partial_{<L>}h^{-1}&\equiv (-1)^{l}(2l-1)!!\hat{\rho}_{<L>}h^{-(l+1)}\text{,}\label{eq:STF_prop1}\\
\hat{Z}^{<L>}\hat{\rho}_{<L>}&\equiv \frac{l!}{(2l-1)!!}P_l(\chi)\text{,}\label{eq:STF_prop2}\\
\hat{Z}^{<L>}\hat{\rho}_{<jL>}&\equiv \frac{l!}{(2l+1)!!}\left(\frac{\dd P_{l+1}}{\dd\chi}\hat\rho_j-\frac{\dd P_l}{\dd\chi}\hat{Z}_j\right)\text{,}\label{eq:STF_prop3}
\end{align}
\end{subequations}
where $\chi=\hat{\bm\rho}\cdot\hat{\mathbf{Z}}$ and $P_l(\chi)$ are the well-known Legendre polynomials of degree $l$, we can rewrite the perturbing gravitational potential in Eq.~\eqref{eq:pot_nonsphe} in terms of the multipole moments $J_l$. Keeping the linear part in $\Phi$ in the development of the index of refraction [cf. Eq.~\eqref{eq:n(Phi)}] with $\Phi=U_0+\sum_lU_l$, we determine an expression for $n$ in term of the non-spherical part of the gravitational potential
\begin{equation}
\label{eq:n_nonsphe}
n=n_0-\frac{\alpha\mu}{h}\sum_{l=2}^{+\infty}J_l\left(\frac{R}{h}\right)^lP_l(\chi)\text{.}
\end{equation}
This profile differs from the spherical one [cf. Eq. \eqref{eq:n0}] by terms of the order $J_l$ when the height of the ray is at the same order of magnitude than the equatorial radius. We can now, determine the $j$th component of the gradient of the index of refraction which is given by
\begin{equation}
\label{eq:gradjn}
\partial_j\delta n=-\alpha\sum_{l=2}^{+\infty}\partial_jU_l\text{.}
\end{equation}
Differentiating Eq. \eqref{eq:pot_nonsphe} and using the previous properties about STF tensors [cf. Eq. \eqref{eq:STF_prop3}], we deduce
\begin{equation}
\label{eq:fj1}
\mathbf{f}_{\text{pert}}=\frac{\alpha\mu}{h^2}\sum_{l=2}^{+\infty}J_l\left(\frac{R}{h}\right)^l\left[\frac{\dd P_{l+1}}{\dd\chi}\hat{\bm\rho}-\frac{\dd P_l}{\dd\chi}\hat{\mathbf{Z}}\right]\text{.}
\end{equation}
In the polar basis the components of $\textbf{f}_{\text{pert}}$ are
\begin{subequations}\label{eq:f_RTS}
\begin{align}
\mathcal{R}&=\frac{\alpha\mu}{h^2}\sum_{l=2}^{+\infty}J_l\left(\frac{R}{h}\right)^l\left[\frac{\dd P_{l+1}}{\dd\chi}-(\hat{\mathbf{Z}}\cdot\hat{\bm\rho})\ \frac{\dd P_l}{\dd\chi}\right]\text{,}\label{eq:f_R}\\
\mathcal{T}&=-\frac{\alpha\mu}{h^2}\sum_{l=2}^{+\infty}J_l\left(\frac{R}{h}\right)^l(\hat{\mathbf{Z}}\cdot\hat{\bm\tau})\ \frac{\dd P_l}{\dd\chi}\text{,}\label{eq:f_T}\\
\mathcal{S}&=-\frac{\alpha\mu}{h^2}\sum_{l=2}^{+\infty}J_l\left(\frac{R}{h}\right)^l(\hat{\mathbf{Z}}\cdot\hat{\bm\sigma})\ \frac{\dd P_l}{\dd\chi}\text{.}\label{eq:f_S}
\end{align}
\end{subequations}
Usually, for modestly deformed bodies (e.g. the planets in the Solar System), the most important term in the developments \eqref{eq:n_nonsphe} and \eqref{eq:fj1} is the one proportional to the quadrupolar moment $J_2$. This parameter measures the rotational flattening due to the proper rotation of the extended body. From now, we restrict our attention to that single parameter.

Using the definition of Legendre polynomials
\begin{subequations}\label{eq:legendre}
\begin{align}
P_2(\chi)&=\frac{1}{2}(3\chi^2-1)\text{,}\label{eq:legendre_2}\\
P_3(\chi)&=\frac{1}{2}(5\chi^3-3\chi)\text{,}\label{eq:legendre_3}
\end{align}
\end{subequations}
with Eqs. \eqref{eq:rhotausigma_XYZ}, we can determine the change in the index of refraction as well as the components of the perturbing gradient
\begin{subequations}\label{eq:f_RTSN_J2}
\begin{align}
\mathcal{R}&=\frac{3}{2}J_2\frac{\alpha\mu}{h}\frac{R^2}{h^3}\big[3\sin^2\iota\sin^2(f+\omega)-1\big]\text{,}\label{eq:f_R_J2}\\
\mathcal{T}&=-\frac{3}{2}J_2\frac{\alpha\mu}{h}\frac{R^2}{h^3}\sin^2\iota\sin 2(f+\omega)\text{,}\label{eq:f_T_J2}\\
\mathcal{S}&=-\frac{3}{2}J_2\frac{\alpha\mu}{h}\frac{R^2}{h^3}\sin 2\iota\sin(f+\omega)\text{,}\label{eq:f_S_J2}\\
\mathcal{N}&=-\frac{1}{2}J_2\frac{\alpha\mu}{np}\frac{R^2}{h^3}\big[3\sin^2\iota\sin^2(f+\omega)-1\big]\text{.}\label{eq:n_J2}
\end{align}
\end{subequations}
Inserting those components into Eqs. \eqref{eq:guauss_pert} and applying Eq. \eqref{eq:int_hyp_elmts}, allows one to find the following estimations which are given at leading order in $1/e$
\begin{subequations}\label{eq:Delta_guauss_J2}
\begin{align}
p(f)&=\frac{\eta J_2}{2e}\frac{R^2}{\alpha\mu}\sin^2\iota\big[3\cos(f+2\omega)+\cos(3f+2\omega)\big]\text{,}\label{eq:Deltap_J2}\\
e(f)&=\frac{\eta^2 J_2}{32e^2}\frac{R^2}{(\alpha\mu)^2}\Big(4\big[1+2\sin^2\iota\cos 2\omega\big(4+3\cos 2f\big)+3\cos 2\iota\big]\cos^3f\nonumber\\
&-\sin^2\iota\sin 2\omega\big[30\sin f+17\sin 3f+3\sin 5f\big]\Big)\text{,}\label{eq:Deltae_J2}\\
\iota(f)&=\frac{\eta^2 J_2}{8e^3}\frac{R^2}{(\alpha\mu)^2}\sin 2\iota\big[3\cos(f+2\omega)+\cos(3f+2\omega)\big]\text{,}\label{eq:Deltai_J2}\\
\Omega(f)&=\frac{\eta^2 J_2}{4e^3}\frac{R^2}{(\alpha\mu)^2}\cos\iota\big[3\sin(f+2\omega)+\sin(3f+2\omega)-6\sin f\big]\text{,}\label{eq:DeltaOmega_J2}\\
\omega(f)&=\frac{\eta^2 J_2}{64e^3}\frac{R^2}{(\alpha\mu)^2}\Big(6\big[11+\cos 2\iota\big(17-8\cos 2\omega\big)+4\cos^2f\cos 2(f+\omega)\big]\sin f+\big(2+6\cos 2\iota\big)\sin 3f\nonumber\\
&-\cos 2\iota\big[42\sin 2\omega\cos f+19\sin(3f+2\omega)+3\sin (5f+2\omega)\big]\Big)\text{,}\label{eq:Deltaomega_J2}\\
\delta s(f)&=\frac{\eta J_2}{4e^2}\frac{R^2}{\alpha\mu}\Big(\big[8-3\sin^2\iota\big(4-\cos 2\omega\big)\big]\sin f+3\sin^2\iota\big[\sin 2\omega\cos f+\sin(3f+2\omega)\big]\Big)\text{,}\label{eq:Deltas_J2}\\
\delta t(f)&=\frac{\eta^2 J_2}{8e^2}\frac{R^2}{\alpha\mu c}\Big(\big[20-\sin^2\iota\big(30-9\cos 2\omega\big)\big]\sin f+\sin^2\iota\big[9\sin 2\omega\cos f+7\sin(3f+2\omega)\big]\Big)\text{,}\label{eq:Deltat_J2}\\
\delta\psi(f)&=\frac{\eta^2 J_2}{32e^3}\frac{R^2}{(\alpha\mu)^2}\Big(\big[1+3\cos 2\iota\big]\sin 3f+6\big[6-\sin^2\iota\big(9-4\cos 2\omega\big)\big]\sin f\nonumber\\
&+2\sin^2\iota\big(1+3\cos 2f\big)\big[3\sin(f+2\omega)+\sin(3f+2\omega)\big]\Big)\text{.}\label{eq:Deltapsi_J2}
\end{align}
\end{subequations}
These equations are defined within a constant which is determined from the initial conditions at the level of the transmitter. They are valid for all $f\geq f_1$.

\subsection{Tidal potential}
\label{subsec:detail_tide}

In the fluid rotating frame, the perturbing tidal potential can be expressed as \cite{2014grav.book.....P}
\begin{equation}
\label{eq:tidal_pot}
U_{\text{tidal}}=-\sum_{l=2}^{\infty}\frac{1}{l!}\left[h^L\mathcal{E}_L(t)-\frac{I^{<L>}}{M}h^j\mathcal{E}_{jL}(t)\right]\text{,}
\end{equation}
where $\mathcal{E}_L(t)$ are the tidal moments being time dependent STF tensors. They are defined by
\begin{equation}
\label{eq:tidal_mom}
\mathcal{E}_L(t)=-\partial_LU_{\neg P}(t,\bm{0})\text{,}
\end{equation}
where the gravitational potential created by the external bodies $U_{\neg P}$ is differentiated $l$ times with respect to $\mathbf{h}$ and the result is evaluated at the central planet's center-of-mass (i.e. $\mathbf{h}=\bm{0}$).

Following \cite{2014grav.book.....P}, $U_{\neg P}$ is given by
\begin{equation}
\label{eq:pot_ext_int}
U_{\neg P}(t,\mathbf{h})=-G\sum_{A\neq P}\int_A\frac{\rho(t,\mathbf{x})}{|\mathbf{h}+\mathbf{r}_P(t)-\mathbf{x}|}\dd^3x\text{,}
\end{equation}
in which the integration must be performed in the vicinity of the external bodies $A$ where $\rho_A(t,\mathbf{x})$ is supposed to be non-null. Thus, the dummy variable of integration will ranges over values close to $\mathbf{r}_A(t)$, and we can introduce $\mathbf{x}=\mathbf{r}_A(t)+[\mathbf{x}-\mathbf{r}_A(t)]$. Then, we are left with two vectors in the denominator, the first one, $\mathbf{h}+\mathbf{r}_{AP}(t)$ with $\mathbf{r}_{AP}(t)\equiv\mathbf{r}_P(t)-\mathbf{r}_A(t)$, representing the separation between the center-of-mass of $A$ and a point $\mathbf{h}$ at which the external potential must be evaluated, and the second one, $\mathbf{x}-\mathbf{r}_A(t)$, representing the separation between the center-of-mass of $A$ and a point lying in the vicinity of $A$. Therefore, because $\mathbf{h}+\mathbf{r}_{AP}(t)$ is usually of the order of the inter-body separation, it is appropriate to express the denominator under the integrand as a Taylor series expansion. After some algebra, we arrive to
\begin{equation}
\label{eq:pot}
U_{\neg P}(t,\mathbf{h})=-G\sum_{A\neq P}\sum_{l'=0}^{\infty}\frac{(-1)^{l'}}{l'!}I_A^{<L'>}\partial_{L'}|\mathbf{r}_{AP}+\mathbf{h}|^{-1}\text{,}
\end{equation} 
We recall that $I_A^{<L'>}$ means the multipole moment STF tensors of body $A$ and that the derivative must be taken with respect to $\mathbf{h}$ and then be evaluated at $P$'s center-of-mass (i.e. ${\mathbf{h}=\bm{0}}$). This notation is shorten noticing that $\partial_{L'}|\mathbf{r}_{AP}+\mathbf{h}|^{-1}=\partial_{L'}r_{AP}^{-1}$, where the derivative is taken at the extremity of the vector $\mathbf{r}_{AP}(t)$.

From Eq. \eqref{eq:pot}, \eqref{eq:tidal_pot}, and \eqref{eq:tidal_mom}, we can define the perturbing external tidal potential
\begin{align}
U_{\text{tidal}}&=-G\sum_{A\neq P}\sum_{l=2}^{\infty}\sum_{l'=0}^{\infty}\frac{(-1)^{l'}}{l!l'!}I_A^{<L'>}\Bigg[h^L\partial_{LL'}r_{AP}^{-1}-\frac{I^{<L>}}{M}h^j\partial_{jLL'}r_{AP}^{-1}\Bigg]\text{.}\label{eq:tidal_pot_gen}
\end{align}
That expression is very general since it takes into account the tidal effects due to the total gravitational potential (all multipole moments included) of all the external bodies acting on the total gravitational potential of the central planet.

At linear order in the generalized potential, the refractive profile and the $j$th component of the gradient of the non-constant change in the index of refraction are explicitly written as
\begin{subequations}\label{eq:ndn_tide}
\begin{align}
n&=n_0+\alpha G\sum_{A\neq P}\sum_{l=2}^{\infty}\sum_{l'=0}^{\infty}\frac{(-1)^{l'}}{l!l'!}I_A^{<L'>}\Bigg[h^L\partial_{LL'}r_{AP}^{-1}-\frac{I^{<L>}}{M}h^j\partial_{jLL'}r_{AP}^{-1}\Bigg]\text{,}\label{eq:n_tidal}\\
\partial_j\delta n&=\alpha G\sum_{A\neq P}\sum_{l=2}^{\infty}\sum_{l'=0}^{\infty}\frac{(-1)^{l'}}{l!l'!}I_A^{<L'>}\Bigg[h^L\partial_{jLL'}r_{AP}^{-1}-\frac{I^{<L>}}{M}h^k\partial_{jkLL'}r_{AP}^{-1}\Bigg]\text{.}\label{eq:gradn_tidal}
\end{align}
\end{subequations}

For the application that we are to consider now, we admit only one perturbing body, $A$, and we define the separation vector between $P$ and $A$ as $\mathbf{r}(t)\equiv -\mathbf{r}_{AP}(t)$. We consider that the central planet is close to spherical symmetry, also the coupling between its higher multipole moments $I^{<L>}$ and the other STF tensors can safely be neglected. In addition, we consider that the perturbing body is a satellite of the main planet $P$, and that $M_A\ll M$, with $M$ the mass of the central planet. Therefore we can safely consider only the monopole contribution of $A$, i.e. $l'=0$. In addition, assuming that inter-body distance is much more larger than the typical scale of the ray closest approach to $P$ (we are interested in the light propagation in the atmosphere of $P$), it is sufficient to keep only the leading quadrupole term in the development, i.e. $l=2$. For a matter of simplicity, we also consider that the satellite follows a circular orbit of radius $r$ which lies in the equatorial plane of the central planet. Then, in the reference frame, the direction of the position vector of the satellite is given at any time $t$ by
\begin{equation}
\label{eq:pos_vec_sat_inert}
\hat{\mathbf{r}}(t)\equiv\mathbf{r}(t)/r=\cos Nt\ \hat{\mathbf{X}}+\sin Nt\ \hat{\mathbf{Y}}\text{,}
\end{equation}
where the mean motion of the satellite around the central body is given by Kepler third law of motion $N\simeq(\mu/r^3)^{1/2}$ (For this application, $N$ is not the refractivity anymore). Making use of Eqs. \eqref{eq:rho_XYZ}-\eqref{eq:sigma_XYZ}, the orbital motion of the satellite is given, in the polar frame, by
\begin{align}
\hat{\mathbf{r}}(t)&=\big[\cos\lambda(t)\cos(f+\omega)+\cos\iota\sin\lambda(t)\sin(f+\omega)\big]\hat{\bm\rho}\nonumber\\
&-\big[\cos\lambda(t)\sin(f+\omega)-\cos\iota\sin\lambda(t)\cos(f+\omega)\big]\hat{\bm\tau}\nonumber\\
&-\sin\iota\sin\lambda(t)\hat{\bm \sigma}\text{,}\label{eq:pos_vec_sat_rot}
\end{align}
in which we have defined $\lambda(t)\equiv Nt-\Omega$ for being the longitude of the perturbing body as measured from the intersection between the propagation plane and the equatorial plane (longitude of the node); this angle remains within the equatorial plane of $P$ since the perturbing body describes an equatorial circular orbit.

Following all the simplifications, the change in the index of refraction and its gradient now read
\begin{subequations}\label{eq:ngradn_tidal_simp}
\begin{align}
\delta n&=\frac{\alpha\mu_A}{2}h^{kl}\partial_{kl}r_{AP}^{-1}\text{,}\label{eq:n_tidal_simp}\\
\partial_j\delta n&=\frac{\alpha\mu_A}{2}h^{kl}\partial_{jkl}r_{AP}^{-1}\text{.}\label{eq:gradn_tidal_simp}
\end{align}
\end{subequations}
Making use of properties about the STF tensors [cf. Eqs. \eqref{eq:STF_prop}], the multi-index derivatives now read
\begin{subequations}\label{eq:ngrad_rAP}
\begin{align}
\partial_{ij}r_{AP}^{-1}&=-\frac{1}{r^3}(\delta_{ij}-3\hat{r}_{ij})\text{,}\label{eq:ngrad_rAP2}\\
\partial_{ijk}r_{AP}^{-1}&=-\frac{3}{r^4}(\delta_{ij}\hat{r}_{k}+\delta_{ik}\hat{r}_{j}+\delta_{jk}\hat{r}_{i}-5\hat{r}_{ijk})\text{.}\label{eq:ngrad_rAP3}
\end{align}
\end{subequations}
In those expressions, the components of the unit vector directed along $\mathbf{r}$ are noted $\hat{r}_j\equiv r_j/r$ and can be inferred from Eq. \eqref{eq:pos_vec_sat_rot} in the polar basis.

We can now express the components of the perturbing gradient and the non-constant change in the index of refraction
\begin{subequations}\label{eq:f_sat_RSTN}
\begin{align}
\textbf{f}_{\text{pert}}&=-\frac{3}{2}\frac{\alpha\mu_A}{r}\frac{h^2}{r^3}\bm{\mathcal{A}}\text{,}\label{eq:f_sat_RST}\\
\mathcal{N}&=-\frac{1}{2}\frac{\alpha\mu_A}{np}\frac{h^2}{r^3}\mathcal{A}_{\mathcal{N}}\text{.}\label{eq:f_sat_N}
\end{align}
\end{subequations}
Here, we have introduced the unitless vector $\bm{\mathcal{A}}$ and the unitless parameter $\mathcal{A}_{\mathcal{N}}$ to simplify the notations. They are explicitly given by
\begin{subequations}\label{eq:A_vecp}
\begin{align}
\bm{\mathcal{A}}&=2(\hat{\mathbf{r}}\cdot\hat{\bm{\rho}})\hat{\bm{\rho}}+\hat{\mathbf{r}}-5(\hat{\mathbf{r}}\cdot\hat{\bm{\rho}})^2\hat{\mathbf{r}}\text{,}\label{eq:A_vec}\\
\mathcal{A}_{\mathcal{N}}&=1-3(\hat{\mathbf{r}}\cdot\hat{\bm{\rho}})^2\text{,}\label{eq:A_p}
\end{align}
\end{subequations}
and once expressed in term of the fundamental angles of the problem they read
\begin{subequations}\label{eq:RTSNC}
\begin{align}
\mathcal{A}_{\mathcal{R}}&=\bm{\mathcal{A}}\cdot\hat{\bm{\rho}}=\big[\cos(f\!+\!\omega)\cos\lambda+\cos\iota\sin(f\!+\!\omega)\sin\lambda\big]\Big(3\!-\!5\big[\cos(f\!+\!\omega)\cos\lambda+\cos\iota\sin(f\!+\!\omega)\sin\lambda\big]^2\Big)\text{,}\label{eq:RC}\\
\mathcal{A}_{\mathcal{T}}&=\bm{\mathcal{A}}\cdot\hat{\bm{\tau}}=\big[\cos\iota\cos(f\!+\!\omega)\sin\lambda-\sin(f\!+\!\omega)\cos\lambda\big]\Big(1\!-\!5\big[\cos(f\!+\!\omega)\cos\lambda+\cos\iota\sin(f\!+\!\omega)\sin\lambda\big]^2\Big)\text{,}\label{eq:TC}\\
\mathcal{A}_{\mathcal{S}}&=\bm{\mathcal{A}}\cdot\hat{\bm{\sigma}}=-\sin\iota\sin\lambda\Big(1\!-\!5\big[\cos(f\!+\!\omega)\cos\lambda+\cos\iota\sin(f\!+\!\omega)\sin\lambda\big]^2\Big)\text{,}\label{eq:SC}\\
\mathcal{A}_{\mathcal{N}}&=1\!-\!3\big[\cos(f\!+\!\omega)\cos\lambda+\cos\iota\sin(f\!+\!\omega)\sin\lambda\big]^2\text{.}\label{eq:NC}
\end{align}  
\end{subequations}

One can now insert Eqs. \eqref{eq:f_sat_RSTN} into Eqs. \eqref{eq:guauss_pert}. Before integrating, we are to further simplify the problem by assuming a ray traveling in the equator of the central planet and passing through the bulge raised by the perturbing body. In such a case, $\iota=0$ and the equation can be further simplified. However, as discussed in Sec.~\ref{subsec:eq_osc}, when the inclination is null the longitude of the node and the argument of closest approach are undefined. Therefore, instead of considering $\omega$ and $\Omega$ separately, we consider the longitude of the closest approach, $\varpi=\omega+\Omega$, which is measured from the $\hat{\textbf{X}}$-direction. Keeping the leading order in $1/e$, and integrating over the true anomaly leads to the following set of equations describing the change in hyperbolic elements following a gravitational tidal stress
\begin{subequations}\label{eq:delta_gauss_sat}
\begin{align}
p(f)&=-\frac{e^{5}}{32\eta^5}\frac{(\alpha\mu)^5}{r^4}\frac{\mu_A}{\mu}\sec^4f\big[15\cos(4f+3Nt-3\varpi)-15\cos(4f-3Nt+3\varpi)\nonumber\\
&+\cos(4f+Nt-\varpi)-\cos(4f-Nt+\varpi)+60\cos(2f+3Nt-3\varpi)+60\cos(2f-3Nt+3\varpi)\nonumber\\
&+4\cos(2f+Nt-\varpi)-4\cos(2f-Nt+\varpi)+90\cos(3Nt-3\varpi)+6\cos(Nt-\varpi)\big]\text{,}\label{eq:delta_gauss_sat_p}\\
e(f)&=-\frac{e^4}{64\eta^4}\frac{(\alpha\mu)^4}{r^4}\frac{\mu_A}{\mu}\sec^4f\big[25\cos(4f+3Nt-3\varpi)-25\cos(4f-3Nt+3\varpi)\nonumber\\
&+\cos(4f+Nt-\varpi)-\cos(4f-Nt+\varpi)+100\cos(2f+3Nt-3\varpi)+80\cos(2f-3Nt+3\varpi)\nonumber\\
&+4\cos(2f+Nt-\varpi)-16\cos(2f-Nt+\varpi)+150\cos(3Nt-3\varpi)-6\cos(Nt-\varpi)\big]\text{,}\label{eq:delta_gauss_sat_e}\\
\varpi(f)&=-\frac{e^3}{16\eta^4}\frac{(\alpha\mu)^4}{r^4}\frac{\mu_A}{\mu}\sec^3 f\big[5\sin(3f+3Nt-3\varpi)+5\sin(3f-3Nt+3\varpi)+6\sin(f+Nt-\varpi)\nonumber\\
&+2\sin(3f+Nt-\varpi)+2\sin(3f-Nt+\varpi)+15\sin(f+3Nt-3\varpi)-15\sin(f-3Nt+3\varpi)\big]\text{,}\label{eq:delta_gauss_sat_opO}\\
\delta s(f)&=-\frac{e^4}{320\eta^5}\frac{(\alpha\mu)^5}{r^4}\frac{\mu_A}{\mu}\sec^5 f\big[15\sin(5f+3Nt-3\varpi)+15\sin(5f-3Nt+3\varpi)\nonumber\\
&+7\sin(5f+Nt-\varpi)+7\sin(5f-Nt+\varpi)+75\sin(3f+3Nt-3\varpi)+75\sin(3f-3Nt+3\varpi)\nonumber\\
&+35\sin(3f+Nt-\varpi)+35\sin(3f-Nt+\varpi)+150\sin(f+3Nt-3\varpi)-150\sin(f-3Nt+3\varpi)\nonumber\\
&+70\sin(f+Nt-\varpi)+10\sin(f-Nt+\varpi)\big]\text{,}\label{eq:delta_gauss_sat_s}\\
\delta t(f)&=\frac{\eta}{c}\delta s(f)\text{,}\label{eq:delta_gauss_sat_t}\\
\delta\psi(f)&=\varpi(f)\text{.}\label{eq:delta_gauss_sat_psi}
\end{align}
\end{subequations}
These equations are defined within a constant which is determined from the initial conditions at the level of the transmitter. They are valid for all $f\geq f_1$. 

In order to account for the non-elastic response of the fluid following a tidal stress, it is shown in \cite{2014grav.book.....P} that the first order effects can be assessed by replacing $\mathcal{E}_L(t)$ by $\mathcal{E}_L(t-\tau)$ where $\tau$ is called the time-lag and is function of the Energy dissipation. A computation of $\mathcal{E}_L$ at the instant $t-\tau$, changes the instantaneous position of the perturbing body [cf. Eq. \eqref{eq:pos_vec_sat_inert}] into a delayed position, $\mathbf{r}(t-\tau)$. Most of the time, $\tau$ is a small quantity compared to orbital characteristic time scale. Consequently, the orbital velocity of the perturbing body is usually too small to change significantly it's apparent position during the time interval $\tau$. In this condition, only internal process occurring on time scale faster than the orbital motion can significantly change $\mathcal{E}_L$. Among these process, the rigid rotation of the atmosphere appears to be a good candidate for fast rotators. The delayed position can therefore be approximated by
\begin{equation*}
\mathbf{r}(t-\tau)=\mathbf{r}(t)+w\tau r\big(\cos Nt\ \hat{\textbf{Y}}-\sin Nt\ \hat{\textbf{X}}\big)\text{,}
\end{equation*}
where $w$ is the magnitude of the angular velocity which has been defined to be aligned with the normal to the equator of reference. The delayed unit vector is given by
\begin{equation*}
\hat{\mathbf{r}}(t-\tau)=\hat{\mathbf{r}}(t)+w\tau\hat{\mathbf{r}}_*(t)\text{,}
\end{equation*}
where the star refers to the non-elastic contribution which is obtained by simple identification with the previous equation.

Then, it is easily shown that the non-elastic part in the components of the perturbing gradient and the non-constant change in the index of refraction are given at first order in $\tau$ by 
\begin{subequations}\label{eq:f_sat_RSTN_ne}
\begin{align}
\textbf{f}_{\text{pert}}^*&=-\frac{3}{2}w\tau\frac{\alpha\mu_A}{r}\frac{h^2}{r^3}\bm{\mathcal{A}}_*\text{,}\label{eq:f_sat_R_ne}\\
\mathcal{N}_*&=-\frac{1}{2}w\tau\frac{\alpha\mu_A}{np}\frac{h^2}{r^3}\mathcal{A}_{\mathcal{N}}^*\text{.}\label{eq:f_sat_N_ne}
\end{align}
\end{subequations}
As before, we have introduced the trigonometric coefficients $\bm{\mathcal{A}}^*$ and $\mathcal{A}_{\mathcal{N}}^*$ which are given by
\begin{subequations}\label{eq:A_vecp_ne}
\begin{align}
\bm{\mathcal{A}}_*&=2(\hat{\mathbf{r}}_*\cdot\hat{\bm{\rho}})\hat{\bm{\rho}}+\hat{\mathbf{r}}_*-5(\hat{\mathbf{r}}\cdot\hat{\bm{\rho}})\Big[(\hat{\mathbf{r}}\cdot\hat{\bm{\rho}})\hat{\mathbf{r}}_*+2(\hat{\mathbf{r}}_*\cdot\hat{\bm{\rho}})\hat{\mathbf{r}}\Big]\text{,}\label{eq:A_vec_ne}\\
\mathcal{A}_{\mathcal{N}}^*&=-6(\hat{\mathbf{r}}\cdot\hat{\bm{\rho}})(\hat{\mathbf{r}}_*\cdot\hat{\bm{\rho}})\text{.}\label{eq:A_p_ne}
\end{align}
\end{subequations}
One can notice from the form of Eqs. \eqref{eq:f_sat_RSTN_ne}, that
\begin{align*}
\cfrac{(\textbf{f}_{\text{pert}}^*\cdot\hat{\bm{\rho}})}{(\textbf{f}_{\text{pert}}\cdot\hat{\bm{\rho}})}&=w\tau\cfrac{\mathcal{A}_{\mathcal{R}}^*}{\mathcal{A}_{\mathcal{R}}}\text{,} & \text{(idem along }\hat{\bm{\tau}}\text{ and }\hat{\bm{\sigma}}\text{),}\\
\cfrac{\mathcal{N}_*}{\mathcal{N}}&=w\tau\cfrac{\mathcal{A}_{\mathcal{N}}^*}{\mathcal{A}_{\mathcal{N}}}\text{.}\\
\end{align*}
Then, it is immediate to show that the change in hyperbolic elements due to the non-elastic response of the atmosphere to tidal stress is
\begin{equation}
\frac{\textbf{C}_*}{\textbf{C}}\propto w\tau\text{,}
\label{eq:comp_tide_elnonel}
\end{equation}
where $\textbf{C}$ represents the vector of the hyperbolic elements.

\subsection{Horizontal gradients}
\label{subsec:detail_HG}

Let us consider the Earth centered frame $(\hat{\textbf{X}},\hat{\textbf{Y}},\hat{\textbf{Z}})$ with the $\hat{\textbf{Z}}$-axis directed through the North pole, the $(\hat{\textbf{X}},\hat{\textbf{Y}})$ plane superimposed within the Earth equator, and the $\hat{\textbf{X}}$-axis pointed through the intersection between the equator and the Greenwich meridian. We assume that the atmosphere is filled with a stationary medium with respect to the frame attached to the Earth, whence Eq.~\eqref{eq:optics} is valid in $(\hat{\textbf{X}},\hat{\textbf{Y}},\hat{\textbf{Z}})$. A station on Earth can be located thanks to its spherical coordinates $(\varphi_2,\lambda_2,R)$, with $R$ the Earth radius, $\varphi_2$ the colatitude (measured from the North pole), and $\lambda_2$ the azimuth of the station\footnote{Here, we consider the case of a receiving station but the same work can be applied for a transmitting station.}. Any point along the light ray trajectory can be located through its spherical coordinates $(\varphi,\lambda,h)$.

We are going to consider a spherical atmosphere with an hyperbolic radial dependency in the refractive profile. In addition to the radial dependency, we assume an horizontal variation of the group refractivity above the transmitting site. At linear order, we can expend the index of refraction around the site of observation as it is done in \cite{1997JGR...10220489C} or \cite{doi:10.1029/2006JB004834}
\begin{equation}
n(h,\varphi,\lambda)=n_0(h)+(\varphi-\varphi_2)n_{\text{NS}}+(\lambda-\lambda_2)n_{\text{WE}}\text{,}
\label{eq:n_hor_gad}
\end{equation}
where $n_0(h)$ is the hyperbolic index of refraction, and $n_{\text{NS}}$ and $n_{\text{WE}}$ are respectively given by $\partial n/\partial\varphi|_{h,\lambda}$ and $\partial n/\partial\lambda|_{h,\varphi}$. In other words, $n_{\text{NS}}$ represents the variation of the horizontal refraction along the North-South direction and $n_{\text{WE}}$ represents the variation of the horizontal refraction along West-East direction. Here, we aim at computing the effects on the hyperbolic elements due to $n_{\text{NS}}$ and $n_{\text{WE}}$.

The perturbative gradient is given by taking the gradient of $\delta n$, with $\delta n(\varphi,\lambda)=n(h,\varphi,\lambda)-n_0(h)$. Expressing the gradient in spherical coordinates we end up with
\begin{equation}
\textbf{f}_{\text{pert}}=\frac{n_{\text{NS}}}{h}\hat{\textbf{u}}_{\varphi}+\frac{n_{\text{WE}}}{h\sin\varphi}\hat{\textbf{u}}_{\lambda}\text{,}
\label{eq:fpert_hor_gad}
\end{equation}
where we have assumed that the radial variations of $n_{\text{NS}}$ and $n_{\text{WE}}$ are of second order. The two unit vectors $\hat{\textbf{u}}_{\varphi}$ and $\hat{\textbf{u}}_{\lambda}$ are part of the usual spherical vectorial basis which is given by the triad of vectors $(\hat{\textbf{u}}_{\varphi},\hat{\textbf{u}}_{\lambda},\hat{\bm{\rho}})$. It is seen that the perturbing gradient does indeed not possess any radial component if we neglect the radial variations of $n_{\text{NS}}$ and $n_{\text{WE}}$.

Now let us write the transverse and the normal components of the perturbing gradient by expressing $\hat{\textbf{u}}_{\varphi}$ and $\hat{\textbf{u}}_{\lambda}$ into the polar basis $(\hat{\bm{\rho}},\hat{\bm{\tau}},\hat{\bm{\sigma}})$.

For any arbitrary point, we can change from the spherical to the Cartesian coordinates thanks to two angles $\varphi$ and $\lambda$
\begin{subequations}\label{eq:rhovarphilambda}
\begin{align}
\hat{\bm{\rho}}&=\sin\varphi\cos\lambda\hat{\bm{X}}+\sin\varphi\sin\lambda\hat{\bm{Y}}+\cos\varphi\hat{\bm{Z}}\text{,}\label{eq:rho_horGrad_XYZ}\\
\hat{\textbf{u}}_{\varphi}&=\cos\varphi\cos\lambda\hat{\bm{X}}+\cos\varphi\sin\lambda\hat{\bm{Y}}-\sin\varphi\hat{\bm{Z}}\text{,}\label{eq:varphi_XYZ}\\
\hat{\textbf{u}}_{\lambda}&=-\sin\lambda\hat{\bm{X}}+\cos\lambda\hat{\bm{Y}}\text{.}\label{eq:lambda_XYZ}
\end{align}
\end{subequations}
The radial direction can be compared to Eq.~\eqref{eq:rho_XYZ} which reveals the following relationships
\begin{subequations}\label{eq:varphilambda}
\begin{align}
\cos\varphi&=\sin\iota\sin(f+\omega)\text{,}\label{eq:cosvarphi}\\
\cos\lambda\sin\varphi&=\cos\Omega\cos(f+\omega)-\cos\iota\sin\Omega\sin(f+\omega)\text{,}\label{eq:coslambda}\\
\sin\lambda\sin\varphi&=\sin\Omega\cos(f+\omega)+\cos\iota\cos\Omega\sin(f+\omega)\text{,}\label{eq:sinlambda}\\
%  \end{align}
%  and
%  \begin{align}
\sin\varphi&=\sqrt{\mathcal{A}_1(f)}\cos(f+\omega)\text{.}\label{eq:sinvarphi}
\end{align}
\end{subequations}
We have introduced the trigonometric function $\mathcal{A}_1(f)$ which is defined in Tab.~\ref{tab:trig_funct}. From Eqs.~\eqref{eq:rhovarphilambda} and \eqref{eq:rhotausigma_XYZ} and by making use of Eqs.~\eqref{eq:varphilambda}, we find
\begin{subequations}\label{eq:rhovarphilambda_rhotausigma}
\begin{align}
\hat{\textbf{u}}_{\varphi}&=-\frac{\sin\iota}{\sqrt{\mathcal{A}_1(f)}}\hat{\bm{\tau}}-\frac{2\cos\iota}{\sqrt{\mathcal{A}_3(f)}}\hat{\bm{\sigma}}\text{,}\label{eq:varphi_rhotausigma}\\
\hat{\textbf{u}}_{\lambda}&=\frac{2\cos\iota}{\sqrt{\mathcal{A}_3(f)}}\hat{\bm{\tau}}-\frac{\sin\iota}{\sqrt{\mathcal{A}_1(f)}}\hat{\bm{\sigma}}\text{,}\label{eq:lambda_rhotausigma}
\end{align}
\end{subequations}
where the $\mathcal{A}_l(f)$ coefficients are given in Tab.~\ref{tab:trig_funct}. Those equations allow us to express the perturbing gradient into the polar basis and allow us to derive its transverse and normal components.

At this level, we also need an equation for the change in the index of refraction in term of hyperbolic elements. That equation can be obtained from Eqs.~\eqref{eq:varphilambda}. Indeed at first order in $f-f_2$ we deduce
\begin{subequations}\label{eq:varphilambda1}
\begin{align}
\varphi-\varphi_2&=-\frac{\sin\iota\cos(f_2+\omega)}{\sqrt{\mathcal{A}_2(f_2)}}(f-f_2)\text{,}\label{eq:varphi1}\\
\lambda-\lambda_2&=\frac{\cos\iota}{\mathcal{A}_1(f_2)\cos^2(f_2+\omega)}(f-f_2)\text{,}\label{eq:lambda1}
\end{align} 
\end{subequations}
which both can be inserted into Eq.~\eqref{eq:n_hor_gad}. We sum up the results into the following relations (the radial component is null)
\begin{subequations}\label{eq:f_RTSN_horGrad}
\begin{align}
\mathcal{T}&=\Big[4n_{\text{WE}}\cos\iota-2n_{\text{NS}}\sqrt{\mathcal{A}_3(f)}\sin\iota\cos(f+\omega)\Big]\frac{(1+e\cos\kappa f)}{p\mathcal{A}_4(f)}\text{,}\label{eq:f_T_horGrad}\\
\mathcal{S}&=-\bigg[\frac{2n_{\text{NS}}\cos\iota}{\sqrt{\mathcal{A}_3(f)}}+\frac{n_{\text{WE}}\sin\iota}{\mathcal{A}_1(f)\cos(f+\omega)}\bigg]\frac{(1+e\cos\kappa f)}{p}\text{,}\label{eq:f_S_horGrad}\\
\mathcal{N}&=\bigg[\frac{n_{\text{WE}}\cos\iota}{\mathcal{A}_1(f_2)\cos^2(f_2+\omega)}-\frac{n_{\text{NS}}\sin\iota\cos(f_2+\omega)}{\sqrt{\mathcal{A}_2(f_2)}}\bigg]\frac{(f-f_2)}{np}\text{.}\label{eq:n_horGrad}
\end{align}
\end{subequations}

\begin{table}
$$ 
\begin{array}{lc}
\hline\hline
\noalign{\smallskip}
l & \mathcal{A}_l(f) \\
\noalign{\smallskip}
\hline
\noalign{\smallskip}
1 & {1+\cos^2\iota\tan^2(f+\omega)} \\
2 & {1-\sin^2\iota\sin^2(f+\omega)} \\
3 & {3+\cos2\iota+2\sin^2\iota\cos2(f+\omega)} \\
4 & 3+\cos2(f+\omega)+2\cos2\iota\sin^2(f+\omega) \\
\noalign{\smallskip}
\hline\hline
\end{array}
$$ 
\caption[]{Definition of the trigonometric functions $\mathcal{A}_l(f)$ which are used in this section. We do not specify the dependencies in $\iota$ and $\omega$, since they are not expected to vary as quickly as $f$ at first order in the components of the perturbation.}
\label{tab:trig_funct}
\end{table}

Finally, the effect on the hyperbolic elements is given by inserting Eqs. \eqref{eq:f_RTSN_horGrad} into Eqs.~\eqref{eq:guauss_pert}. Before integrating we are to simplify the problem by assuming source tracking at the meridian of the site of observation. For this geometry, the propagation plane of photons is orthogonal to the plane attached to Earth's equator ($\iota=\pi/2$). In addition, the azimuth of the source is the same as the azimuth of the receiving station, and because the inclination is $\pi/2$, the longitude of the node is the same as the azimuth of both the source and the station ($\Omega=\lambda$). Keeping the leading order in $1/e$, and integrating over the true anomaly leads to the following expressions
\begin{subequations}\label{eq:delta_gauss_HG}
\begin{align}
p(f)&=-\frac{2e^2}{\eta^2}\alpha\mu n_{\text{NS}}\tan f\text{,}\label{eq:delta_gauss_HG_p}\\
e(f)&=-\frac{e}{\eta}n_{\text{NS}}(f+\tan f)\text{,}\label{eq:delta_gauss_HG_e}\\
\iota(f)&=-\frac{n_{\text{WE}}}{\eta}\tan f\text{,}\label{eq:delta_gauss_HG_i}\\
\Omega(f)&=\frac{n_{\text{WE}}}{\eta}\Big[\cot\omega\tan f+\Big(\ln|\cos( f+\omega)|-\ln|\cos f|\Big)\csc^2\omega\Big]\text{,}\label{eq:delta_gauss_HG_O}\\
\omega(f)&=\frac{n_{\text{NS}}}{\eta}\ln|\cos f|\text{,}\label{eq:delta_gauss_HG_o}\\
\delta s(f)&=-\frac{e}{2\eta^2}\alpha\mu n_{\text{NS}}\Big[1-(f_2-f)\sin 2f+(1+\cos 2f)\ln|\cos f|\Big]\sec^2f\text{,}\label{eq:delta_gauss_HG_s}\\
\delta t(f)&=-\frac{e}{2\eta c}\alpha\mu n_{\text{NS}}\Big[\sec^2f+4\Big(\ln|\cos f|-(f_2-f)\tan f\Big)\Big]\text{,}\label{eq:delta_gauss_HG_t}\\
\delta\psi(f)&=\omega(f)\text{.}\label{eq:delta_gauss_HG_psi}
\end{align}
\end{subequations}
These equations are defined within a constant which is determined from the initial conditions at the level of the transmitter. They are valid for all $f\geq f_1$. 

\section{Doppler shift and refractive bending}
\label{app:Doppler_bend}

In this chapter, we show how the Doppler frequency shift can be related to the refractive bending.

\subsection{Deviation from vacuum contribution}

We start by expressing Eq.~\eqref{eq:doppler_instant_sv} in an alternative way making use of the Doppler frequency shift in a vacuum, namely
\begin{equation*}
\left[\frac{\nu_2}{\nu_1}\right]_{\text{vac}}=\frac{\Gamma_2}{\Gamma_1}\frac{(1-\hat{\textbf{u}}\cdot\textbf{s}_{\text{vac}})_2}{(1-\hat{\textbf{u}}\cdot\textbf{s}_{\text{vac}})_1}\text{.}
\end{equation*}
Inserting the definition \eqref{eq:S} into Eq. \eqref{eq:doppler_instant_sv}, we arrive to
\begin{equation}
\label{eq:doppler_instant_sv_bis}
\frac{\nu_2}{\nu_1}=\left[\frac{\nu_2}{\nu_1}\right]_{\text{vac}}\frac{\left[1-n\cfrac{\hat{\textbf{u}}\cdot\delta \textbf{s}}{1-\hat{\textbf{u}}\cdot\textbf{s}_{\text{vac}}}-(n-1)\cfrac{\hat{\textbf{u}}\cdot\textbf{s}_{\text{vac}}}{1-\hat{\textbf{u}}\cdot\textbf{s}_{\text{vac}}}\right]_2}{\left[1-n\cfrac{\hat{\textbf{u}}\cdot\delta \textbf{s}}{1-\hat{\textbf{u}}\cdot\textbf{s}_{\text{vac}}}-(n-1)\cfrac{\hat{\textbf{u}}\cdot\textbf{s}_{\text{vac}}}{1-\hat{\textbf{u}}\cdot\textbf{s}_{\text{vac}}}\right]_1}\text{,}
\end{equation}
where $\textbf{s}_{\text{vac}}$ is the light beam direction in a vacuum and $\delta \textbf{s}\equiv \textbf{s}-\textbf{s}_{\text{vac}}$ is the deviation of the actual direction of the ray with respect to $\textbf{s}_{\text{vac}}$. In vacuum, $\delta \textbf{s}=\bm{0}$, and $n-1=0$, thus $\nu_2/\nu_1$ reduces to $[\nu_2/\nu_1]_{\text{vac}}$.

For application within the Solar System, Eq. \eqref{eq:doppler_instant_sv_bis} can be simplified considering the small velocity approximation ($|\hat{\textbf{u}}|\ll 1$) together with the low refractivity limit ($n-1\ll 1$). The later condition imposes that the deviation with respect to the light ray direction in a vacuum is small, also $|\delta\textbf{s}|\ll 1$, and substitutions into Eq. \eqref{eq:doppler_instant_sv_bis} leads to the following approximation
\begin{align*}
\frac{\nu_2}{\nu_1}\bigg/\left[\frac{\nu_2}{\nu_1}\right]_{\text{vac}}-1&\approx n_1\beta_1(\hat{\textbf{u}}_1\times\textbf{s}_{\text{vac}})\cdot\hat{\bm{\sigma}}_1+n_2\beta_2(\hat{\textbf{u}}_2\times\textbf{s}_{\text{vac}})\cdot\hat{\bm{\sigma}}_2+(n_1-1)\hat{\textbf{u}}_1\cdot\textbf{s}_{\text{vac}}-(n_2-1)\hat{\textbf{u}}_2\cdot\textbf{s}_{\text{vac}}\text{.}
\end{align*}
For an hyperbolic path, we can consider that $\hat{\bm{\sigma}}_1=\hat{\bm{\sigma}}_2=\hat{\bm{\sigma}}$ since $\textbf{K}$ is constant in direction overall the light path. The angle $\beta$ represents the elevation of the photon path above from the straight line connecting the transmitter and the receiver (see Fig. \ref{fig:path}).

For spherical symmetry, $\beta_2$ is proportional to $\beta_1$ by means of a trigonometric factor: $\beta_2=-\Lambda\beta_1$. This $\Lambda$-factor tends to be respectively zero or unity in the limiting cases where i) the receiver is at infinity or ii) the receiver and the transmitter are at equal distance from the center of mass ($\Lambda\rightarrow \{0;1\}$ when $h_2\rightarrow\{+\infty;h_1\}$). Moreover, from Fig. \ref{fig:path} it is seen that the total refractive bending can be expressed as $\epsilon\equiv \beta_2-\beta_1$, also we end up with
\begin{align}
\frac{\nu_2}{\nu_1}\bigg/\left[\frac{\nu_2}{\nu_1}\right]_{\text{vac}}-1&\approx-\frac{\epsilon n_1}{1+\Lambda}(\hat{\textbf{u}}_1\times\textbf{s}_{\text{vac}})\cdot\hat{\bm{\sigma}}+\frac{\Lambda\epsilon n_2}{1+\Lambda}(\hat{\textbf{u}}_2\times\textbf{s}_{\text{vac}})\cdot\hat{\bm{\sigma}}+(n_1-1)\hat{\textbf{u}}_1\cdot\textbf{s}_{\text{vac}}-(n_2-1)\hat{\textbf{u}}_2\cdot\textbf{s}_{\text{vac}}\text{.}\label{eq:DopplerShift_bis}
\end{align}
This equation shows that in first approximation, the Doppler shift due to atmospheric effects is controlled by the total refractive bending, $\epsilon$. As discussed in appendix \ref{sec:hyp}, $\epsilon$ is defined as the net change in the argument of the refractive bending, $\psi$, between the transmitter and the receiver ($\epsilon\equiv\psi_2-\psi_1$), where the exact expression of $\psi$ has been given in Eq. \eqref{eq:psi(F)}.

\subsection{Simplifications}

For one-way Earth AO experiments \cite{1997JGR...10223429K,1999AnGeo..17..122S}, the index of refraction at the level of the two spacecrafts orbiting the Earth can be considered for being unity ($n_1=n_2=1$), also the previous equation reduces to
\begin{equation}
\frac{\nu_2}{\nu_1}\bigg/\left[\frac{\nu_2}{\nu_1}\right]_{\text{vac}}-1\approx-\frac{\epsilon}{1+\Lambda}\Big([\hat{\textbf{u}}_1-\Lambda\hat{\textbf{u}}_2]\times\textbf{s}_{\text{vac}}\Big)\cdot\hat{\bm{\sigma}}\text{,}\label{eq:DopplerShift_bis_eqrth_occ}
\end{equation}
when $\hat{\bm{\sigma}}=\hat{\bm{\sigma}}_1=\hat{\bm{\sigma}}_2$.

For one-way planetary AO experiments (see e.g. \cite{1971AJ.....76..123F,1973P&SS...21.1521E}), the index of refraction at the level of the spacecraft can be taken for being unity ($n_1=1$). In addition, the receiver (DSN antenna on Earth) can safely be considered at infinity, also $\Lambda\rightarrow 0$, and the simplified Doppler shift expression becomes
\begin{equation}
\frac{\nu_2}{\nu_1}\bigg/\left[\frac{\nu_2}{\nu_1}\right]_{\text{vac}}-1\approx-\epsilon(\hat{\textbf{u}}_1\times\textbf{s}_{\text{vac}})\cdot\hat{\bm{\sigma}}-(n_2-1)\hat{\textbf{u}}_2\cdot\textbf{s}_{\text{vac}}\text{.}\label{eq:DopplerShift_bis_plan_occ}
\end{equation}
Most of the time, the second term in the right-hand side of the equation is neglected since it can be absorbed by fitting a constant offset in the Doppler frequency shift measurements.

\end{document}